\documentclass[twocolumn]{aastex6}

\graphicspath{./}

\newcommand{\Spitzer}{{\em Spitzer}}
\newcommand{\Chandra}{{\em Chandra}}
\newcommand{\um}{$\mu$m}
\newcommand{\hii}{\ion{H}{2}~}
\newcommand{\Msun}{\rm M_{\odot}}

\newcommand{\Lsun}{\rm L_{\odot}}

\newcommand{\lya}{\ifmmode {\rm Ly}\alpha \else Ly$\alpha$\fi}
\def\msunyr{\ifmmode M_{\odot} {\rm yr}^{-1} \else M$_{\odot}$ yr$^{-1}$\fi}

\begin{document}

\slugcomment{Accepted for publication in {\em The Astrophysical Journal}---17 January 2017.}

\shorttitle{X-ray Emitting Candidate OB Stars in MYStIX}

\title{Candidate X-ray-Emitting OB Stars in the MYStIX Massive Star-Forming Regions}

\author{Matthew S. Povich}

\affil{Department of Physics and Astronomy, California State Polytechnic University, 3801 West Temple Ave, Pomona, CA 91768 USA; mspovich@cpp.edu}

\author{Heather A. Busk, Eric D. Feigelson, and Leisa K. Townsley}

\affil{Department of Astronomy \& Astrophysics, The Pennsylvania State University, 
525 Davey Lab, University Park, PA 16802 USA}

\author{and Michael A. Kuhn}

\affil{Millennium Institute of Astrophysics, Vicu\~{n}a Mackenna 4860, 7820436 Macul, Santiago, Chile  \\ and Instituto de Fisica y Astronom\'{i}a, Universidad de Valpara\'{i}so, Gran Breta\~{n}a 1111, Playa Ancha, Valpara\'{i}so, Chile}

\begin{abstract}
Massive, O and early B-type (OB) stars remain incompletely catalogued in the nearby Galaxy due to high extinction, bright visible and infrared nebular emission in \hii regions, and high field star contamination. These difficulties are alleviated by restricting the search to stars with X-ray emission. Using the X-ray point sources from the Massive Young star-forming complex Study in Infrared and X-rays (MYStIX) survey of OB-dominated regions, we identify 98 MYStIX candidate OB (MOBc) stars by fitting their 1--8~\um\ spectral energy distributions (SEDs) with reddened stellar atmosphere models.  We identify 27 additional MOBc stars based on $JHK_S$ photometry of X-ray stars lacking SED fitting.
These candidate OB stars indicate that the current census of stars earlier than B1, taken across the 18 MYStIX regions studied, is less than 50\% complete.
We also fit the SEDs of 239 previously-published OB stars to measure interstellar extinction and bolometric luminosities, revealing six candidate massive binary systems and five candidate O-type (super)giants.
As expected, candidate OB stars have systematically higher extinction than previously-published OB stars.
Notable results for individual regions include: identification of the OB population of a recently discovered massive cluster in NGC 6357; an older OB association in the M17 complex; and new massive luminous O stars near the Trifid Nebula. In several relatively poorly-studied regions (RCW 38, NGC 6334, NGC 6357, Trifid, and NGC 3576), the OB populations may increase by factors of ${\ga}2$. 
\end{abstract}


\section{Introduction}

Massive ($M>8~\Msun$) stars have a profound influence on galactic processes  over cosmic history. Through their ultraviolet radiation, winds, and supernova remnants, massive stars ionize and stir up the interstellar medium (ISM), disrupt or trigger star formation, and drive galactic winds \citep{trigger1, Leitherer99, Smith06}.  They are likely the principal source of reionization in the early universe, and metal enrichment of the ISM from the first massive stars allowed the first generation of lower mass stars to form \citep{Bromm04, Fan06}.  The star formation rate (SFR) in other galaxies are estimated from spatially averaged measures of OB stellar radiation, such as recombination lines from \hii regions and infrared (IR) emission of heated dust \citep{Kennicutt12}.  Within our Galaxy, it is important that the census of massive ionizing stars be accurate; small systematic errors in massive stellar models or uncertainties in the high-mass tail of the initial mass function (IMF) can lead to large errors in derived SFRs \citep{chomiukpovich}.  

Our understanding of the properties of massive stars and their effect on their environments has been hampered by a highly incomplete census of Galactic OB stars, even in relatively nearby star forming regions.  Most of the known Galactic OB populations are either based on visible-light surveys that place stars on the upper main sequence of the Hertzsprung-Russell diagram,  or on indirect  measures such as the radio emission of \hii regions  \citep{Dalgarno72, WC89, Massey95, GOSC04}.  But visible-light surveys are severely limited by dust extinction while observed \hii region properties depend on the ionizing luminosity and (poorly-known) mass-loss rates of the ionizing stars, possible mechanical energy injected by supernovae, and the density distribution of nearby molecular clouds and photo-dissociation regions relative to the ionizing stellar population.  Surveys limited to active star forming regions may miss a significant fraction of massive stars that are ``runaways'' spatially dispersed from young stellar clusters by dynamical ejection \citep{K+01,runaway,Gvaramadze12,bowshocks_cat16}.  Accurate OB populations are also needed for comparative study of IMFs under different conditions and for understanding mass segregation in young stellar clusters \citep{IMF_2010ARAA}.  

The Massive Young stellar complex Study in Infrared and X-rays (MYStIX) project \citep{overview} provides an opportunity to improve the census of massive young stars in 18 star-forming regions within 4~kpc of the Sun.  X-rays are produced by shocks in the winds of OB stars; small scale shocks from wind instabilities produce soft X-rays while magnetically-confined wind shocks and colliding winds in close binaries often produce harder X-rays \citep{lucywhite80, mcws, collide}.  MYStIX X-ray point sources detected in archival {\em Chandra X-ray Observatory} data with bright IR counterparts can be candidate OB stars that escaped notice in previous visible-light surveys of star forming regions. Previous tabulations of X-ray selected candidate OB stars include ${\sim}30$ stars in RCW 38 \citep{Wolk06,Winston11}, 16 stars in the Trumpler 16 region of the Carina Nebula \citep{Sanchawala07},  24 stars in NGC 6357 \citep{wang_ngc6357}, 12 stars in M17 \citep{Broos07}, 41 stars in RCW~108 \citep{Wolk08}, 94 stars in the entire Carina Nebula complex \citep{povich11}, and 9 stars in the obscured cluster HM~1 \citep{Naze13}.   In most cases, the stars are identified photometrically as spatial coincidences of X-ray point sources with IR-bright stars (NIR; e.g.\ $K < 10$), but some studies have applied additional constraints on X-ray luminosities or IR spectral energy distributions (SEDs).  

Our primary catalog of 98 candidate OB stars was constructed using the SED-fitting methodology developed by \citet[][hereafter P11]{povich11} for the {\it Chandra} Carina Complex Project \citep{Townsley11}. We also present a secondary catalog of 27 OB candidates produced by analyzing the $JHK_S$ near-infrared (NIR) colors and magnitudes that lacked sufficient {\em Spitzer Space Telescope} mid-infrared (MIR) photometric data for SED fitting.
The 18 regions in the MYStIX project examined here vary greatly in their distance, degree of contamination by unrelated field stars, richness of stellar population, and thoroughness of previous study, so the numbers of new OB candidates relative to the size of the known OB population to vary strongly between the regions.  
The rest of this paper is organized as follows: The MYStIX data products and methodology for identifying candidate OB stars are presented in Section~\ref{method.sec}. The resulting lists of OB candidates, along with new, SED-based measurements of extinction and luminosity for previously-cataloged MYStIX OB stars, are presented in Section~\ref{results.sec}, with a general discussion and summary in Section~\ref{discussion.sec}. Details of the results for individual MYStIX star-forming complexes are discussed in Appendix~\ref{appendix}.

\section{Methodology} \label{method.sec}

\subsection{The MYStIX X-ray and Infrared Point Source Lists}
X-ray point source catalogs for the 18 MYStIX regions studied here were produced by \citet{Kuhn10}, \citet{mikexcat}, and \citet{MOXC} from archival {\em Chandra} Advanced CCD Imaging Camera \citep[ACIS,][]{ACIS} observations using the ACIS extract software package \citep{acisextract}. The point-source resolution varies from ${<}1\arcsec$ within a few arcmin of the {\em Chandra} image axis to several arcsec at larger off-axis angles toward the edges of the ACIS-I field. A summary of the various X-ray datasets, including number of $17\arcmin \times 17\arcmin$ ACIS-I images and integration times for each, is given by \citet{overview}.

MYStIX $JHK_s$ NIR photometry was obtained from images taken with the United Kingdom InfraRed Telescope (UKIRT) Wide-Field Camera or from the Two-Micron All Sky Survey (2MASS) point-source catalog \citep{2MASS}. Some of the data were obtained as part of the UKIRT Infrared Deep Sky Survey \citep[UKIDSS;][]{ukidss,UKIDSS-GPS}. The UKIRT data analysis, modified from the UKIDSS pipeline to treat crowding and nebulosity, is described by \citet{nirengland}. Compared to 2MASS, the UKIRT photometry provides higher spatial resolution and sensitivity but saturates on bright sources. 

{\em Spitzer} MIR photometry at 3.6, 4.5, 5.8, and 8.0~\um\ was provided either by the Galactic Legacy Mid-Plane Survey Extraordinaire \citep[GLIMPSE;][]{glimpse} or by \citet{mikemir}. The ${\sim}2\arcsec$ spatial resolution of the {\em Spitzer} Infrared Array Camera \citep[IRAC;][]{IRAC} is well-matched to the 2MASS resolution, while the ${<}1\arcsec$ UKIRT resolution is comparable to the on-axis ACIS spatial resolution.

Averaged across all MYStIX star-forming regions, 62\% of X-ray sources were matched to predominantly faint NIR or MIR point-sources with ${>}0.80$ counterpart probability \citep{xnmatch,Broos13}. Of these X-ray/IR matches, the vast majority (95\% of X-ray to NIR matches and 91\% of X-ray to MIR matches) have apparent separations of ${<}1\arcsec$. This gives projected physical separations of ${\le}2000$~AU for X-ray/IR matches in the majority of MYStIX regions that have heliocentric distances ${\le}2$~kpc (Table~\ref{clusterloc.tbl}). Chance alignments between physically unrelated X-ray sources and IR stars bright enough to be selected as OB candidates are thus expected to be exceedingly rare, and likely occur only in the cores of the densest MYStIX clusters or toward the edges of the ACIS-I fields where the {\em Chandra} resolution is degraded. It is possible that in unequal-mass binary/multiple systems, for example a B-type star with low-mass, T Tauri companions, the X-ray emission could be produced by the lower-mass component while the IR source is dominated by the primary \citep[e.g.][]{Evans_CCCP11}. Such cases may enable the identification of a new candidate OB star based on X-ray emission from a lower-mass companion.
While the large majority of NIR counterpart photometry to MYStIX X-ray sources comes from UKIRT, the majority of published OB stars and OB candidates presented here have 2MASS photometry replacing saturated UKIRT sources.

Applying a Bayesian statistical framework to the combined X-ray and IR datasets, \citet{Broos13} calculated the posterior probability that each MYStIX X-ray source was a young star associated with its parent star-forming region versus a foreground or background contaminant. The final product was the MYStIX Probable Complex Member (MPCM) catalog, which combined three overlapping samples: (1) X-ray sources with spatial distributions, variability, and/or IR counterpart properties consistent with pre-main-sequence (pre-MS) stars \citep{CCCP_class}; (2) young stellar objects (YSOs) with MIR excess emission consistent with circumstellar disks and/or envelopes \citep[regardless of an associated X-ray source; MIRES catalog from][]{pov_mires}; and (3) stars with published OB spectral types.\footnote{For the known OB stars with published spectral types but no X-ray detection in the MYStIX data, \citet{Broos13} did not list IR counterpart photometry. We have therefore identified 2MASS and \Spitzer\ sources corresponding to these massive stars using a 1\arcsec\ matching radius to the coordinates of the star listed in the MPCM catalog. This photometry was used only to analyze the known OB population (see Section \ref{pubOB.sec}); obviously it is irrelevant to our selection of {\em X-ray detected} OB candidates.}

Because the MPCM criteria were based on assumptions of X-ray source properties and NIR brightnesses for {\em lower-mass,} pre-MS stars \citep{CCCP_class,Broos13}, it is possible that some unpublished {\em massive} stars were not included as MPCMs or were misclassified as foreground stars because they are systematically brighter in the NIR and often have softer X-ray spectra.
We therefore searched the SED Classification of IR counterparts to MYStIX X-ray sources (SCIM-X) catalog \citep[Appendix A of][]{pov_mires}, for candidate OB stars, rather than the MPCM catalog. A few of the published OB stars are massive YSOs deeply embedded in their natal clouds and hence present themselves as very bright MIR excess sources because the OB stars themselves heat the surrounding dust. Only SCIM-X SEDs that did {\em not} show evidence of excess IR emission above the stellar photosphere at $\lambda \le 4.5~$\um\ (identified by SED\_flg $=-2$ or $-1$) were included in the search.
The resulting OB candidate lists hence will not include new candidate {\em embedded} massive stars, but will include {\em obscured} OB stars observed through high foreground dust extinction.


The target massive star-forming regions for this study are listed in Table~\ref{clusterloc.tbl}. These are 18 of the 20 MYStIX regions with Galactic coordinates, distance from the Sun, and earliest published spectral type obtained from \citet{overview}.  The Orion Nebula Cluster is omitted because it has been well-characterized, and the Carina Nebula OB populations are discussed by P11 and \citet{gagne}. 

\begin{deluxetable}{lrrccc}
\tablecaption{MYStIX Cluster Data \label{clusterloc.tbl}}
\tabletypesize{\footnotesize}
\centering
\tablehead{
\colhead{Name} & \colhead{l} & \colhead{b} & \colhead{Distance} & \colhead{Earliest} & \colhead{Max}  \\
 \colhead{ } & \colhead{(deg)} & \colhead{(deg)} & \colhead{(kpc)} &  \colhead{Sp.Ty.} & \colhead{$A_{V}$}
 }
\startdata
Flame Nebula & 206.53 & $-$16.35 & 0.414 & O8 & 35 \\ 
W40          & 28.79   & $+$03.48 & 0.5                & late O & 35 \\
RCW 36     & 265.08 & $+$01.40 & 0.7  & O8 & 25 \\
NGC 2264  & 202.96 & $+$02.22 & 0.913              & O7 & 30 \\
Rosette  Nebula & 206.31 & $-$02.08 & 1.33 & O4 & 35 \\
Lagoon   Nebula & 5.96     & $-$01.17  & 1.3        & O4 & 39 \\
NGC 2362 & 238.20 & $-$05.58  & 1.48                & O9 I & 12 \\
DR 21       & 81.68   & $+$00.54 & 1.5                & \nodata & 42 \\
RCW 38    & 268.03 & $-$00.98  & 1.7 & O5 & 21 \\
NGC 6334 & 351.16 & $+$00.70 & 1.7               & O6/7: & 40 \\
NGC 6357 & 353.01 & $+$00.89 & 1.7               & O3.5 & 40 \\
Eagle  Nebula & 16.95   & $+$00.79  & 1.75       & O4 & 40 \\
M17         & 15.05    & $-$00.67  & 2.0 & O4 & 40 \\
W3           & 133.95  & $+$01.07 & 2.04 & O5 & 30 \\
W4           & 134.73   & $+$00.92 & 2.04            & O4 & 15 \\
Trifid Nebula & 7.00     & $-$00.25 & 2.7                & O7.5 & 20 \\
NGC 3576 & 291.27 & $-$00.71 & 2.8                & O3-6?  & 20 \\
NGC 1893 & 173.58 & $-$01.68 & 3.6 & O5 & 15 \\
\enddata
\end{deluxetable}

\subsection{Selection of Candidate OB Stars Via SED Fitting}\label{sec:OBc_SED}


Our primary OB candidate selection methodology follows the procedures described in P11. \citet{pov_mires} fit a family of \citet{kurucz} ATLAS9 stellar atmosphere models to the IR SEDs using the weighted least-squares (minimum $\chi^2$) fitting tool of \cite{ysomodel2}. The extinction law of \citet{I05}, parameterized by the visual extinction $A_V$, was used to apply interstellar reddening to the model spectra as part of the SED fitting process.
To be considered for OB candidacy, each star must have measurement in $N_{\rm data} \ge 4$ of the 7 available photometric bands ($J$, $H$, $K_s$, [3.6], [4.5], [5.8], and [8.0]) including $K_s\le 14.6$~mag. The fractional uncertainties on [5.8] and [8.0] flux densities were set to a minimum 10\%, and the other bands to a minimum 5\%, to avoid underestimating systematic errors in the fitting process \citep{pov_mires}.

The statistical criterion for acceptable model fits was $\chi^2_i-\chi^2_0 \leq N_{\rm data}$, with $\chi^2_0 \leq N_{\rm data}$ \citep{pov_mires}, where $\chi^2_i$ and $\chi^2_0$ are the goodness-of-fit parameter for the $i$th and the best-fit models, respectively. This is a more stringent criterion than the $\chi_i^2-\chi_0^2 \leq 2N_{\rm data}$ used by P11, and would have removed six stars from their list of 94 OB candidates. All of the models that fit the data with $\chi_i^2$ values within this limit were kept, so a given SED will have a set of many well-fit models. The range of the $A_{V}$ parameter accepted for our analysis was further restricted below a maximum value for each MYStIX region (Table~\ref{clusterloc.tbl}), which we estimated from $JHK_s$ color-color diagrams. 
Bolometric luminosities are based on SED fits to the IR photometry and the assumed distance to each MYStIX star forming complex (Table~\ref{clusterloc.tbl}).   The results of this fitting process are a set of self-consistent and well-fit models parameterized by effective temperature ($T_{\rm eff}$), bolometric luminosity ($L_{\rm bol}$) and their corresponding $A_{V}$ values.  

\begin{figure}
\epsscale{1.}
  \plotone{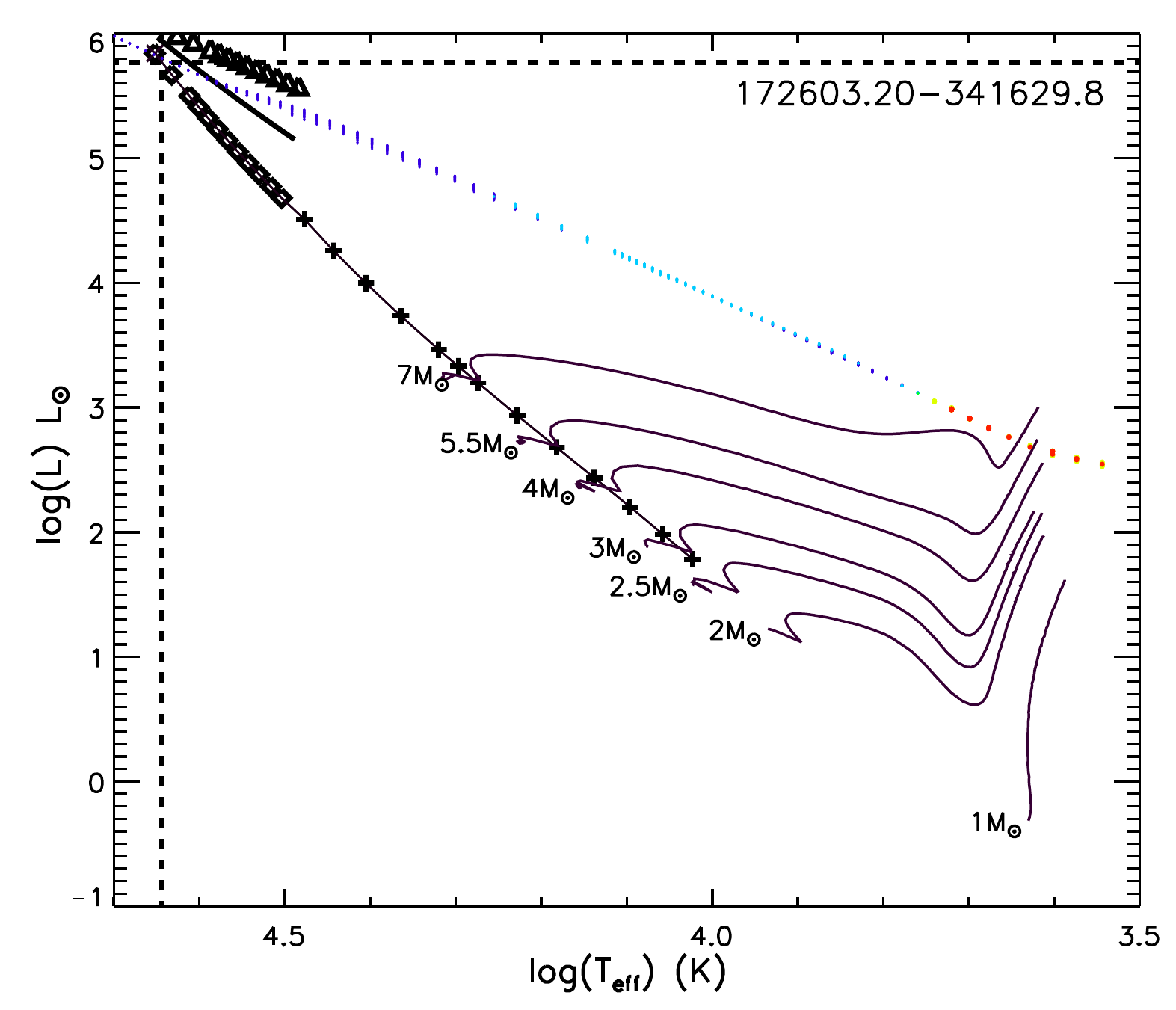} \\
  \plotone{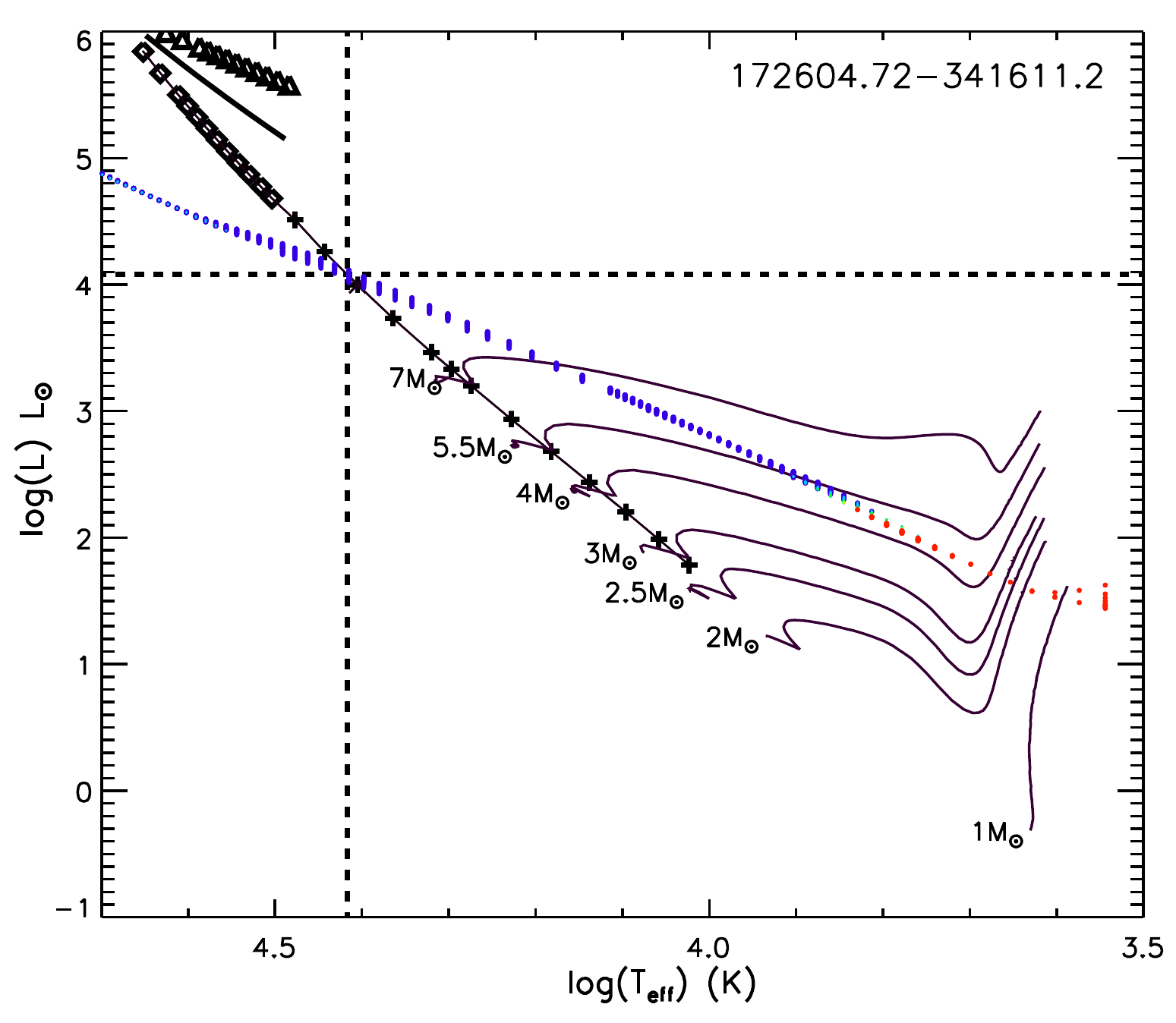}  
\caption{Two examples of the H-R diagram analysis used for selecting candidate OB stars from SED fitting results. Each panel shows the loci of all acceptable model fits to a single IR SED, plotted as overlapping points color-coded by lowest (blue) to highest (red) $\chi^2$ values. The top panel shows MOBc NGC 6357 21, one of the most luminous candidates in our sample, while the bottom panel shows MOBc NGC 6357 25, which exemplifies a star meeting the minimum luminosity criterion for selection.
The intersection point of these models with the theoretical massive ZAMS (solid curve with diamonds and $+$ symbols marking the O and B spectral subclasses, respectively) is marked with dashed cross-hairs. The thick, black line and triangles trace the theoretical O giant and supergiant loci,  respectively, from \citet{MSH05}. The solid purple curves are pre-MS evolutionary tracks from \citet{Siess}. 
\label{hr} }
\end{figure}

The set of well-fit $T_{\rm eff}$ and $L_{\rm bol}$ model parameters for each SED in the SCIM-X catalog was plotted on a theoretical H-R diagram, as illustrated in Figure~\ref{hr}. The intersection point of the locus of model parameters with the theoretical OB zero-age main sequence (ZAMS) \citep{dJN87,MSH05} is marked with cross-hairs. A star is selected as an OB candidate if the luminosity at the intersection point is greater than our ``cutoff luminosity'' of $1\times10^4~ \Lsun$, corresponding approximately to a main-sequence spectral type of B1 V or earlier (see also Figure 2 of P11).  

Two caveats must be borne in mind when interpreting these SED fitting results. The inferred $L_{\rm bol}$ depends on the assumption that the star lies at the (possibly incorrect) distance of the parent MYStIX complex listed in Table~\ref{clusterloc.tbl}. A minor, systematic bias is introduced by the adoption of the ATLAS9 spectral models, which assume plane-parallel, stationary atmospheres in local thermodynamic equilibrium and do not include the effects of massive stellar winds. This can lead the model fits to return values that overestimate the true bolometric luminosities, and the effects should be most pronounced for evolved, OB supergiants with strong mass loss (see Section 3.2 of P11). The magnitude of this systematic bias reported by P11 is, however, small enough to be completely irrelevant to OB candidate selection and, as we will show, unimportant compared to other sources of uncertainty in measuring the extinction and luminosity of spectroscopically classified OB stars.


The locus of fits may intersect both the ZAMS and pre-MS evolutionary tracks (Figure~\ref{hr}). Models significantly to the left of the ZAMS can be ruled out on physical grounds. At higher luminosities, the model locus may also intersect the OB giant/supergiant tracks, and this is a plausible alternative.   
When the locus of models intersects the ZAMS above the cutoff luminosity, it only crosses very early stages of the pre-MS tracks. The exclusion of IR-excess sources therefore reduces the likelihood of misidentification, because very young pre-MS stars are expected to be surrounded by dusty circumstellar disks or protostellar envelopes (P11). In most cases, the models with the lowest $\chi^2$ values (colored blue in the example diagrams) fell near the ZAMS, bolstering our confidence that they are actually hot, massive stars.

\begin{figure*}[p]
\centering
\includegraphics[height=0.4\textheight]{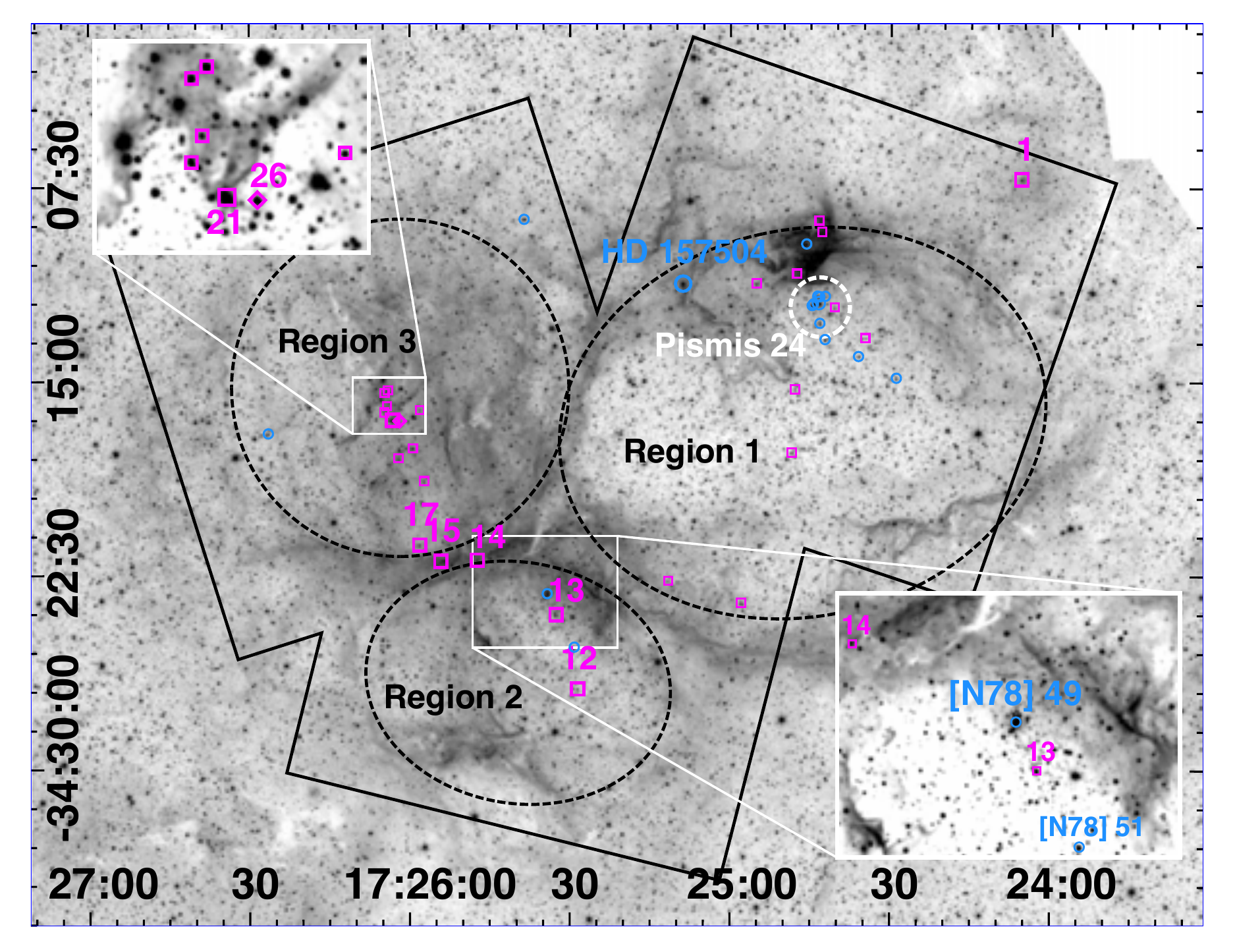}\\
\includegraphics[height=0.35\textheight]{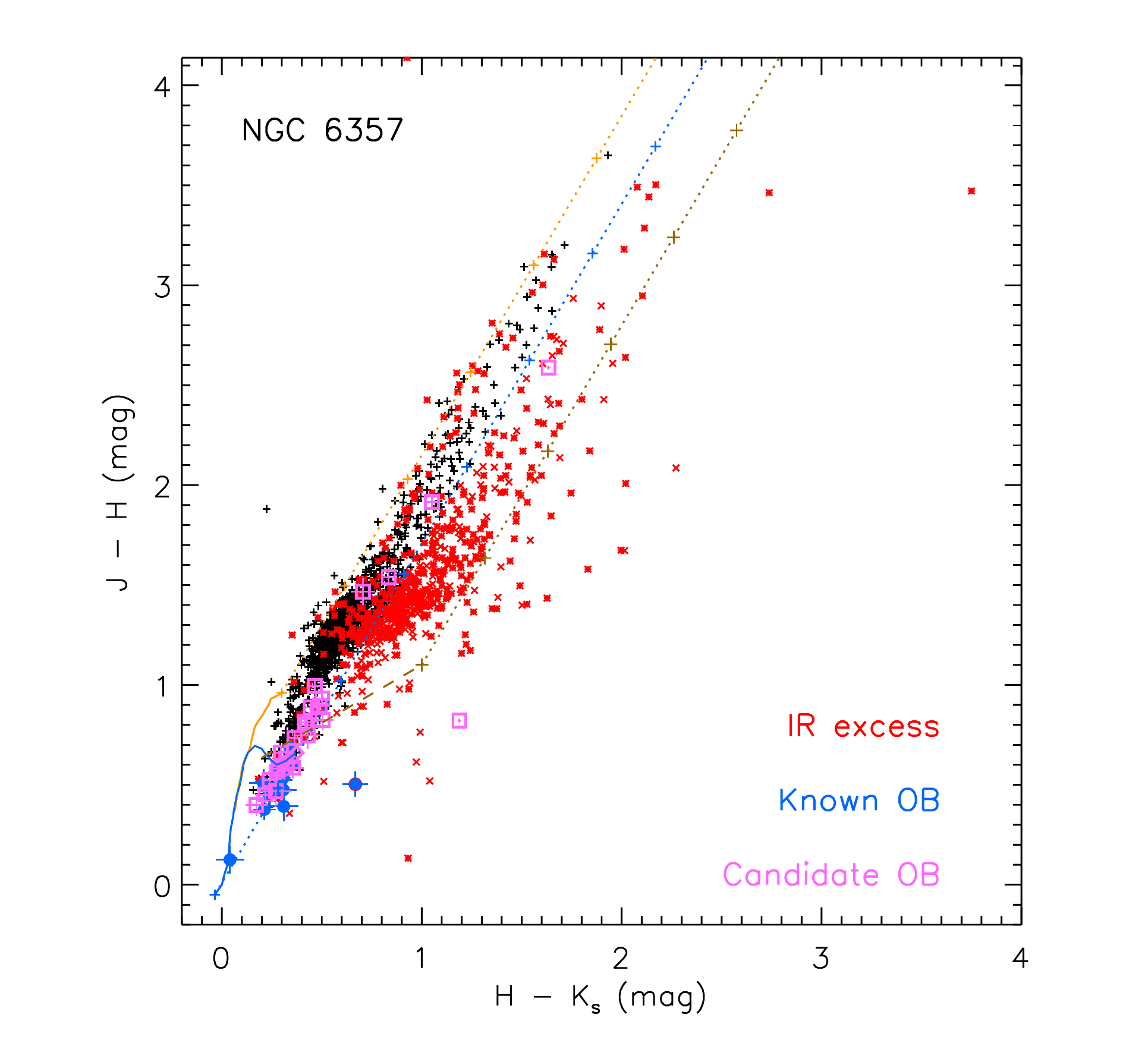}
\includegraphics[height=0.35\textheight]{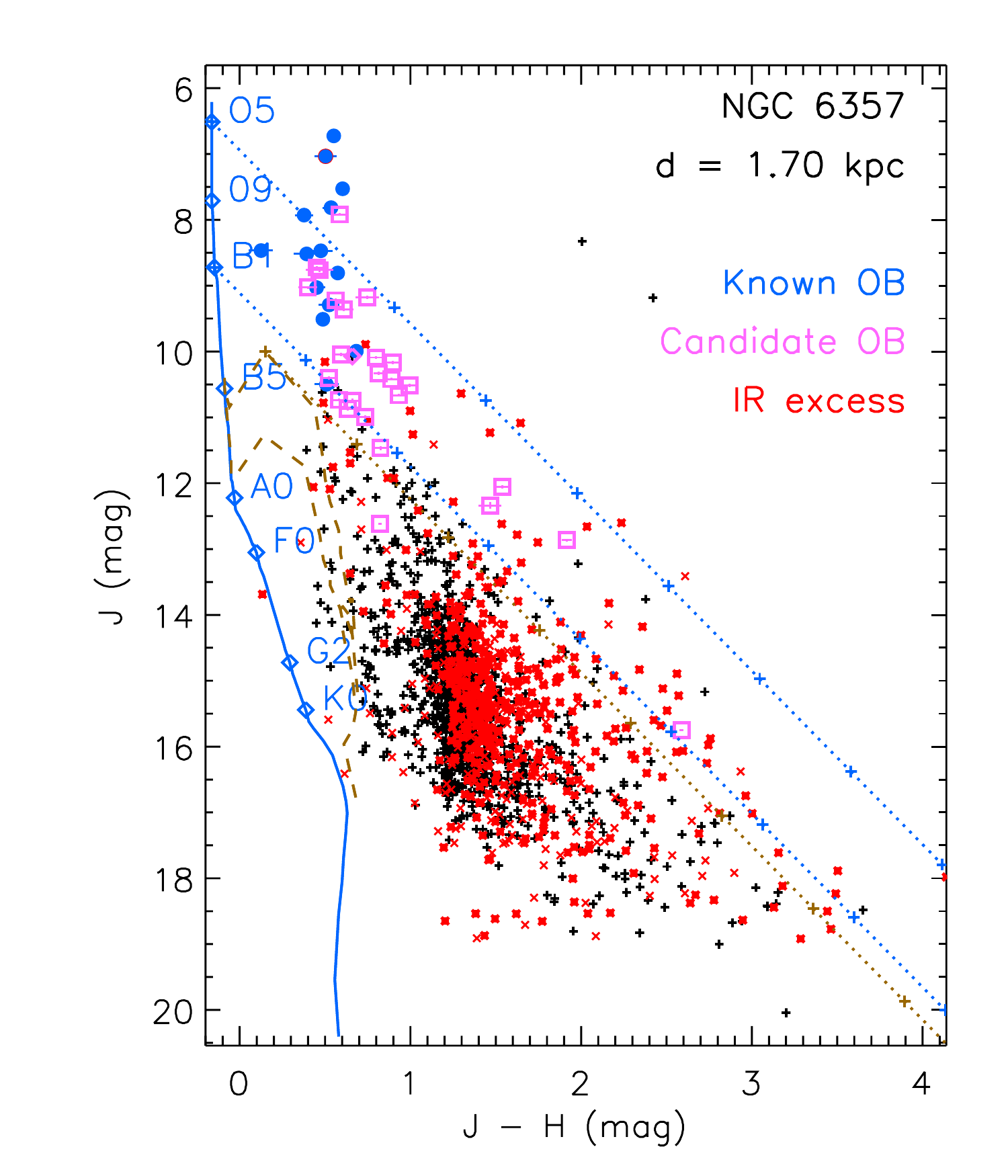}
\caption{\label{fig:each} \small {\em Top:} \Spitzer/IRAC 3.6~\um\ images (inverted, logarithmic grayscale) for each of the 18 MYStIX star-forming regions used in this study, with positions of published and candidate OB stars marked. {\em In all panels} published OB stars (Table~\ref{pubOB.tbl}) are marked by blue circles/dots, OBc from SED fitting (Table~\ref{tab:OBc_SED}) by magenta squares, and additional OBc from NIR color-magnitude analysis (Table~\ref{tab:OBc_NIR}) by magenta diamonds. Notable published and candidate OB stars are labeled by their names from the relevant Table. {\em Bottom:} NIR $JHK_s$ color-color diagram (CCD, left) and $J$ versus $J-H$ color-magnitude diagram (CMD, right) of all MYStIX MPCMs plus candidate OB stars (with MIR excess sources from \citealp{pov_mires} and $K_S$-excess sources from the CCD indicated in red). In each CCD, the main sequence, red giant locus, and classical T Tauri locus are plotted as blue, orange, and dashed brown curves, respectively, with reddening vectors (dotted lines) marked (with $+$ symbols) at $A_V=5$~mag intervals \citep{RL85} extending from each. In each CMD, the main sequence is plotted as a blue curve (with labeled spectral types marked by diamonds), 1 and 3 Myr \cite{Siess} isochrones are plotted as dashed brown curves, with 3 representative reddening vectors extending from the main sequence and 1 Myr isochrone.
({\it k}).--- NGC 6357 (the War and Peace Nebula). (The rest of this figure set is included at the end of this document.)}

\end{figure*}

\subsection{Secondary Selection of Candidate OB Stars Via NIR Colors and $J$ Magnitudes}\label{sec:OBc_NIR}

Inspection of the NIR color-magnitude diagrams (CMDs) for each MYStIX region, plotted in the lower-right panels of Figure~\ref{fig:each}, reveals that most regions contain X-ray detected MPCMs that have bright $J$-band counterparts but are neither previously classified OB stars (blue dots), IR excess sources (red symbols), nor OB candidates selected by our SED-fitting criteria (magenta boxes). SED fitting reveals that the majority of these NIR-bright stars are inconsistent with hot photospheres, ruling them out as OB candidates. The remaining NIR-bright, X-ray detected stars constitute a less reliable, secondary sample of possible candidate OB stars that lacked MIR photometry of sufficient quality for SED fitting. There are two main reasons why a legitimate X-ray OB candidate with a good NIR counterpart could lack high-quality MIR photometry: (1) MIR nebulosity can be extremely bright in the MYStIX star-forming complexes, obscuring point sources at longer wavelengths, and (2) the \Spitzer/IRAC point-source resolution is poorer than that of \Chandra\ or UKIRT, so in crowded regions the X-ray/NIR combination may resolve stars that are confused in the MIR (provided they do not saturate the UKIRT images).

We used the NIR color-color diagrams (CCDs) and CMDs (Figure~\ref{fig:each}) to identify additional OB candidates without SED fitting results in the SCIM-X catalog, based on the NIR colors and $J$ magnitudes of X-ray detected stars. Our simple procedure applied the following steps to each of the 18 MYStIX star-forming regions listed in Table~\ref{clusterloc.tbl}:
\begin{enumerate}
  \item Identify all MYStIX MPCMs in the region that were not listed as previously-published OB stars by \citet{Broos13} and had no SED fitting result reported in the SCIM-X catalog \citep[SED\_FLG = $-99$;][]{pov_mires}. All sources thus selected automatically have \Chandra\ X-ray detections but cannot have been identified as OB candidates by our primary SED fitting methodology.
  \item Select the subset of sources from the previous step that fell into the locus of normally-reddened main-sequence stars on the NIR CCD (between the blue and orange reddening vectors in Figure~\ref{fig:each}, lower-left panels). This step ensures that all secondary OB candidates are detected in all three NIR bands ($JHK_s$). To be conservative, sources falling within this locus but with photometric uncertainties overlapping the reddened red-giant locus (orange) were excluded from consideration as OB candidates.
  \item Identify the sources from the previous two steps that, when dereddened to the OB main sequence (as defined by \citealp{PM13}) on the $J$ vs. $J-H$ CMD (Figure~\ref{fig:each}, lower-right panels), have absolute magnitudes $M_J \le -2.4$ (equivalent to a B1 V star or brighter). 
\end{enumerate}
The 27 sources in 10 different MYStIX regions satisfying the above NIR criteria are designated as secondary OB candidates and marked by magenta diamonds in all panels of Figure~\ref{fig:each}. We emphasize that these secondary candidates are very likely {\em less} reliable than the primary OB candidates identified via SED fitting. As is evident in the CCDs, the locus of reddened main-sequence stars is strongly contaminated by IR excess sources, some of which are bright enough to fall within the region occupied by reddened OB stars on the CMDs. Hence with only NIR photometry available, we cannot rule out the presence of circumstellar disks\footnote{This is a general rule applicable to NIR photometric studies of young stellar populations: A $K_s$-excess color observed on the CCD is a reliable indicator of the presence of warm, circumstellar dust, but the converse is {\em not} generally true. See \citet{Whitney03} for a discussion of inclination and other effects governing the observed colors of young stellar objects.} that could make a lower-mass, pre-MS star appear brighter in the NIR (although we have attempted to minimize this effect by using $J$ to estimate luminosities, as it is less likely to be contaminated by disk emission than $H$ or $K_s$). Strong photometric variability is frequently observed among YSOs \citep{YSOVar}, and a flaring, low-mass YSO could conceivably be classified as an OB candidate in the absence of MIR photometry that could establish the presence of a disk or reveal a flare by providing a second epoch of photometry. The inferred luminosity of secondary OB candidates (here represented by $M_J$ rather than $L_{\rm bol}$) is more dependent on the dereddening calculation than for the SED-fitting candidates, because in highly-obscured regions MIR photometry is less affected by dust extinction compared to NIR photometry.

\section{Results}\label{results.sec}

We first present results from fitting the 1--8~\um\ SEDs of known MYStIX OB stars, to provide context for readers unfamiliar with the strengths and limitations of this method for identifying candidate OB stars.

\subsection{SED Fitting Results for Previously-Published MYStIX OB Stars}\label{pubOB.sec}
The 328 previously-published OB stars included in the MPCM catalog \citep{Broos13} within the 18 MYStIX regions analyzed for this study are listed in Table~\ref{pubOB.tbl}. The first three columns displayed in this table\footnote{Each row in Table~\ref{pubOB.tbl} begins with a ``hidden'' column, available in the electronic version of the table, giving the MYStIX MPCM CLASS\_NAME from \citet{Broos13}.} give the name, spectral type, and citation source of the spectral type for each star; in the vast majority of cases spectral types come from optical or NIR spectroscopy (the only exceptions being a few extremely embedded stars with surrounded by MIR-bright nebulosity, for which spectral types were estimated from radio continuum or IR luminosity of the associated UC \hii regions).\footnote{All spectral types and references listed come from \citet{Broos13}, with the exception of a few notable O stars that were recently re-typed by \citet{GOSSS16}.} While this list is not complete, we adopt it as the best available compilation of ``known'' OB stars across all of these well-studied regions against which to compare and evaluate our X-ray selected OB candidates. We later note (in Tables~\ref{tab:OBc_SED} and \ref{tab:OBc_NIR}) the handful of cases where we have discovered that one of our OB candidates had a published spectral type in the literature that was not included in this compilation.

The provenance (GLIMPSE Catalog/Archive or \citealp{mikemir}) and name (Galactic coordinates) for each \Spitzer/IRAC source associated with an OB star are provided in Column 7 of Table~\ref{pubOB.tbl}. NIR and MIR point-source magnitudes used for our SED-fitting analysis are provided in Columns 8--14, and the [T]woMASS/[U]KIRT provenance of each magnitude in column 15. In the typset table the MIR photometry is hidden for compactness, but the $JHK_s$ magnitudes are displayed to facilitate comparisons with the OB candidates in Tables~\ref{tab:OBc_SED} and \ref{tab:OBc_NIR} as well as the CCDs and CMDs in Figure~\ref{fig:each}. In Column 16 we indicate the presence of a MYStIX X-ray detection (X) and/or marginal/significant IR excess emission (M/Y).

We have performed {\em two} analyses of the SED fitting results on all previously-published OB stars in Table~\ref{pubOB.tbl} with sufficient NIR and MIR photometric detections. First we applied the OB candidate search criteria described in Section~\ref{sec:OBc_SED} to identify those known OB stars that {\em would have been selected} as OB candidates by SED fitting if they had not already been spectroscopically classified; these 67 O and early B-type stars are indicated by checkmarks in Column 17. 

The second SED modeling analysis fixed the $T_{\rm eff}$ parameter to a value $T_{\rm eff}({\rm ST})$ based on the published spectral type, using the calibrations of \citet{MSH05} for O stars and the \citet{Crowther05} extension to early B stars  (column 4 of Table~\ref{pubOB.tbl}). For the O-type stars in Table~\ref{pubOB.tbl} we used the different \citet{MSH05} temperature scales for supergiants (classes I or II), giants (classes III or IV), or dwarfs (class V, or the default if no published luminosity class was available); for B-type stars of all luminosity classes we used a single temperature scale. For unresolved binary/multiple systems, $T_{\rm eff}({\rm ST})$ was chosen based on the spectral type of the most luminous component. With temperature fixed, the extinction $A_V^{\rm SED}$ and bolometric luminosity $L_{\rm bol}^{\rm SED}$ become tightly constrained by the SED modeling; these parameter values are presented in columns 5 and 6, respectively, of Table~\ref{pubOB.tbl}. As P11 showed, the random uncertainties on the model parameters are typically ${\la}0.3$~mag for $A_V^{\rm SED}$ and ${<}5\%$ for $L_{\rm bol}^{\rm SED}$ (or ${<}0.02$~dex in $\log{L_{\rm bol}^{\rm SED}}$), although the uncertainties will be greater (particularly on $A_V^{\rm SED}$) for stars missing NIR photometry. Blank data values in columns 5 and 6 indicate either that there were too few SED points for SED fitting or that {\em no} acceptable model fits to the SED were achieved, while values of $-99$ indicate that there were well-fit models but none consistent with the $T_{\rm eff}({\rm ST})$ for that star in column 4.

\begin{figure}[thp]
\centering
\plotone{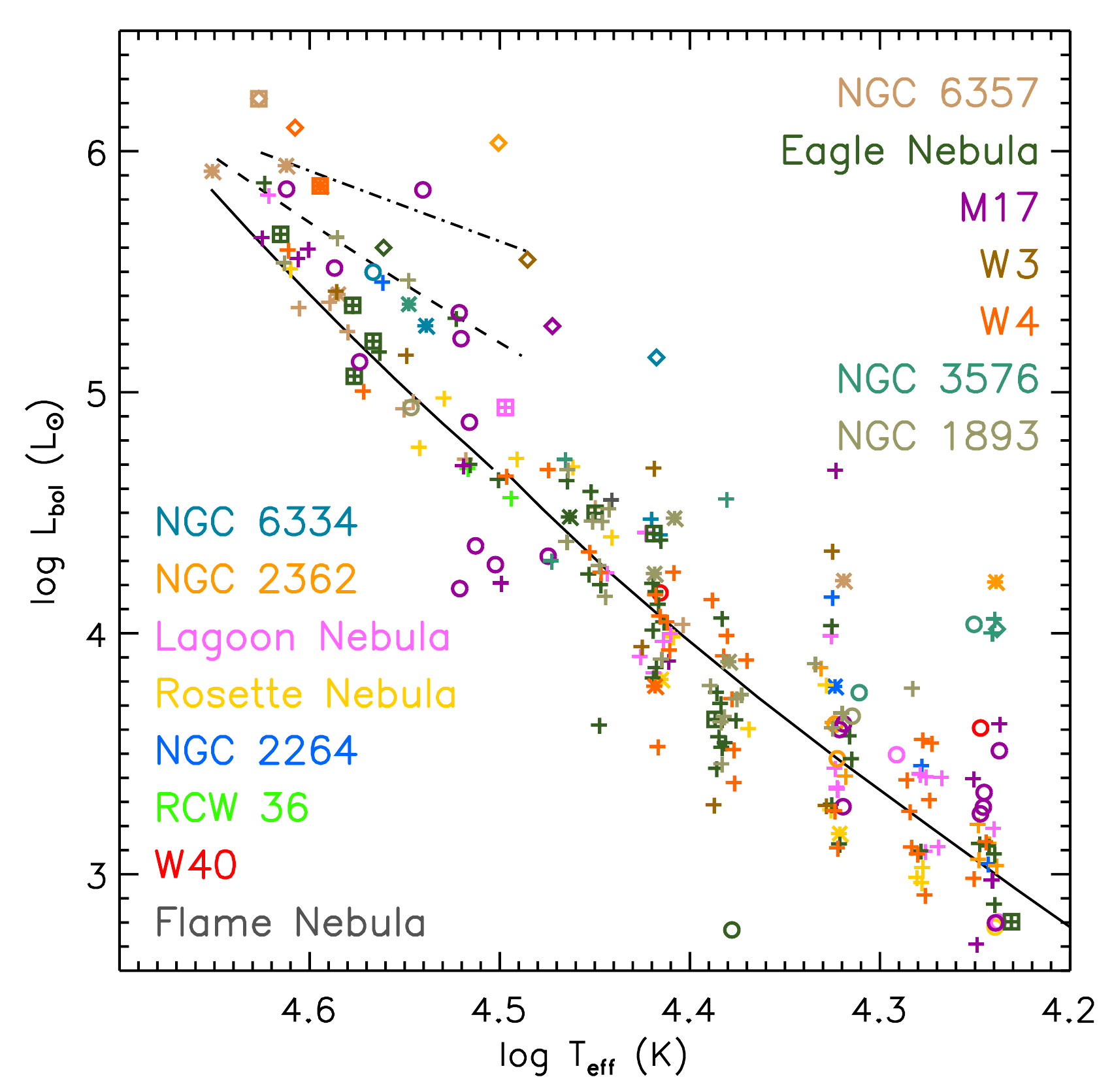}
\caption{HR diagram for spectroscopically classified OB stars in all MYStIX regions with successful SED fitting results (Table~\ref{pubOB.tbl}). Symbols encode published luminosity classes: $+$ signs for dwarfs (V), asterisks for (sub)giants (III/IV), diamonds for supergiants or bright giants (I/II), and circles for no published luminosity class. Known (spectroscopic) binaries are additionally marked with boxes around the symbols. A small random shift in temperature, equivalent to a fraction of a spectral subclass, had been given to each data point to avoid strongly overlapping the symbols in the plot. Random uncertainties are typically smaller than the plotting symbols, and possible systematic errors are discussed in the main text. The theoretical ZAMS for O and B-type stars are plotted as the two solid black curves, while dashed and dash-dotted curves mark the O giant and supergiant sequences, respectively, from \citet{MSH05}.} \label{fig:OB_HRD}
\end{figure}

Fifteen of the 18 MYStIX regions contained one or more OB stars with SEDs that we could model, and we plot $\log{L_{\rm bol}^{\rm SED}}$ versus $\log{T_{\rm eff}({\rm ST})}$ for all of these stars on a single H-R diagram in Figure~\ref{fig:OB_HRD}. This figure can be compared with similar H-R diagrams published by \citet[][their Figure 2]{gagne} for Carina and \citet[][their Figure 11]{K+15w3} for W3. 
In spite of differences in methodology for measuring $L_{\rm bol}$ and $A_V$, when examining the dispersion of the stars with respect to the ZAMS both of these {\em single-region} H-R diagrams appear qualitatively similar to our H-R diagram of the OB stars across {\em different} MYStIX regions.

Among spectroscopically-classified O-type dwarfs, there is generally good correlation with the theoretical ZAMS \citep{MSH05}, with the majority of stars falling within 0.2 dex in $\log{L_{\rm bol}}$ of the ZAMS for their spectral type. This supports the conclusion of P11 that even fitting {\em only} the 1--8~\um\ IR SEDs can effectively constrain the luminosities of OB stars in the absence of $UBV$ photometry.\footnote{Perhaps counterintuitively, it could be advantageous to use {\em only} IR photometry in regions with significant dust obscuration, as the reddening correction in visible-light is {\em very} sensitive to the adopted extinction law (P11). Note that \citet{gagne} adopted an extinction law with reddening parameter $R_V=4.0$ for Carina, while \citet{K+15w3} adopted $R_V=3.6$ for W3; in reality the reddening parameter can vary even among different stars in the same star-forming region.} 







\begin{deluxetable*}{hllccccrcrrrhhhhccc}
\tablecaption{Previously Cataloged OB Stars among MYStIX MPCMs \label{pubOB.tbl}}
 \tabletypesize{\tiny}
 \tablewidth{\textwidth}
 \setlength{\tabcolsep}{0.025in}
\tablehead{
\nocolhead{MYStIX} & \colhead{Star}  &\colhead{} & \colhead{Ref.} & 
\colhead{$T_{\rm eff}({\rm ST})$\tablenotemark{a}} &\colhead{$A_V^{\rm SED}$} & \colhead{$\log{L_{\rm bol}^{\rm SED}}$} & 
\multicolumn{2}{c}{{\em Spitzer} origin \& name}  & 
\colhead{$J$} & \colhead{$H$} & \colhead{$K_S$} & \nocolhead{[3.6]} & \nocolhead{[4.5]} & \nocolhead{[5.8]} & \nocolhead{[8.0]} & \colhead{Origin\tablenotemark{c}} & \colhead{X-ray/} & \colhead{SED} \\ 
\nocolhead{MPCM} &  \colhead{Name} & \colhead{ST}  & \colhead{ST} & \colhead{(kK)} & \colhead{(mag)} & \colhead{($L_{\sun}$)} & \multicolumn{2}{c}{(GLIMPSE/K13)\tablenotemark{b}} & 
\colhead{(mag)} &\colhead{(mag)} & \colhead{(mag)} & \nocolhead{(mag)} & \nocolhead{(mag)} & \nocolhead{(mag)} & \nocolhead{(mag)} & \colhead{$JHK_S$} & \colhead{IRE?\tablenotemark{d}} & \colhead{Cand?\tablenotemark{e}} \\
\nocolhead{} & \colhead{(1)} & \colhead{(2)} & \colhead{(3)} & \colhead{(4)} & \colhead{(5)} & \colhead{(6)} & \colhead{(7a)} & 
\colhead{(7b)} & \colhead{(8)} & \colhead{(9)} & \colhead{(10)} &\nocolhead{(11)} &\nocolhead{(12)} &\nocolhead{(13)} &\nocolhead{(14)} &\colhead{(15)} & \colhead{(16)} & \colhead{(17)} 
}
\startdata
& \multicolumn{1}{c}{Flame Nebula} & \multicolumn{17}{c}{} \\
\cline{2-3} \noalign{\smallskip}
054137.85-015436.5  & NGC 2024 1                     & B0.5Vp                         & 1                                                                        & 28.0 &   8.1 & 4.55 & K & G206.5282-16.3772 &  7.34 &  6.52 &  5.94 &  5.76 &  5.62 &  5.49 &  5.10 & TTT & X & $\checkmark$ \\
054145.48-015428.7  & IRS 2b                         & O8/B2V                         & 1                                                                        & 35.0 &   \nodata &  \nodata & K & G206.5416-16.3480 &   \nodata &   \nodata &   \nodata &  6.11 &  5.65 &   \nodata &   \nodata & ... & X & \\
\noalign{\smallskip}
\hline
& \multicolumn{1}{c}{W40} & \multicolumn{17}{c}{} \\
\cline{2-3} \noalign{\smallskip}
183114.83-020349.9  & IRS 5                          & B1                             & 2                                                                      & 26.0 &   7.3 & 4.17 & K & G028.7867+03.5452 &  8.28 &  7.51 &  7.08 &  6.86 &  6.71 &  6.59 &  5.85 & TTT &  XM & \\
183123.95-020410.7  & IRS 3A                         & B3                             & 2                                                                      & 17.5 &   8.4 & 3.61 & K & G028.7990+03.5087 &   \nodata &   \nodata &   \nodata &  7.18 &  6.93 &  6.45 &  5.13 & TTT &  M & \\
183127.83-020523.4  & IRS 1A South                   & O9.5                           & 2                                                                      & 31.5 &   \nodata &  \nodata &  &  &   \nodata &   \nodata &   \nodata &   \nodata &   \nodata &   \nodata &   \nodata & ... &  X  & \\
\noalign{\smallskip}
\hline
& \multicolumn{1}{c}{RCW 36} & \multicolumn{17}{c}{} \\
\enddata
\end{deluxetable*}

We mark known spectroscopic binaries (or higher-multiple systems) with boxed symbols in Figure~\ref{fig:OB_HRD}. Since multiplicity is common among massive stars \citep{S+13_O-multi}, but the fraction of MYStIX OB stars classified as binaries is low, it is probable that a significant fraction of the stars listed in Table~\ref{pubOB.tbl} have undiscovered multiplicity. An unresolved, equal-mass binary system misclassified as a single star will be over-luminous by 0.3 dex, placing it significantly above our observed symmetrical scatter of 0.2 dex around the O-type ZAMS on the H-R diagram. This extra luminosity is sufficient to move a star from the O-type ZAMS to the O giant sequence \citep{MSH05}. We hence identify six candidate equal-mass binary systems, or alternatively single O-type giants, that may have been misclassified as single O dwarfs (+ symbols without boxes located along the O giant sequence in Figure~\ref{fig:OB_HRD}). These possible multiple systems are mentioned in the discussions of their respective parent regions in Appendix~\ref{appendix}.

The O-type giants and supergiants lie significantly above the ZAMS, as expected, with the exception of two O giants in NGC 6357 (see Appendix~\ref{ngc6357.app}). 
We identify five candidate O (super)giants with no published luminosity class (open circles located on or above the O giant sequence in Figure~\ref{fig:OB_HRD}), four in M17 (Appendix~\ref{m17.app}) and the most luminous known O star in NGC 6334 (Appendix~\ref{ngc6334.app}). Our SED fitting measured $\log{L_{\rm bol}^{\rm SED}/\Lsun}>6$ for three O supergiant systems that constitute the the dominant ionizing stars in NGC 2362, W4, and the Pismis 24 cluster of NGC 6357. We discuss each of these systems in the appendix, and we reiterate here the caveat that our $L_{\rm bol}^{\rm SED}$ and $A_V^{\rm SED}$ values must be regarded as upper limits, especially for stars with very strong mass loss (P11).

Some individual OB stars are strong outliers in luminosity for their published spectral types, and these are noted in Table~\ref{pubOB.tbl}. Two stars in the Eagle Nebula and four in M17 are among the most extreme examples of SED fitting results that are {\em under-luminous} compared to their published spectral types, as discussed in the appendix (\ref{eagle.app} and \ref{m17.app}).

Compared to the O stars, the MYStIX B-type stars show far more scatter about the theoretical ZAMS, nearly one order of magnitude in $L_{\rm bol}$. Similar large scatter on the H-R diagram is evident for early B stars both in Carina \citep{gagne} and in W3; in the latter case \citet{K+15w3} point out that this scatter is dominated by uncertainties in spectral types. While in general we have no information about uncertainties on the spectral types from the literature, we suspect that the difficulty in assigning accurate spectral types to (often obscured) early B stars dominates the scatter in our sample as well. We therefore cannot accurately identify new candidate binaries of evolved (super)giants among the MYStIX B stars, and we caution the reader that published early-B spectral types in obscured star-forming regions should be viewed with some skepticism.

Comparing the individual MYStIX regions to each other, Figure~\ref{fig:OB_HRD} reveals no regions that show significant, {\em systematic} shifts to higher or lower luminosities for their {\em ensemble} of OB stars compared to the expectations from their spectral types. This suggests that the assumed distance to each region (Table~\ref{clusterloc.tbl}) is accurate to within ${\sim}25\%$ to 50\%, the relatively loose constraints provided by the observed luminosity dispersion about the OB ZAMS.

\subsection{MYStIX Candidate X-ray-Emitting OB Stars}\label{sec:cand_results}

The 98 candidate MYStIX OB stars satisfying our SED-based selection criteria (Section \ref{sec:OBc_SED}) are listed in Table~\ref{tab:OBc_SED}, while Table~\ref{tab:OBc_NIR} lists the additional 27 candidate OB stars identified via our secondary NIR selection criteria (Section~\ref{sec:OBc_NIR}).
The first two columns in each table give the  the MYStIX X-ray source designation, which encodes the source position in equatorial (J2000; sexigesmal) coordinates. Within each MYStIX region, candidates are sorted and numbered first by method of identification and then by increasing right ascension, hence, for example, in NGC 6357 the first of the 25 primary X-ray OB candidates from SED fitting is MYStIX OB candidate (MOBc) NGC 6357 1, while the first of the secondary X-ray OB candidates from NIR photometry is MOBc NGC 6357 26.

Columns 3--5 in Table~\ref{tab:OBc_SED} give results from the SED fitting: the interstellar extinction (magnitudes $A_V^{\rm MS}$), stellar effective temperature ($T_{\rm eff}^{\rm MS}$ in kilo-degrees Kelvin) and bolometric luminosity ($\log{L_{\rm bol}^{\rm MS}/\Lsun}$), {\em assuming} the stars lie on the ZAMS (Figure~\ref{hr}). Columns 3 and 4 in Table~\ref{tab:OBc_NIR} give the analogous information for our secondary X-ray NIR candidates: 
$A_V^{\rm MS}$ and $M_J^{\rm MS}$
from dereddening observed $J-H$ colors and $J$ magnitudes to the ZAMS (Figure~\ref{fig:each}, CMDs in bottom right panels). Formally, all of these ``MS'' extinction, luminosity, and temperature values are {\em upper limits only}, pending follow-up spectral typing.







\begin{deluxetable*}{lcrrrcrrrhhhhcl}
\tablecaption{Candidate X-ray-emitting OB Stars from SED Fitting \label{tab:OBc_SED}}
\tablehead{
\colhead{MOBc} & \colhead{}   &  \colhead{A$_{V}^{\rm MS}$ \tablenotemark{b}} & \colhead{$T_{\rm eff}^{\rm MS}$ \tablenotemark{b}} & \colhead{$\log{L_{\rm bol}^{\rm MS}}$ \tablenotemark{b}} & \colhead{{\em Spitzer}}  & \colhead{$J$} & \colhead{$H$} & \colhead{$K_S$} & \nocolhead{[3.6]} & \nocolhead{[4.5]} & \nocolhead{[5.8]} & \nocolhead{[8.0]} & \colhead{Origin} & \colhead{Other ID/ST (Ref);\tablenotemark{e}} \\ 
\colhead{Name} & \colhead{X-ray ID\tablenotemark{a}} & \colhead{(mag)} & \colhead{(kK)}  & \colhead{$\Lsun$ } & \colhead{Name\tablenotemark{c}} & \colhead{(mag)} & \colhead{(mag)} & \colhead{(mag)} & \nocolhead{(mag)} & \nocolhead{(mag)} & \nocolhead{(mag)} & \nocolhead{(mag)} & \colhead{NIR\tablenotemark{d}} & \colhead{IRE?} \\
\colhead{(1)} & \colhead{(2)} & \colhead{(3)} & \colhead{(4)} & \colhead{(5)} & \colhead{(6)} & \colhead{(7)} & \colhead{(8)} & \colhead{(9)} &  \nocolhead{(10)} &  \nocolhead{(11)} &  \nocolhead{(12)} &  \nocolhead{(13)} & \colhead{(14)} & \colhead{(15)}
}
\startdata
Lagoon 1 &  180307.97-242355.1 &       5.8 &       28.7 &       4.4 & G005.8929-01.0766   &       10.09 &       9.49 &       9.20 &       9.04 &       8.92 &       8.89 &       8.97 & T & \\ 
Lagoon 2 &  180322.59-242204.9 &       3.7 &       33.2 &       4.8 & G005.9468-01.1098   &        8.72 &       8.31 &       8.14 &       8.16 &       8.15 &       8.04 &       8.09 & T & \\ 
Lagoon 3 &  180400.80-242334.6 &       7.3 &       28.1 &       4.3 & G005.9963-01.2484   &       10.63 &       9.82 &       9.41 &       9.18 &       9.12 &       8.98 &       9.09 & T & \\ 
Lagoon 4 &  180423.55-242116.8 &       4.8 &       25.8 &       4.0 & G006.0721-01.3048   &       10.29 &       9.74 &       9.54 &       9.42 &       9.41 &       9.28 &       9.29 & T & OB (Bik15) \\ 
Lagoon 5 &  180438.91-242414.1 &       4.3 &       39.2 &       5.4 & G006.0578-01.3799   &        7.99 &       7.46 &       7.29 &       7.28 &       7.25 &       7.20 &       7.17 & T & NGC 6530 95/K0III (2) \\ 
\noalign{\smallskip}
\hline 
RCW 38 1 &  085859.19-472658.0 &       11.0 &       26.8  &       4.2 & G351.1645+00.6753 &       12.22 &       10.99 &       10.32 &       9.89 &       9.82 &       9.58 &  \nodata  & T &  OBc (1) \\ 
RCW 38 2 &  085901.68-473037.1 &       7.9 &       27.2  &       4.2 & G351.2109+00.6680  &       11.14 &       10.33 &       9.89 &       9.55 &       9.52 &  \nodata  &  \nodata  & T &  OBc (1) \\ 
\noalign{\smallskip}
\hline
NGC 6334 1  &  172003.34-355824.8 &       4.0 &       30.2 &       4.5 & G351.1645+00.6753  &       9.44 &       9.04 &       8.81 &       8.80 &       8.83 &       8.70 &  \nodata  & T & \\
\enddata 
\vspace{1Em}
\begin{minipage}{\textwidth}
\tablenotetext{a}{{\em Chandra} X-ray source designation from \citet{Broos13}.}
\tablenotetext{b}{Values in Columns 3--5 were derived from SED fitting under the assumption of single masstive stars on the theoretical zero-age main sequence (see the text). Formally, $T_{\rm eff}^{\rm MS}$, $A_V^{\rm MS}$, and $\log{L_{\rm bol}^{\rm MS}}$ reported here are upper limits and should be used only for comparison purposes among this sample (and to the Carina OBc sample from P11).}
\tablenotetext{c}{MIR source designation from \citet{mikemir} or GLIMPSE point-source archive.}
\tablenotetext{d}{NIR source from (T)woMASS or (U)KIRT.} 
\tablenotetext{e}{Notes in Column 15: YSOc = YSO/T Tauri candidate, OBc = photometric massive star candidate, OB (Bik15) = spectroscopically confirmed OB star (A. Bik, private communication, 2015), (WIRO) = new spectral type reported from spectroscopic follow-up with WIRO, and (Kuhn16) spectroscopic followup by Kuhn et al.\ (2016, in preparation). References in Column 15: (1) \citet{Wolk06}, (2) \citet{R+02lagoon}, (3) \citet{Gvaramadze11}, (4) \citet{wang_ngc6357}, (5) \citet{povichm17b}, (6) SIMBAD \citep{SIMBAD}, (7) \citet{Reed03OBcat}, (8) \citet{K+15w3}, 
(9) \citet{Neckel84}, (10) \citet{pris2011}, (11) \citet{R+M09}. As in Table~2, ``M'' denotes marginal IR excess sources for which available [5.8] and [8.0] photometry was excluded from SED fitting.}
\end{minipage}  
\end{deluxetable*}


\begin{deluxetable*}{lcCCCCCCCCcl}
\tablecaption{Secondary Candidate X-ray-emitting OB Stars from NIR Photometry \label{tab:OBc_NIR} }
\tablehead{
\colhead{MOBc Name} & \colhead{X-ray ID\tablenotemark{a}}   &  \colhead{A$_V^{\rm MS}$} & \colhead{$M_J^{\rm MS}$}  & \colhead{$J$} & \colhead{$H$} &
\colhead{$K_S$}  & \colhead{$\sigma_J$} & \colhead{$\sigma_H$} & \colhead{$\sigma_{K_S}$} & \colhead{Origin\tablenotemark{b}} & \colhead{Other ID/ST (Ref)\tablenotemark{c}}
}
\colnumbers
\startdata
Rosette 1 & 063302.88+043617.0 &   6.4 & -4.14 &  8.29 &  7.77 &  7.68 &  0.02 &  0.05 &  0.02 & T &  HD 259513/K2?  \\
\noalign{\smallskip}
\hline
DR 21 1 & 203828.79+421038.5 &  37.3 & -4.03 & 17.36 & 13.55 & 11.37 &  0.01 &  0.01 &  0.01 & U &   \\
DR 21 2 & 203900.04+421936.5 &  33.4 & -2.53 & 17.77 & 14.32 & 12.38 &  0.03 &  0.01 &  0.01 & U & [DS96] I1 (1)  \\
\enddata
\tablecomments{Columns 3 -- 10 list data values in magnitude
  units. Magnitudes in Columns 3 and 4 were derived dereddening the observed $J-H$ color and $J$ magnitude to the main sequence, assuming the distance to each region listed in Table~1 (see the text).}
\tablenotetext{a}{{\em Chandra} X-ray source designation from \citet{Broos13}.}
\tablenotetext{b}{NIR source from (T)woMASS or (U)KIRT.} 
\tablenotetext{c}{Notes in Column 12: 
YSOc = YSO/T Tauri candidate, OBc = photometric massive star
candidate. References in Column 12: (Bik15) A. Bik, private
communication (2015), 
(1) \citet{DS96-DR21}, (2) \citet{Winston11}, (3) \citet{Wolk06}, 
(4) \citet{R-I+11-W3} (5) \citet{K+15w3}
}
\tablenotetext{d}{Dereddening to the main sequence would produce $M_J$ more luminous than an O3 V star. This could occur for evolved, luminous stars or unresolved binary/multiple systems, hence no intrinsic source colors can be assumed for the NIR source and no $M_{J}^{\rm MS}$ or $A_{V}^{\rm MS}$ are reported.}
\tablenotetext{e}{NIR spectrum shows CO absorption bands (Bik15).}

\end{deluxetable*}

We also provide the IR point source information and photometry used in our analysis and plotted in the CCDs and CMDs of Figure~\ref{fig:each}. For the SED-fitting candidates (Table~\ref{tab:OBc_SED}) the \Spitzer\ MIR source name, $JHK_s$ and IRAC point-source photometry values (the IRAC photometry values are included in the full online version of the table), and provenance of the NIR photometry (T = 2MASS or U = UKIRT) are listed in columns 9--14.\footnote{\citet{pov_mires} used only high-quality NIR photometry for SED fitting, as defined by certain 2MASS and UKIRT source quality flags. Hence a lack of reported photometry in Table~\ref{tab:OBc_SED} does not necessarily indicate a non-detection but simply a datapoint withheld from the SED analysis.}  For the NIR photometry candidates we give simply the $JHK_S$ magnitudes and provenance (Table~\ref{tab:OBc_NIR}).

The final column in each table lists, to the best of our knowledge, the most relevant independent identification, spectroscopic classification, or {\em photometric} selection as an OB candidate (OBc) for each star, as well as the citation for the source of this information. 
We also indicate (with an ``M'' in column 15 of Table~\ref{tab:OBc_SED}) the five OB candidates showing evidence for marginal MIR excess emission in the SCIM-X catalog \citep{pov_mires}. 

Four of our candidates in W3 (MOBc W3 1, 3, 5, \& 8) were independently identified and confirmed, one as an O-type binary, two as B stars, and one as a B-type binary, by \citet{K+15w3}.
Targeted, NIR spectroscopic followup of our candidates using the KMOS spectrometer on the Very Large Telescope (A. Bik 2015, private communication) confirmed one new OB star in Lagoon (MOBc Lagoon 4), and 12 new OB stars in NGC 6357 (MOBc NGC 6357 12, 14, 15, 16, 17, 18, 20, 22, 23, 24, 25, \& 30). Spectroscopic followup using the ARCoIRIS instrument of the 4-m Blanco Telescope (Kuhn et al. 2016, in preparation) has confirmed three new OB stars near the Trifid Nebula (MOBc Trifid 1, 2, \& 3).
Visible-light spectroscopic followup of SED-selected OB candidates in M17 the 2.3-m Wyoming Infrared Telescope (WIRO) identified five new early-type stars (MOBc M17 7, 11, 12, 17, \& 18) but classified four others (MOBc M17 2, 9, 10, \& 14) as late-type stars. 
Six other OB candidates in various regions (MOBc Rosette 1, Lagoon 5, W3 4, and NGC 1893 4 \& 5) had previous spectroscopic classifications indicating that they are {\em cool} stars.  We have chosen to list spectroscopically classified cool stars, which are not actual OB candidates, in Tables~\ref{tab:OBc_SED} and \ref{tab:OBc_NIR} as examples of ``false positives'' selected by our SED fitting criteria, and some of them could still prove to be evolved, massive yellow or red supergiants. Overall, 20 of the 125 candidates have already been spectroscopically confirmed as OB stars, while nine are late-type interlopers.


\subsection{Interstellar Extinction to Known and Candidate MYStIX OB Stars}

\begin{figure}[thb]
\epsscale{1.25}
\plotone{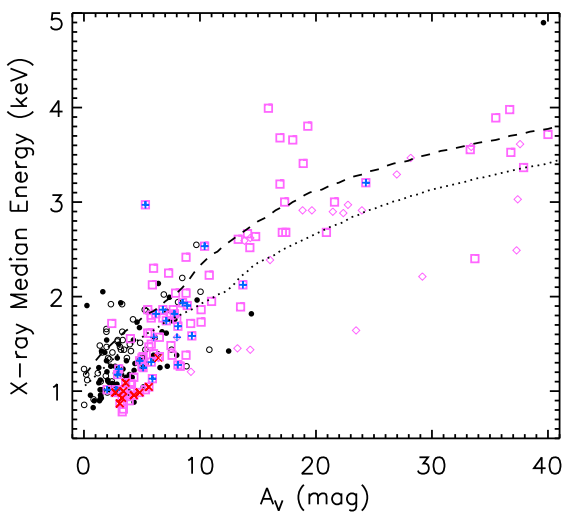}
\caption{Plot of total-band (0.5--8 keV) X-ray median energy versus interstellar extinction showing 
all X-ray detected MYStIX OB stars with SED fitting results (black dots: spectral types O through B1, open circles: B1.5 and later types) and X-ray emitting OB candidates (magenta boxes = SED-selected, diamonds = NIR-selected). Candidates that have been spectroscopically confirmed or rejected as early-type stars are additionally marked by blue $+$ or red $\times$ symbols, respectively. Two calibration curves for low-mass, pre-MS stars from \citet[][their Table 3]{XPHOT} for X-ray absorbing column density (converted to $A_V$ assuming $N_H/A_V=2\times 10^{21}~{\rm cm^{-2}~mag^{-1}}$) versus $E_{\rm med}$ are plotted as dotted (absorption-corrected, hard-band X-ray luminosity $\log{L_{h,c}=28.5}$ model) and dashed ($\log{L_{h,c}=30.0}$ model) curves. 
\label{evav.fig}}
\end{figure}

In Figure~\ref{evav.fig} we plot X-ray median energy ($E_{\rm med}$, measured across the total 0.5--8 keV \Chandra/ACIS-I passband; \citealp{mikexcat,MOXC}) versus interstellar extinction ($A_{V}^{\rm SED}$ from Table~\ref{pubOB.tbl} or $A_V^{\rm MS}$ from Tables~\ref{tab:OBc_SED} and \ref{tab:OBc_NIR}) as a simple basis for comparing the known and candidate OB samples. We have further marked the several candidate OB stars that have already been confirmed (blue $+$ symbols) or rejected (red $\times$ symbols).
$E_{\rm med}$ is affected by both the extinction (since the soft X-rays are preferentially obscured, hardening the radiation) and the intrinsic X-ray spectrum, which can vary greatly across individual massive stars due to differences in the wind-shock properties \citep[e.g.,][]{gagne}. Notwithstanding the expected variation in intrinsic X-ray hardness, median energy is clearly correlated with extinction, with both known and candidate OB stars following the same general trend. The relatively large scatter about the trend is commonly seen in plots of X-ray versus infrared measures of stellar extinction in star forming regions (e.g., \citealp{Vuong03, Feigelson05}; P11). 

Published O through early B-type stars and spectroscopically confirmed OB candidates show characteristically lower $E_{\rm med}$ over the exinction range of $0 < A_V^{\rm SED}\la 6$~mag compared to calibration curves of $E_{\rm med}$ versus X-ray absorbing column density $N_H$ (converted to $A_V$ assuming $N_H/A_V=2\times 10^{21}~{\rm cm^{-2}~mag^{-1}}$ \citealp{Vuong03}) for low-mass, pre-MS stars \citep{XPHOT}. For relatively lightly-absorbed sources, $E_{\rm med}$ hence reflects the canonical, softer intrinsic X-ray spectra for massive stars compared to the young, low-mass stars that dominate the MYStIX source populations.
Notwithstanding the scatter, stars classified as B1.5 or later appear systematically harder in $E_{\rm med}$ compared to earlier OB stars and OB candidates in this same extinction range (recall that our selection criteria only identify the equivalent of B1 V or earlier/more luminous stars), which indicates intrinsically harder X-ray spectra compared to earlier-type massive stars. Many of the X-ray-detected MYStIX B stars could therefore be intrinsically fainter, weak-wind sources with unresolved, low-mass pre-MS companions producing the observed X-ray emission \citep[see, e.g.][]{Evans_CCCP11}. A detailed analysis of the X-ray spectral properties of the known and candidate MYStIX OB stars is beyond the scope of this study, and our past experience has shown that X-ray spectral modeling does not provide sufficient evidence to confirm or reject candidate OB stars (see Section~\ref{contam.sec}).

OB candidates with higher extinction but softer X-ray median energy compared to the trend for published and confirmed OB stars may be late-type contaminants or incorrect matches between the X-ray source and the IR counterpart. All 8 of our primary OB candidates from SED fitting that have been spectroscopically classified as late-type stars (Table~\ref{tab:OBc_SED}) cluster near $E_{\rm med}\approx 1$~keV for $2 \la A_V^{\rm MS} \la 6$~mag.  Since several previously-published MYStIX OB stars also lie in or near this region in Figure~\ref{evav.fig}, we cannot recommend any simple cut on extinction and median energy to flag probable contaminants.

Several candidates appear to be extreme outliers. Three NIR-selected candidates have $A_V^{\rm MS}> 20$~mag and $E_{\rm med}<2.6$~keV (DR 21 1 and Trifid 5 \& 6). All are located well outside of dense or embedded clusters, and they could be chance alignments between an X-ray source and an unassociated, IR field star. For these three sources the MPCM classification by \citet{Broos13} may be incorrect. MOBc NGC 6334 11 is the sole SED-fitting candidate with similar soft median energy but high extinction in Figure~\ref{evav.fig}, but it is centrally located among the embedded, massive clusters and \hii regions, increasing the likelihood that it is actually a member of the NGC 6334 complex. It is possible that our SED fitting overestimated the reddening and hence luminosity for this star (see Figure~\ref{fig:each}j and Appendix \ref{ngc6334.app}), or that there was a mismatch between the X-ray and MIR source in this dense cluster.

Six OB candidates identified via SED fitting (MOBc Eagle 4, M17 4, and NGC 6334 3, 7, 9 \& 10) and three identified via NIR photometry (MOBc DR21 2, M17 23 \& 24) have very high $A_V^{\rm MS}>30$~mag and commensurately hard $E_{\rm med} > 3$~keV in Figure~\ref{evav.fig}. All nine of these stars have projected locations within or behind dense molecular clouds, so their extreme reddening is reasonable, but given the uncertainties in the extinction laws used for the selection, there is an increased chance that an obscured OB candidate could instead be an embedded, lower-mass protostar. Deep, NIR follow-up spectroscopy should reveal the true nature of these candidates. 

Another outlying group of seven SED-fitting candidates (MOBc NGC 6357 11, Eagle 2, 6 \& 7, M17 6, W3 2, and NGC 3576 4) has moderately high extinction ($15~{\rm mag} < A_V^{\rm MS}<20$~mag) but disproportinately hard median energy ($E_{\rm med} > 3$~keV); these could be OB stars with intrinsically hard spectra or perhaps background contaminants (for example, X-ray-detected active galactic nuclei that were erroneously matched to IR-bright Galactic field stars). A similar outlier with intriguing, hard X-ray emission is the confirmed OB star NGC 6357 12, with $E_{\rm med}=2.97$~keV, $A_V^{\rm MS}=5.3$~mag. The most extreme X-ray star plotted in Figure~\ref{evav.fig} is W3(OH) in the far upper-right corner, which, unlike the other stars in this plot, we fit using embedded YSO models (Table~\ref{pubOB.tbl}). Clearly $E_{\rm med} > 3$~keV is possible for certain massive stars, particularly when they are highly absorbed.

\section{Discussion and Summary} \label{discussion.sec}

Table~\ref{sum.tbl} summarizes the results of our OB candidate search in each MYStIX region. The MOB1c(SED) and MOB1c(NIR) columns give the number of new OB candidates identified via SED fitting and NIR photometry, respectively. The OB(pub), OB1(pub), and OB1X(pub) columns give the numbers of previously published OB stars, previously published stars of spectral type B1 or earlier, and previously published B1 or earlier stars that were detected as MYStIX X-ray sources and hence, in principle, would be selected as OB candidates by our SED and/or NIR search criteria. The average interstellar extinction measured from SED fitting of published OB stars in each region (Section~\ref{sec:OBc_SED} and Table~\ref{pubOB.tbl}) is reported as $\langle A_{V}^{\rm SED}\rangle$, while the average extinction (assuming stars on the ZAMS) for all OB candidates in each region is $\langle A_{V}^{\rm MS}\rangle$. As expected, the candidate OB stars are observed through systematically higher extinction than the published OB stars, $\langle A_{V}^{\rm MS}\rangle \ge \langle A_{V}^{\rm SED}\rangle$ in each of the 11 MYStIX regions for which the extinction of published and candidate OB stars can be directly compared.

\subsection{Implications for the Massive Stellar Populations in the MYStIX Regions}
Recalling that our OB candidate selection criteria were designed to identify stars more luminous than the equivalent of a B1 V star, we can estimate the potential multiplicative factor $F_{\rm inc}$ by which the inferred massive stellar population of each region would increase due to the addition of newly-confirmed OB candidates using
\begin{displaymath}
  F_{\rm inc} = \frac{f_{\rm conf}[{\rm MOB1c(SED)} + {\rm MOB1c(NIR)}]}{\rm OB1(pub)} + 1,
\end{displaymath}
where $f_{\rm conf}$ is the confirmation rate of candidate OB stars. The confirmation rate is a measure of the {\em reliability}  of our OB candidate search (high $f_{\rm conf}$ implies few false-positive OB candidates that turn out to be late-type stars), and it will certainly vary from region to region.\footnote{We also anticipate a lower confirmation rate for secondary NIR-selected candidates than SED-selected candidates.} In the (unrealistic) limit that {\em all} OB candidates in every region are eventually confirmed ($f_{\rm conf}\rightarrow 1$), we would have $F_{\rm inc}$ ranging from zero for the three MYStIX regions with no OB candidates, to ${\ga}3$ for regions such as NGC 6334, RCW 38, and Trifid, which have relatively few published OB stars compared to their numbers of identified OB candidates. 

\begin{deluxetable*}{lrrrrccc}
\tabletypesize{\small}
\tablewidth{0pt}
\centering
\tablecaption{Summary of MYStIX OB Populations\label{sum.tbl}}
\tablehead{
\colhead{Region} & \colhead{MOB1c(SED)\tablenotemark{a}} & \colhead{MOB1c(NIR)\tablenotemark{a}} & \colhead{OB(pub)} & \colhead{OB1(pub)\tablenotemark{a}} & \colhead{OB1X(pub)\tablenotemark{a}} & \colhead{$\langle A_{V}^{\rm SED}\rangle$\tablenotemark{b}} & \colhead{$\langle A_{V}^{\rm MS}\rangle$\tablenotemark{c}} 
}

\startdata
Flame Nebula   & 0 & 0 & 2  & 2  & 2  & 8.1 & \nodata \\
W 40           & 0 & 0 & 3  & 2  & 2  & 7.9 & \nodata  \\
RCW 36         & 0 & 0  & 2  & 2  & 2  & 7.2 & \nodata  \\
NGC 2264       & 0 & 0  & 7  & 1  & 1  & 1.3 & \nodata \\
Rosette Nebula & 0 & 1 & 21 & 10 & 6  & 1.7 & 6.4  \\
Lagoon Nebula  & 5 & 0 & 28 & 10 & 7  & 2.2 & 5.2  \\
NGC 2362       & 0 & 0 & 12 & 1  & 1  & 0.5\tablenotemark{d} & \nodata  \\ 
DR 21          & 0 & 2 & 1  & 0  & 1  & \nodata & 35.3  \\
RCW 38         & 2 & 3 & 1  & 1  & 1  & 0.3 & 13.2  \\
NGC 6334       & 14 & 3 & 8  & 8  & 6  & 3.6 & 18.9  \\
NGC 6357       & 25 & 1 & 16 & 15 & 13  & 7.6 & 8.8 \\
Eagle Nebula   & 7  & 1 & 56 & 29 & 21 & 3.4 & 15.6 \\
M 17           & 18 & 7 & 67 & 36 & 26 & 7.2 & 15.5 \\
W 3            & 5  & 3 & 24 & 13 & 8  & 10.5 & 12.0 \\
W 4            & 1  & 0 & 36 & 17 & 9  & 2.4 & 2.4  \\
Trifid Nebula  & 4  & 4 & 2  & 2  & 2  & \nodata & 12.0  \\
NGC 3576       & 8  & 2 & 10 & 5  & 3  & 1.5 & 8.4  \\
NGC 1893       & 9  & 0 & 32 & 18 & 8  & 2.1 & 3.9 \\
\noalign{\smallskip}
\cline{1-6} 
ALL            & 98 & 27 & 328 & 172  & 119 &  \multicolumn{2}{c}{}
\enddata
\tablenotetext{a}{These columns tally the following: ``MOB1c(SED)'' = candidate O through B1 stars identified via SED fitting (Table~\ref{tab:OBc_SED}), ``MOB1c(NIR)'' = candidate O through B1 stars identified via NIR colors (Table~\ref{tab:OBc_NIR}),  ``OB1(pub)'' = all stars with published spectral types of B1 or earlier (Table~\ref{pubOB.tbl}; these are, in principle, luminous enough to have been selected as OB candidates by our methodology), and ``OB1X(pub)'' = published OB stars in each region that were detected in X-rays (indicated in column 16 of Table~\ref{pubOB.tbl}).}
\tablenotetext{b}{Visual extinction derived from SED fitting with frozen $T_{\rm eff}$ parameter for all OB stars with previously published spectral types (average values from column 6 of Table~2).}
\tablenotetext{c}{Main-sequence equivalent visual extinction for the new OB candidates (average of values from column 3 of Tables~3 and 4); this places an {\em upper bound} on the true average extinction to these stars.}
\tablenotetext{d}{Excluding the unreliable SED fitting results for HD 57061.}
\end{deluxetable*}

We also consider {\em completeness}, the (false-negative) rate at which we fail to identify true OB stars as candidates. 
Examining the published OB stars in Table~\ref{pubOB.tbl}, all stars earlier than O9 were detected in X-rays, as well as all O giants and supergiants.  But a decreasing fraction of O9 to B1 main sequence and B giant stars are detected by \Chandra\ in the MYStIX fields. This is consistent with the long-established relationship between X-ray and bolometric luminosities in massive stars attributed to a trend in wind shock properties \citep{gagne}.
Hence one estimate of completeness is the fraction of known O through B1 stars that have MYStIX X-ray detections, 67 of which satisfied our SED fitting search criteria (column 17 of Table~\ref{pubOB.tbl}), while the virtually all of the remaining 55 satisfied our NIR search criteria (Figure~\ref{fig:each}). This fraction, $f_{X} = $ OB1X(pub)/OB1(pub), typically ranges between 0.5 and 1 among studied MYStIX regions (with the exception of NGC 1893, the most distant region, for which $f_X\sim 0.4$), and averages to $\bar{f}_{X} = 119/172 = 70\%$ across all regions (Table~\ref{sum.tbl}).
The observed variation in $f_X$ among the regions is caused primarily by differences in distance and depth of the X-ray observations.



Correcting for the incompleteness in our X-ray search for new OB1 stars, we estimate the increase in the inferred, total massive stellar populations in the MYStIX regions studied here as $F^{\prime}_{\rm inc}=F_{\rm inc}/f_X$. The key unknown parameter governing the potential impact of our OB candidates on the inferred massive stellar population is, unsurprisingly, the confirmation rate.





















\subsection{Spectroscopic Confirmation of OB Candidates}\label{sec:M17conf}

We can begin to constrain $f_{\rm conf}$ by considering the five MYStIX regions (including Carina) that have recent spectroscopic follow-up of a subset of their OB candidates (Section \ref{sec:cand_results}): 

\begin{itemize}

\item {\em NGC 6357}---All twelve OB candidates in Region 3 (see Figure~\ref{fig:each}k) followed up by NIR spectroscopy using VLT/KMOS were confirmed (A. Bik 2015, private communication). An additional two, brighter OB candidates (MOBc NGC 6357 19 and 21) are intermingled with these confirmed candidates and have a high probability of eventual confirmation as newly-discovered, O-type stars (Appendix~\ref{ngc6357.app}). With 26 OB candidates in NGC 6357, we can report $0.5 \la f_{\rm conf} \le 1$. Adding the 14 confirmed and high-probability candidates to the 15 OB1(pub) stars (Table~\ref{sum.tbl}), we have already {\em doubled} ($F_{\rm inc}\sim 2$) the known massive stellar content of this rich star-forming complex.

\item {\em M17}---A subset of the candidates in M17 were observed with the long slit spectrograph of the Wyoming Infrared Observatory (WIRO). The types were determined using the red (5300~\AA--6700~\AA) portion of the spectra, since this includes several \ion{He}{1} lines and one \ion{He}{2} line, and because the blue portion was often not usable, given the high extinctions. Many of the candidates were still too obscured to detect---candidates with $J \ga 10$~mag could not be classified, while all candidates brighter than this limit were assigned spectral types.
Five relatively unobscured candidates were hence confirmed as OB stars, while four others were classified late-type stars (Table~\ref{tab:OBc_SED}).  With 25 OB candidates in M17, we can place only a loose constraint on the confirmation rate, $0.2 \la f_{\rm conf} \la 0.8$. With 36 pubOB1 stars (Table~\ref{sum.tbl}), we have already increased the cataloged massive stellar population by 10\%, with the potential for up to a 60\% increase ($1.1 \la F_{\rm inc} \la 1.6$). High-resolution, NIR spectroscopy is needed to classify the numerous obscured OB candidates located near the crowded NGC 6618 cluster and the M17 giant molecular cloud.

\item{\em Trifid}---Three the OB candidates in Trifid have been spectroscopically confirmed as early-type stars (Kuhn et al. 2016, in preparation), while the remaining five candidates have not yet been classified spectroscopically, hence $0.38 \le f_{\rm conf} \le 1$.

\item {\em W3}---\citet{K+15w3} independently identified and spectroscopically classified four of our candidates as OB stars, while one candidate was previously classified as a late-type star. With eight candidates in W3, we have $0.5 \le f_{\rm conf} \la 0.9$, similar to NGC 6357. 

\item {\em Carina}---\citet{CCCP-AAT} obtained visible-light spectra for 71 of the 94 OB candidates in the Carina Nebula identified by P11, classifying 21 as OB stars and 50 as late-type stars. This constrains the confirmation rate for the Carina OB candidates to $0.2 \le f_{\rm conf} < 0.5$. These relatively low confirmation and high rejection rates for Carina may not be representative of the MYStIX regions. The large CCCP X-ray mosaic surveyed by P11 included a much greater portion of the surrounding Galactic field than was typical for the MYStIX X-ray observations \citep{overview}, and the majority of the {\em rejected} OB candidates are relatively unobscured and located in the outlying regions, away from the principal star-forming clusters.

\end{itemize}

Taken together, the above results favor an average $\bar{f}_{\rm conf}\approx 0.5$ across all MYStIX regions, predicting fractional increases of $\bar{F}_{\rm inc}\approx 0.5(125/172)+1=1.4$ in the cataloged massive stellar populations and $\bar{F}^{\prime}_{\rm inc}=\bar{F}_{\rm inc}/0.7=1.9$.
We therefore conclude that even a conservative 50\% confirmation rate for our OB candidates would double the inferred massive stellar content averaged across the 18 MYStIX star-forming complexes studied here.

\subsection{Possible Contaminating Sources}\label{contam.sec}

P11 predicted that the most likely populations of contaminating sources in their CCCP sample of X-ray selected candidate OB stars would be F or G stars, either field giants or (very nearby) foreground dwarfs. The same predictions can be extended to our MYStIX OB candidates, but the recent spectroscopic followup of the CCCP OB candidates by \citet{CCCP-AAT} allows us to refine our understanding of the most likely types of contaminants. The 50 CCCP OB candidates classified as later-type stars consisted of 32 G through K5 (super)giants, 1 F supergiant, a single K dwarf, and 15 F or A stars for which no luminosity classification was made. None of the CCCP candidates was classified as a late K or M giant or dwarf, confirming the prediction of P11 that the coolest stars, which dominate the Galactic field populations, are not selected as X-ray emitting OB candidates. These results indicate that the most common type of contaminating source in our MYStIX OB candidates sample will be F through early K giants matched to X-ray sources. We note that some of these stars could be evolved massive stars with strong intrinsic X-ray emission.

Seven of the CCCP OB candidates classified as (super)giant stars by \citet{CCCP-AAT} had X-ray spectra of sufficient quality that P11 were able to fit one- or two-component thermal plasma models to measure their absorbing columns and plasma temperatures. All of these sources exhibited at least one soft plasma component with $kT \la 1$~keV (OBc 1, 3, 10, 12, 18, 32, and 41 in Table 5 of P11), consistent with the soft X-ray emission from symbiotic binary systems in which an evolved giant that transfers mass to a white dwarf companion \citep{M+97_symbiotic,L+13_symbiotic}. Symbiotic binaries can produce nova outbursts accompanied by spectacular X-ray flares that decay rapidly as the ejecta interact with the extended wind region surrounding an evolved giant \citep{S+06_Nature}.
In light of its new G0-K0 I spectral classification, such an outburst now provides the best explanation for the unusual X-ray variability observed for CCCP OBc 41 \citep{Townsley11}.

The soft X-ray emission from quiescent, symbiotic binary systems originates in accretion shocks or colliding winds \citep{M+97_symbiotic}, and it is generally not possible to distinguish these systems from OB stars by analysis of the relatively low-resolution MYStIX X-ray spectra. P11 compared $A_V^{\rm MS}$ from SED fitting to absorbing column $N_{\rm H}^{\rm Xfit}$ from X-ray spectral fitting in an attempt to evaluate which of their OB candidates had X-ray spectra consistent with reddened OB stars. Among the nine candidates identified as having X-ray spectra consistent with IR SED-fitting, the results were ambiguous, with three confirmed new OB stars (OBc 39, 55, and 57), three classified as late-type stars (OBc 2, 12, and 67), and three that were not detected by AAT optical spectroscopy \citep{CCCP-AAT}.

Young, yellow giants or exotic, symbiotic binary systems are far less common than red giants in the Galactic field, hence the contamination they may introduce to our samples should not be severe. We do not see systematically increased numbers of candidates in regions that are projected near the Galactic center, as might be expected if our candidate lists were dominated by contamination from IR-bright field giants.  For example, the Lagoon Nebula, Trifid Nebula, NGC~6334 and NGC~6357 all have longitudes within 10$^\circ$ of the Galactic Center, yet the former two regions have few new candidates while the latter host numerous candidates. By contrast, the Carina Nebula is relatively far from the inner Galaxy yet contained nearly as many OB candidates as all 18 of our MYStIX regions combined (P11).

\subsection{Concluding Remarks}

We have identified 125 candidate X-ray-emitting O or early-B (MOBc) stars through a systematic search of the combined X-ray and IR point sources catalogs for 18 of the 20 MYStIX star-forming complexes, supplementing the 328 previously identified by spectroscopy (Table~\ref{sum.tbl}). Follow-up or independent spectroscopy has already confirmed 24 of our candidates as newly-discovered Galactic OB stars. We also modeled the 1--8~\um\ SEDs of 239 previously-published OB stars, with effective temperature fixed based on the spectral type, to measure their interstellar reddening and bolometric luminosities and to provide additional context for our candidate-selection methodology. This analysis identified six candidate massive binary/multiple systems (the principal ionizing stars in NGC 2264 and the Lagoon Nebula, two stars in the Eagle Nebula, and two stars in NGC 1893) and five candidate O-type (super)giants with no published luminosity class (four in M17 and one in NGC 6334).

Perhaps the clearest trend in Table~\ref{sum.tbl} is an anticorrelation with historical research. We find few to no new candidate OB stars in lightly-obscured complexes that have been heavily studied in the past using visible-light photometry and spectroscopy.  The stellar populations of regions such as NGC~2264, the Rosette Nebula, the Lagoon Nebula, the Eagle Nebula, and W4 have been surveyed for decades from northern hemisphere observatories.  In contrast, nebulous and heavily obscured regions accessible only to southern hemisphere observatories (such as RCW 38, NGC~6334, NGC~6357 and NGC~3576) have been less carefully examined.  The majority of previously-cataloged MYStIX OB stars have $A_V\la 3$~mag, while the large majority of candidate OB stars have $A_V \ga 5$~mag, assuming de-reddening to the OB ZAMS (CMDs in Figures~\ref{fig:each}).
Previous work has also typically focused on the most prominent parts of large star-forming complexes, whether a rich cluster or bright \hii region, but now in several cases, notably NGC 6357, M17, Trifid, and Carina (P11; \citealp{CCCP-AAT}), new spectroscopically-confirmed OB stars have been discovered by surveying the less-studied yet often lightly-obscured surrounding regions.

Notable results for individual regions include: identification of the OB population of a recently discovered massive cluster in NGC 6357; an older OB association in the M17 complex; three new luminous O stars near the Trifid Nebula; doubling of the known massive stellar population in NGC 6357 and the potential to more than double the known OB populations in RCW 38, NGC 6334, and NGC 3576. 
In a few cases, we have probably identified the principal ionizing star of an \hii region.  MOBc NGC 6334 11 is a high-luminosity, highly-obscured ($A_{V}^{\rm MS} \simeq 34$) star located at the center of the radio \hii region NGC~6334~D, with associated nebular emission and masers (Appendix~\ref{ngc6334.app}).  
MOBc NGC~6357 21, a very luminous candidate consistent with an O3~V star on the ZAMS, appears to be the dominant member of the NGC 6357 F subcluster, which contains the largest concentration of newly-discovered OB stars in our sample and ionizes the G353.24+0.64 \hii region (Appendix~\ref{ngc6357.app}).  
MOBc Trifid 3, a newly-confirmed O-type giant, likely ionizes the PAH-bright MIR bubble in the far-north region of the Trifid Nebula complex (Appendix~\ref{trifid.app}). Additional discussion of individual regions is presented in the appendix.

The 18 MYStIX regions studied here represent great diversity of heliocentric distance, obscuration and size of the young massive stellar population, and degree of previous study. For one-third of them, existing catalogs of massive stars appear largely complete. For the rest, the priority should be continued spectroscopic follow-up, preferably in the NIR, to assess the true nature of these OB candidates and move ever closer to completing the census of massive stars in these prominent Galactic star-forming complexes.

\acknowledgements
We are grateful to A. Bik for performing VLT spectroscopic follow-up of select OB candidates in NGC 6357 and the Lagoon Nebula and sharing preliminary spectral classifications for inclusion in this work. We are similarly grateful to H. A. Kobulnicky and J. E. Andrews for performing spectroscopic follow-up and classification of select OB candidates in M17 using WIRO.
We thank the anonymous referee for comments and suggestions that improved this work.
M.S.P. is supported by the NSF through award CAREER-1454333.
The scientific results reported in this article are based on observations made by the {\em Chandra X-ray Observatory} and published previously in cited articles. This work is based in part on archival data obtained with the {\em Spitzer Space Telescope}, which is operated by the Jet Propulsion Laboratory, California Institute of Technology under a contract with NASA. 
This work is based in part on data obtained from the United Kingdom Infrared Telescope (UKIRT) as part of the UKIRT Infrared Deep Sky Survey, when UKIRT was operated by the Joint Astronomy Centre on behalf of the Science and Technology Facilities Council of the UK.
This publication makes use of data products from the Two Micron
All-Sky Survey, which is a joint project of the University of
Massachusetts and the Infrared Processing and Analysis
Center/California Institute of Technology, funded by NASA and the NSF. 
This research has made use of the SIMBAD database,
operated at CDS, Strasbourg, France.

\facility{CXO (ACIS)} \facility{Spitzer (IRAC)}  \facility{CTIO:2MASS} \facility{FLWO:2MASS} \facility{UKIRT (WFCAM)}


\appendix

\section{Candidate OB Star Samples in Each MYStIX Region}\label{appendix}

Figure~\ref{fig:each} visualizes the known and candidate OB populations in each MYStIX region in three panels.  The top panel shows the published (blue) and new candidate (magenta) stars superposed on a \Spitzer\ 3.6~\um\ mosaic of the MYStIX region with the \Chandra\ field of view outlined. Notable known and candidate OB stars discussed below are labeled. 
The bottom panels show a $J-H$ vs. $H-K_s$ color-color diagram and a $J$ vs. $H-K_s$ color-magnitude diagram of all MPCM stars in the region, with known and candidate OB stars marked by larger symbols (see figure caption for details). Stars with infrared excess emission are marked or highlighted in red. Together, these figures summarize much of the information listed in Tables~\ref{pubOB.tbl}--\ref{tab:OBc_NIR}. 

\subsection{MYStIX Regions Without OB Candidates} \label{without.sec}

No new OB candidates were found in five of the 18 examined MYStIX star forming regions: 
\begin{itemize}

\item {\em The Flame Nebula}---The NIR color-color and color-magnitude diagrams (Figure~\ref{fig:each}a) do not reveal any bright candidates lacking IR excess emission in this well-studied, very young star-forming region \citep{flame_08book}. The most luminous star in the cluster, IRS 2b, is an extended, embedded IR source. The most luminous infrared point source is NGC~2024~1 (B0.5 V), which reveals its photosphere at wavelengths shorter than 4.5~\um. We modeled this source with A$_{\rm V} = 8.1$ mag (Table~\ref{pubOB.tbl}), close to the averaged cluster extinction $A_{V} \simeq 10$~mag reported by \citet{S+2003_Flame}. 
 
\item {\em W 40}---Of the three OB stars identified with infrared spectroscopy \citep{Shuping-W40}, two were infrared point sources for which we were able to measure $A_V$ and $L_{\rm bol}$ with SED fitting (Table~\ref{pubOB.tbl}), but the most luminous, IRS 1A South, is an embedded, extended MIR source (Figure~\ref{fig:each}b). The other bright stars appearing in the NIR color-magnitude diagram are IR excess sources and hence not selected as candidate OB stars.

\item {\em RCW 36}---This embedded \hii region has two spectroscopically identified OB stars, and the more luminous of the two is the more highly obscured (Figure~\ref{fig:each}c). 

\item {\em NGC 2264}---This very well-studied region \citep {ngc2264_08book} contains no new OB candidates, which is reassuring.
HD~261782 is a bright, evolved intermediate-mass giant star \citep[G5 IIIp;][]{D+07-NGC2264} that would satisfy our secondary NIR selection criteria (Figure~\ref{fig:each}d) but fitting its full SED ruled out an OB star. 
This demonstrates that yellow and red (super)giants can be erroneously identified as OB stars by our NIR photometry criteria, if they happen to have an X-ray detection in the MYStIX sourcelists but lack sufficient MIR photometry for SED fitting. Our SED fitting for HD 47839, the principal ionizing star, returns about twice the luminosity expected for its O7 V((f)) spectral type (Figure~\ref{fig:OB_HRD}). As mentioned in Section~\ref{pubOB.sec}, this could indicate multiplicity.

\item {\em NGC 2362}---Unsurprisingly, we found no OB candidates in the oldest MYStIX cluster \citep{ngc2362_08book}, which lacks any obscuring molecular material (Figure~\ref{fig:each}g). For all of the known OB stars our SED fitting measures very low extinction ($\langle A_V^{\rm SED}\rangle = 0.5$ mag, Table~\ref{sum.tbl}), as expected, except for the bright O giant, HD 57061 (O9 II), for which SED fitting returned anomalously high $A_V^{\rm SED}=2.3$ mag and $\log{L_{\rm bol}^{\rm SED}/\Lsun=6.03}$ (Table~\ref{pubOB.tbl} and Section~\ref{pubOB.sec}). The IRAC [3.6] photometry for HD 57061 is saturated and was not used for SED fitting.
IR excess emission due to non-LTE effects in a strong stellar wind could explain the anomalously high $A_V^{\rm SED}$ for HD 57061, leading to a corresponding overestimate of $L_{\rm bol}^{\rm SED}$ (P11).

\end{itemize}

\subsection{The Rosette Nebula}
In the Rosette Nebula and associated molecular cloud \citep{rosette_08book}, five well-established early-type stars were independently identified by our SED fitting selection:  HD~46150, HD~46056, HD~46149, HD~46106, and HD~259012 (Table~\ref{pubOB.tbl}). No new OB candidates were found via SED fitting, but one candidate, HD 259513, was selected using our secondary NIR criteria (Table~\ref{tab:OBc_NIR}). This star has a late spectral type of unknown provenance listed in SIMBAD.

\subsection{The Lagoon Nebula}

The Lagoon Nebula (M8) is composed of an \hii region, an open cluster dominated by B-type stars, and a giant molecular cloud with a few O stars and dozens of B stars at a distance around 1.3~kpc \citep{ladam8, m8book}. Relatively few of the 28 spectroscopically confirmed OB stars were recovered by our SED fitting analysis because only 15 were matched to X-ray point sources, some of which had poor $JHK_s$ photometry. 

SED fitting identified four new candidate OB stars with $4<A_{V}^{\rm MS} < 7$ (Table~\ref{tab:OBc_SED}), including MOBc Lagoon 4 in the middle of the open cluster (Figure~\ref{fig:each}f), which has been followed up spectroscopically in the NIR using the VLT and confirmed as an early-type star (A. Bik, private communication, 2015). MOBc Lagoon 5 is actually a K-type giant. Some of the OB candidates appear in earlier optical, infrared and X-ray studies of M8 members without being identified as possible massive stars.  

Our SED fitting for HD 164794, the principal ionizing star, returns nearly double the luminosity expected for its O4((f)) spectral type (Figure~\ref{fig:OB_HRD}). As mentioned in Section~\ref{pubOB.sec}, this could indicate undiscovered muliplicity, with the caveat that the presence of a strong wind in this early O-type emission-line star might cause our LTE models to overestimate its $L_{\rm bol}$ (see Section~\ref{sec:OBc_SED}).

\subsection{DR 21}

SED fitting identified no candidate OB stars in the very young complex DR 21, which lies within the much larger Cygnus~X complex \citep{cygnus_08book}. This is not surprising, given that no large \hii region is present in DR21 (Figure~\ref{fig:each}h), and the most luminous IR point sources are protostars and pre-main sequence stars with dusty disks \citep{pov_mires,MYStIXprotostars}. Bright, extended MIR emission associated with embedded massive stars also foils the MYStIX MIR point-source photometry. Two OB candidates were identified via NIR photometry. MOBc DR 21 2 is associated with the extended infrared source [DS96] I1 \citep{DS96-DR21} and may be the ionizing source for this (ultra)compact \hii region. If confirmed, DR 21 2 would be among the most highly-obscured new OB stars in MYStIX, behind $A_V = 33$~mag of extinction (Table~\ref{tab:OBc_NIR}). The other OB candidate, DR 21 1, lies on the edge of the MYStIX X-ray field and is not actually associated with DR 21.

\subsection{RCW 38}

RCW 38 is a centrally concentrated rich cluster producing a very bright \hii region at an uncertain distance around 1.7~kpc \citep{Wolk06}. SED fitting identified two new candidates with luminosities consistent with early B stars (Table~\ref{tab:OBc_SED}). Both are inferred to have high absorptions ($A_V=8$--11~mag). These two stars were previously identified by \citet{Wolk06} among their list of 31 candidate OB stars in RCW 38 (their sources \#107 and \#143 ), but with lower absorptions and thus lower inferred bolometric luminosities. 
We identified three additional candidates via NIR photometry, two of which were previously noted as OB candidates by \citet{Wolk06}. An additional candidate, RCW 38 5, was classified as either a YSO or an OB candidate by \citet{Winston11}. 
The large majority of the 31 OB candidates identified by \citet{Wolk06} have not been selected here. While their methodology is similar to our secondary NIR selection, they include late B-type stars and stars with disks, which are much more numerous (particularly the latter in a very young, embedded region such as RCW 38), than the O through early B-type stars targeted by our search.

\subsection{NGC 6334}\label{ngc6334.app}

NGC~6334, the Cat's Paw Nebula, is a complex of rich clusters and \hii regions in the Sagittarius--Carina spiral arm \citep{ngc6334_08book,feig6334}. Its stellar population has been poorly characterized in the infrared and optical due to heavy and variable extinction, extremely high field star contamination (at a projected location less than $10^{\circ}$ from the Galactic center), and bright nebular emission. 
The most luminous known OB star is 2MASS J17203178-3551111 (O7 with no luminosity class specified; Table~\ref{pubOB.tbl}) with $\log{L_{\rm bol}^{\rm SED}/\Lsun}=5.5$, consistent with an unresolved, O-type binary (or multiple) system, which seems a more plausible interpretation than a single, evolved star, given its location in the midst of the very young, highly embedded clusters (Figure~\ref{fig:each}j). 

Our analysis reveals a rich population of 21 candidate massive stars, mostly concentrated along the filamentary ``backbone'' of the giant molecular cloud that runs parallel to the Galactic plane (Figure~\ref{fig:each}j).
The most luminous OB candidates, if they were on the main sequence, surpass the most luminous stars currently known in the region:  

\begin{itemize}

\item {\em MOBc NGC 6334 6} 
($A_V^{\rm MS}=10$, $\log{L_{\rm bol}^{\rm MS}/\Lsun}=5.8$; Table~\ref{tab:OBc_SED}) lies North of the principal embedded clusters of NGC 6334 (Figure~\ref{fig:each}j), only $20\arcsec$\ from CD~$-35^\circ 11482$, a possible Herbig Be star with MIR excess emission detected (B0.5:Ve, Table~\ref{pubOB.tbl}). 
Both stars appear to be members of the absorbed massive subcluster NGC~6334~G \citep{pov_mires,mikexclus}.

\item {\em MOBc NGC 6334 11} 
is extremely obscured, with $A_V^{\rm MS}\simeq 34$ mag (Table~\ref{tab:OBc_SED}). This star was undetected in $J$ by 2MASS, which is consistent with its extremely high reddening, hence it does not appear on the NIR CCD or CMD Figure~\ref{fig:each}j, but SED modeling indicates a very high luminosity of $\log{L_{\rm bol}^{\rm MS}/\Lsun}=5.6$, equivalent to an early O star on the ZAMS. The infrared source was previously noted as FIR-II~24 by \citet{Straw89}, who classified it as a less absorbed, lower-mass pre-MS star. Its rising MIR SED may indeed indicate some reprocessing of stellar radiation by circumstellar dust, and this, along with its relatively low X-ray median energy ($E_{\rm med} = 2.4$~keV; Figure~\ref{evav.fig}), may indicate that our SED fit with a normal stellar photospheric model overestimated the reddening and hence bolometric luminosity (assuming that the matching between the X-ray and MIR source is correct). It lies near the center of the radio \hii region NGC~6334~D, which is about 2$^\prime$ (1~pc) in diameter \citep{rodriguez}, and at the base of a much larger MIR bubble blowing out to the east (dark, round nebulosity in Figure~\ref{fig:each}j). These nebular structures lie on the southeast edge of the large heavily absorbed subcluster NGC~6334~J \citep{mikexclus}.  Two other OB candidates are located east of NGC 6334 11, inside the MIR bubble.   

\item {\em MOBc NGC 6334 15 and 16} are secondary NIR candidates that would be brighter than any single, O-type dwarf if dereddened to the colors of the ZAMS. While we could not measure $A_V^{\rm MS}$ for these stars, their locations on the CMD (Figure~\ref{fig:each}j) would indicate $A_V\sim 17$ and 25 mag for 15 and 16, respectively. Of the two, NGC 6334 15 is the better candidate because it is located close to the massive, obscured cluster at the southwestern end of the complex.

\end{itemize}

\citet{russeil} recently reported about 15 new candidate OB stars in the region of NGC~6334 covered by MYStIX, based on deep optical $UBV$ color-color diagrams and distance cuts. As this study did not provide clear identifications or coordinates to the individual stars, we cannot compare their results with our sample.

\subsection{NGC 6357}\label{ngc6357.app}

NGC~6357, the War and Peace Nebula, is a large star-forming complex in the Sagittarius--Carina spiral arm near NGC~6334, with three rich clusters. The historical Pismis 24 cluster is observed near the edge of the large annular \hii region G353.1+0.9 (which we will call Region 1, see Figure~\ref{fig:each}k), and two other, more obscured rich clusters are found in the nearby molecular cloud with associated \hii regions G353.1+0.6 (Region 2) and G353.24+0.64 \citep[Region 3;][]{6357cappa2011, MOXC}.  Among the MYStIX regions examined in this study, our SED fitting procedure has found the largest number of candidate OB stars in NGC 6357, 25 (Table~\ref{tab:OBc_SED}), with one additional candidate found by secondary NIR photometric analysis (Table~\ref{tab:OBc_NIR}). \citet{wang_ngc6357} presented a list of 24 candidate OB stars in Region 1 based on high X-ray luminosity or bright NIR emission ($K_s < 10$~mag), and we have independently identified eight of these plus one additional candidate (MOBc NGC 6357 5) that was not previously selected. Only three of the high-luminosity X-ray candidates have been recovered here, and all three had $K_s < 10.2$ mag. Conversely, our SED fitting and NIR photometry criteria have excluded five of the NIR-bright \citet{wang_ngc6357} candidates. Simple, single-band bright magnitude cutoffs, even when combined with X-ray detection, are vulnerable to identifying YSOs with $K_s$-excess emission from disks or late-K through M (super)giants that are excluded by our selection criteria.
 NGC 6357 has also been surveyed using deep $UBVR$ images combined with DENIS/2MASS $IJHK$ photometry by \citet{russeil}.  Candidate massive stars with spectral types O to B3 were identified from $U-B$ and $B-V$ colors after dereddening from multi-band color relations (allowing for spatial variation in the $R_V$ ratio, normally 3.1) and removal of $K$-excess objects.  They report 458 candidate OB stars in their Zone 2 which is similar, but not identical, to the extent of the MYStIX region for NGC~6357.  As they do not provide a catalog of their OB candidates, we cannot make a star-by-star comparison, but our OB candidate list is obviously more selective, and limited to more luminous stars.

Pismis~24 in Region 1 is centered on an unusual concentration of spectroscopically established extremely massive and luminous OB stars including O3 (super)giants \citep{Walborn02}. Cl Pismis 24 17, with spectral type O3.5 III(f*), is the earliest star in the entire MYStIX sample for which we were able to measure a bolometric luminosity ($\log{L_{\rm bol}^{\rm SED}/\Lsun}=5.92$; Table~\ref{pubOB.tbl}), and given that theoretical luminosities for giants and dwarfs converge at the upper end of the ZAMS, our SED-based luminosity appears reasonable (in spite of the likely presence of non-LTE wind effects in this star that could bias our models toward higher luminosities). Cl Pismis 24 1 SW+NEab, the very massive binary system with O4 III(f+) and O3.5 If* components \citep{M-A07-Pismis24}, was the highest-luminosity system in the entire MYStIX sample, as measured by SED fitting, with $\log{L_{\rm bol}^{\rm SED}/\Lsun}=6.22$ (Table~\ref{pubOB.tbl}).
The components of Cl Pismis 24 1 SW+NEab are unresolved by our NIR and MIR photometry. This confusion caused a failure of the automated matching of the 2MASS counterpart to the X-ray source \citep{xnmatch}, so we substituted the 2MASS photometry included with the associated GLIMPSE Archive source. The IRAC [3.6] and [4.5] magnitudes for this star are above the GLIMPSE saturation limits, so we did not include them in our SED fitting.


Region 2 contains two published O-type stars, [N78] 49 and 51. [N78] 49, recently reclassified as an O5.5 IV(f) star \citep{GOSSS16}, is the most promising candidate for the principal ionizing source of the G353.1+0.6 \hii region, and it lies at the center of the very compact, dense NGC 6357 D subcluster \citep{mikexclus} (see lower-right inset image in the top panel of Figure~\ref{fig:each}k).
The star was detected in all four IRAC bandpasses and saturated at [3.6]. The SED rises toward longer MIR wavelengths, which could be due to free-free emission in a strong, optically thick wind, consistent with the observed emission-line spectrum of this star \citep{GOSSS16}. It was strongly saturated in all three UKIRT images, and confusion with neighboring stars in the subcluster created an offset in the 2MASS point source position, causing the MYStIX automated matching algorithm \citep{xnmatch} to report no 2MASS counterpart to the X-ray source. We retrieved the 2MASS photometry matched to the GLIMPSE Catalog source and were able to achieve good SED fits to [N78] 49. The extremely high luminosity ($\log{L_{\rm bol}^{\rm SED}/\Lsun}=5.87$, the fourth most luminous star for which we were able to fit an SED; see Table~\ref{pubOB.tbl} and Figure~\ref{fig:OB_HRD}) is inconsistent with a single, O-type subgiant, suggesting that multiple O stars contribute to the IR source, the luminosity class was underestimated, or both.

There are two OB candidates in Region 2, both of them less luminous than [N78] 49.
MOBc NGC 6357 12 has been confirmed as an OB star by follow-up VLT NIR spectroscopy (A. Bik 2016, private communication). 
NGC 6357 13 is optically bright, with $V = 12.3$ and a photometrically estimated spectral type of O9.5 \citep{Neckel84}; this star was too bright in the NIR for VLT follow-up. Three other OB candidates, 14, 15, and 17, located on the boundary between Regions 2 and 3 have all been spectroscopically confirmed as OB stars using the VLT (Figure~\ref{fig:each}k and Table~\ref{tab:OBc_SED}).
 

Region 3 of NGC 6357 has proven to be the most fruitful hunting ground for new OB stars (see the upper-left inset image in the top panel of Figure~\ref{fig:each}k), with ten candidates clustered within the MYStIX subcluster NGC 6357 F \citep{mikexclus}. {\em All but two of these candidates have already been confirmed as OB stars by VLT follow-up}, including MOBc NGC 6357 26, which was selected via our secondary NIR criteria. The two remaining unconfirmed OB candidates, NGC 6357 19 and 21, could not be observed spectroscopically using VLT because they are too {\em bright} in the NIR ($K_s \le 8$; Table~\ref{tab:OBc_SED}).\footnote{It is recommended that these two bright stars be classified spectroscopically using a smaller telescope than VLT (A. Bik 2016, private communication).}
Most remarkable is the unstudied star MOBc NGC 6357 21, 
appearing very bright in the IR ($K=7.0$, $[5.8]=6.7$) with $\log{L_{\rm bol}^{\rm MS}}=5.8$, equivalent to an O3 V star on the main sequence observed through $A_V^{\rm MS}=6$~mag of extinction.  This star is the best candidate for the principal ionizing source for the G353.24+0.64 \hii region.
The median extinction in Region 3 is $\simeq 8$ mag compared to $5-6$ mag in Regions 1 and 2; one star has inferred $A_{V} = 14$ and another has $A_{V} = 24$.  We thus provide the best census available of candidate massive stars in this easternmost cluster of the NGC 6357 complex.  The large population of probable massive stars paints a picture of a cluster that is nearly as rich as Pismis 24, but hidden behind higher extinction.

LS 4151, on the northern outskirts of NGC 6357, appears to have an uncertain spectral type (O6/7 III, $\log{L_{\rm bol}^{\rm SED}/\Lsun}=5.41$; Table~\ref{pubOB.tbl}), and its SED fitting results place it near the ZAMS (Figure~\ref{fig:OB_HRD}) suggesting a dwarf (or subgiant) rather than a true giant star.

On the far northwestern outskirts of NGC 6357, 
MOBc NGC 6357 1 is associated with an arcuate 24~\um\ nebula that is likely a bow shock caused by its ejection from the compact cluster core \citep{Gvaramadze11}.  These authors estimated a photometric spectral type of O9-O9.5, which is consistent with our SED fitting results (Table~\ref{tab:OBc_SED}).  

\subsection{The Eagle Nebula}\label{eagle.app}

The Eagle Nebula (M16) lies near the Omega Nebula (M17), about 2~kpc from the Sun in the inner Galactic plane \citep{eagle08_book}.  Field star contamination is very high.  The ionizing cluster, NGC~6611, has a well-studied population of 13 O stars and over 50 B stars \citep{Walker61}; the earliest is HD~168076 with spectral type O4 V((f+)) (Table~\ref{pubOB.tbl}). SED fitting for HD 168076 and another star, NGC 6611 161 (O8.5 V), gives nearly double the luminosity expected for their spectral types (Figure~\ref{fig:OB_HRD}). This could indicate undiscovered multiplicity (Section~\ref{pubOB.sec}) or that the stars are actually O-type giants that have been misclassified as dwarfs. In the case of HD~168076 the presence of a strong wind might cause our LTE models to overestimate its $L_{\rm bol}$ (see Section~\ref{sec:OBc_SED}). In contrast, NGC 6611 421 and 283, both classified as Be stars, have significantly lower $L_{\rm bol}^{\rm SED}$ compared to expectations for their spectral types.  Both stars are located on the outskirts of the main NGC 6611 cluster (Figure~\ref{fig:each}l). Their lack of MIR excess emission suggests that they could be evolved, classical Be stars rather than Herbig Be stars with dust disks. They could be background stars unassociated with Eagle, or their spectral types are actually later than reported. 

We report eight new candidate OB stars, seven from SED fitting and one extreme candidate from secondary NIR selection (Table~\ref{tab:OBc_SED} and \ref{tab:OBc_NIR}).  
The most luminous SED-fitting candidate, MOBc Eagle 6, lies 16\arcmin\/ east of HD~168076 but could be reddened by as much as $A_V^{\rm MS} = 19$ mag of visual extinction. The luminous, secondary NIR candidate MOBc Eagle 8 is dubious, as its NIR colors are close to the reddened giant locus in the CCD, it is too bright in $J$ for dereddening to the OB main sequence, and it is located on the far northern end of the MYStIX X-ray field, where the \Chandra\ off-axis resolution is relatively poor. This increases the likelihood of a chance match between the X-ray point source and bright IR star in this very crowded field (Figure~\ref{fig:each}l). 

Most of the NGC~6611 cluster stars are lightly absorbed with $\langle A_V^{\rm SED}\rangle \sim 3$~mag, 
while the new candidates appear to probe much more deeply into the cloud, with mean $\langle A_V^{\rm MS}\rangle \sim 15.6$~mag (Table~\ref{sum.tbl}). The Eagle Nebula displays the greatest discrepancy in extinction between published and candidate OB stars (14~mag) of all the MYStIX regions except NGC 6334.\footnote{NGC 6334 has been far less studied than Eagle, and its population of published OB stars is far smaller (Table~\ref{sum.tbl}).}

\subsection{M17}\label{m17.app}

The ionizing cluster of M17, NGC~6618, is extremely rich, dominated by a pair of O4+O4 spectroscopic binaries, a dozen other O stars, and many B stars \citep{Townsley03,Hoffmeister08_M17}. Due to the combination of extremely bright MIR nebulosity and crowding of stars within the dense cluster, \Spitzer/IRAC photometry is unavailable for a large fraction of the published OB stars in NGC~6618 (Table~\ref{pubOB.tbl}). We would therefore not expect our SED fitting selection, which depends on MIR photometry, to be very effective at identifying new candidates in NGC 6618 even if the cluster were not already very well studied. Nevertheless, our SED fitting did find three new OB candidates on the outskirts of NGC 6618, MOBc M17 4, 5, and 6 (Figure~\ref{fig:each}m), that may have gone unidentified as hot, massive stars due to very high extinction ranging from $A_V^{\rm MS}=14$ to 40~mag (the latter value, for M17 4, was the maximum extinction allowed for the SED fitting). The brightest of these, M17 6, would be an early O star on the main sequence. Five of the seven OB candidates identified through secondary NIR selection in M17 are located near or within NGC 6618, the most notable of these being MOBc M17 23, with a projected location near the center of the cluster but behind an estimated extinction of $A_V^{\rm MS}=38$~mag (Table~\ref{tab:OBc_NIR}); this candidate could be another highly-obscured mid-O star.

MOBc M17 3 is located within a subgroup of known OB stars associated with the M17~X subcluster immediately north of the main NGC 6618 cluster \citep{Broos07}, which is dominated by the previously identified O6 star NGC~6618 B~260 \citep{Hoffmeister08_M17}.\footnote{NGC 6618 B 260 drives the large stellar-wind bow shock M17-S2 \citep{Povich08_bowshocks}, and two other candidate bow shocks are associated with less luminous stars in this subcluster.}
Two NIR-selected candidate OB stars, MOBc M17 24 \& 25, are found within the highly obscured X-ray cluster M17 North \citep{Broos07}, which is embedded within an area of MIR nebulosity extending north from the main \hii region between two IR dark clouds that run approximately north-south (Figure~\ref{fig:each}m). MOBc M17 24 lies behind as much as $A_V^{\rm MS}=37$~mag of extinction (Table~\ref{tab:OBc_NIR}).

Half of the OB candidates identified via SED fitting are found in a loose group far to the northeast of the main NGC 6618 cluster. Five of these, MOBc M17 7, 11, 12, 17, and 18, have already been spectroscopically followed up using WIRO and confirmed as OB stars (Table~\ref{tab:OBc_SED} and Figure~\ref{fig:each}m). These new OB stars are part of an older, more dispersed massive cluster or OB association, NGC 6618PG, which likely ionizes the extended \hii region and MIR bubble M17 EB described by \citet{povichm17b}. Two massive, luminous members of NGC 6618PG have been previously classified spectroscopically, these are BD-16 4816 (O5.5 V((f))z) and LS 4957 (also known as BD-16 4831; B0 Ib), several others have been previously identified as OB candidates based on spectral types estimated from photometry only \citep{Reed03OBcat,povichm17b}.
Four previously-classified O-type stars with no published luminosity class (Cl* NGC 6618 B 137, Cl* NGC 6618 SLS 373, Cl* NGC 6618 SLS 17, and BD-16 4818) have high luminosity from SED fitting, consistent with evolved giants (Figure~\ref{fig:OB_HRD}). In spite of their historical designations suggesting membership in the very young, embedded NGC 6618 cluster, these four stars may also be part of the earlier generation of star formation in M17. The most extreme of these, SLS 373, could be an O-type supergiant ($\log{L_{\rm bol}^{\rm SED}/\Lsun}=5.83$; Table~\ref{pubOB.tbl}).

In contrast to the NGC 6618PG field, only three OB candidates and no known OB stars are found to the east of NGC 6618, where the diffuse X-ray outflow from the \hii region is observed \citep{Townsley11b}. The \Chandra/ACIS total exposure on the eastern region was 85~ks compared to 40~ks for the NGC 6618PG field \citep{MOXC}. Comparing the lack of X-ray detected massive stars in this deeper observation of a similarly unobscured field to the relative wealth of newly-confirmed OB stars in the NGC 6618PG field reinforces our conclusion that NGC 6618PG is a distinct, loose OB cluster or association and not, for example, a uniformly-distributed group of misclassified field stars or runaway OB stars from NGC 6618.

As mentioned in Section~\ref{pubOB.sec}, four previously identified O stars in M17 have $L_{\rm bol}^{\rm SED}$ that is significantly lower than expected for their published spectral types. These stars (Cl* NGC 6618 B 272, B227, B213, and B 136) all have projected locations within the obscured \hii region, so they are very likely members of NGC 6618, but spectral classification is especially difficult in regions of high obscuration and nebular contamination, hence they may be B-type stars rather than O9/9.5 stars as reported by \citet{Hoffmeister08_M17}. Two of the four lack MYStIX X-ray detections, which is unusual among the published O-type stars but more common among B-type stars (Table~\ref{pubOB.tbl}).

\subsection{W3 and W4}

The huge W3-4-5 complex near Galactic longitude 135$^\circ$ has been extensively studied \citep{Heyer98,w345_08book}.  W3~Main is a rich embedded cluster with many small \hii regions, W3(OH) is a clumpy group of nascent stars, and an isolated O star is found in W3~North \citep{Feigelson08}.  The lightly-obscured and older IC~1795 cluster \citep{R+11-IC1795} lies between W3 Main and W3(OH) (Figure~\ref{fig:each}n). The W4 complex (Figure~\ref{fig:each}o) lies one degree east of W3.  It has one well-studied, young massive cluster located near the center of a bipolar superbubble. 

In W3, we report seven OB candidates in the vicinity of the W3~Main cluster, plus one candidate near W3(OH), MOBc W3 8, that has been spectroscopically classified as an early B-type binary system \citep{R-I+11-W3} but was omitted from the MYStIX compilation of OB stars \citep{Broos13}.  
The most luminous OB candidate, MOBc W3 1, has been recently identified independently and spectroscopically classified as an O-type binary system by \citet{K+15w3}, and these authors also classified two of our other candidates, MOBc W3 3 and 5, as B-type stars (Table~\ref{tab:OBc_SED}). The most highly obscured OB candidate, MOBc W3 6, comes from our secondary NIR selection with an estimated $A_V^{\rm MS}=27$~mag. This star is located among the known OB population and (ultra)compact \hii regions in W3 Main and was classified as a candidate YSO by \citet{R-I+11-W3}. Our available NIR photometry does not indicate a $K_s$-excess from circumstellar material (6 is the topmost star in the NIR CCD of Figure~\ref{fig:each}n), but we cannot rule out MIR excess emission for this star, which is bright at 3.6~\um.
The absorptions to the other MYStIX candidate OB stars range over $4 < A_{V}^{\rm MS} < 17$~mag, generally higher than the $A_V \sim 4$~mag for optically bright members studied by \citet{Oey05}.  It is possible that some are outlying members of the W3~Main cluster rather than the IC~1795 cluster. 

The X-ray properties and visible-light spectra of the massive stellar population of W4 were recently studied in detail by \citet{RN16_W4}. Our SED fitting of HD 15570, the dominant ionizing supergiant star with spectral type O4 If+, returned an anomalously high $A_V^{\rm SED}=3.7$ mag compared to the average $\langle A_V\rangle = 2.4$ mag for W4 (Table~\ref{sum.tbl}) and a correspondingly high $\log{L_{\rm bol}/\Lsun}=6.10$ (Table~\ref{pubOB.tbl}). The MIR photometry for HD 15570 \citep{mikemir} does not appear heavily affected by saturation, but the rising SED toward [5.8] and [8.0] triggered a marginal IR excess flag \citep{pov_mires}, so these two longest-wavelength bandpasses were not used in our SED fitting for this star. This MIR excess could be produced by an optically thick wind, which would cause the static, LTE stellar atmosphere models used in our SED fitting to overestimate both luminosity and exctinction (P11). Indeed, \citet{B+12-Osupergiants} derived a significantly lower $\log{L_{\rm bol}/\Lsun}=5.94$ for HD 15570 using a more appropriate, non-LTE wind model for the FUV--visible spectrum. 

We identify only one candidate OB star in W4, at the extreme western corner of the ACIS-I field (Figure~\ref{fig:each}o). This candidate may be spurious, because while it satisfied our NIR--MIR SED fitting criteria, its NIR colors and magnitudes alone, while uncertain, are inconsistent with a reddened OB star. It could also be a mismatch of the  X-ray source with the IR counterpart, given the reduction in positional accuracy for far off-axis \Chandra\ point sources.

\subsection{The Trifid Nebula}\label{trifid.app}

The Trifid Nebula (M20) \hii region in the inner Galactic plane \citep{rhotrif} is illuminated by a young cluster dominated by the O7.5 Vz star HD 164492A, and the surrounding field suffers from extreme field star contamination (Figure~\ref{fig:each}p).  Due to the combination of bright nebulosity and filamentary cloud obscuration, optical study of the cluster is poor; only a few B stars, and no additional O stars are known in the cluster.  The northern component of the nebula $\sim 10$\arcmin\ north of HD 164492 appears as a blue reflection nebula in the optical band.  The A7~Iab supergiant HD~164514 lies in the middle of the reflection nebula, but it is debated whether it is the source of radiation \citep{Voshchinnikov75, Lynds86}.  A third nebular component lying 5--10\arcmin\ further to the north consists of a MIR bubble (CN 95 in the \citealp{bubbles2} catalog) not observed in visible light. \citet{Rho06} suggested this third component is an independent \hii region with active star formation and an unidentified principal ionizing star. 

Our SED-fitting search uncovered four candidate massive stars, including the newly-confirmed early-type stars MOBc Trifid 1, 2, and 3, located 1\arcmin\ apart inside the CN 95 bubble in the extreme northeast corner of the MYStIX X-ray field (Kuhn et al. 2016, in preparation). Trifid 2 and 3 are among the most luminous OB candidates in our sample, equivalent O3 V stars on the main sequence. 
The spectral classification and SED fitting luminosity for Trifid 2 indicate an evolved O6--O8 (super)giant, and this star is likely the principal ionizing source of the northern portion of the Trifid Nebula/CN 95 bubble \citet{Rho06,bubbles2}.  
These three stars are spatially associated with and have similar extinction ($A_{V} \simeq 5$--6.5~mag) to Trifid~D, a subcluster of lower-mass pre-main sequence stars identified by \citet{mikexclus}.  We note the possibility that this cluster and associated MIR bubble is only coincidentally associated with the Trifid Nebula to the south, however 
this region is unlikely to lie in front of the blue reflection nebula seen in visible light.

Four candidate OB stars identified by our secondary NIR criteria and one other candidate identified via SED fitting are found elsewhere in the Trifid field, most of them near the main nebula (Figure~\ref{fig:each}p) and none, to the best of our knowledge, previously noted in the literature.
 
\subsection{NGC 3576}

NGC 3576 is an embedded \hii region in the Sagittarius-Carina spiral arm, projected near the more distant NGC~3603 young rich cluster and \hii region. Its stellar content is relatively poorly studied; most of the ionizing OB stars appear to be highly obscured and are not yet identified \citep{figueredo2002}.  The MYStIX field includes the embedded cluster and a second \Chandra/ACIS-I pointing to the north \citep{Townsley11b} that reveals an older cluster dominated by HD~97319 (O9.5 Ib) and EM Car/HD~97484 (O8+O8 V binary). 

We report 10 new OB candidates in NGC~3576, all but one in the southern ACIS-I pointing containing the embedded \hii region (Figure~\ref{fig:each}q).  The southern obscured stars range in estimated extinction over $5.5 < A_V^{\rm MS}<16$~mag. The most luminous candidate is MOBc NGC 3576 3, which would be an O7.5~V star if dereddened by $A_V^{\rm MS}=6.8$~mag to the ZAMS (Table~\ref{tab:OBc_SED}), and it appears to ionize a second, smaller \hii region a few arcmin southwest of the main embedded \hii region. The sole new candidate in the northern, lightly obscured cluster is TYC 8959-1106-1 (MOBc NGC 3576 7 with $A_{V}^{\rm MS} = 3.4$~mag).

\subsection{NGC 1893}

NGC 1893 is the most distant MYStIX star-forming region, located toward the Galactic anticenter with low obscuration.  The low-mass stellar population has been studied mostly based on the deep \Chandra\ observation used by MYStIX \citep{pris2011}. 
There are nine new candidate late-O and B-type stars, although two of them (MOBc NGC 1893 4 \& 5) may be foreground A-type stars (Table~\ref{tab:OBc_SED}).  None lie in dense cluster cores or are associated with optical nebulosity, and most appear to trace the extended OB population reported by \citet{negue2007}. The most luminous OB candidate is MOBc NGC 1893 9, located toward the edge of the field beside a prominent pillar that does not appear to know about this star (Figure~\ref{fig:each}r). This unremarkable location, coupled with the low extinction ($A_V^{\rm MS}\simeq 3$~mag; Table~\ref{tab:OBc_SED}) increases the odds that this candidate is actually an unassociated field star.

SED fitting for HD 242926 returns approximately double the luminosity expected for its O8((f)) spectral type (Figure~\ref{fig:OB_HRD}). This could indicate undiscovered multiplicity (Section~\ref{pubOB.sec}), with the caveat that the presence of a strong wind in this  emission-line O star might cause our LTE models to overestimate its $L_{\rm bol}$ (see Section~\ref{sec:OBc_SED}).


\clearpage
\newpage

\clearpage

\begin{figure*}[p]
\centering
\includegraphics[height=0.4\textheight]{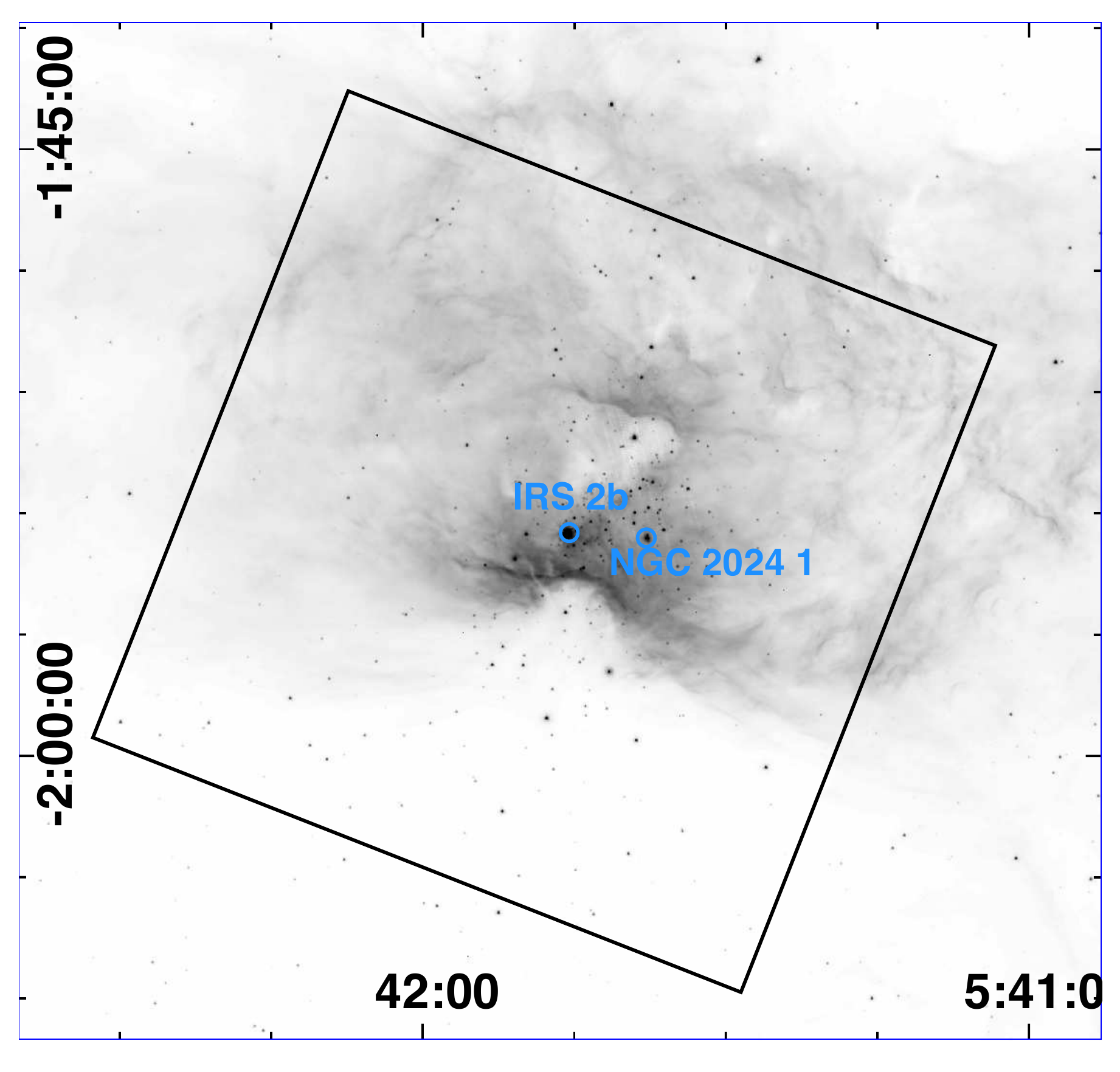}\\
\includegraphics[height=0.3\textheight]{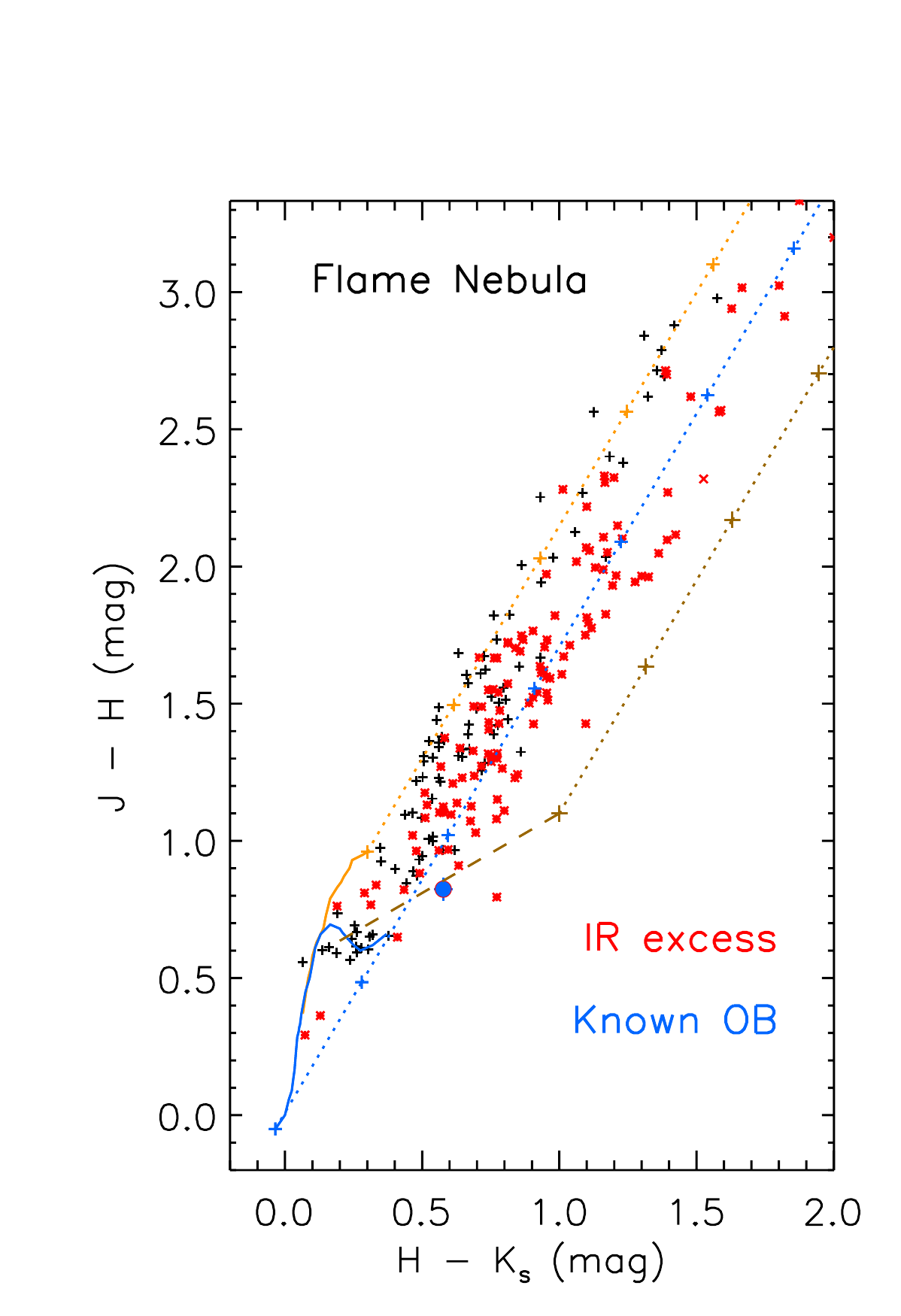}
\includegraphics[height=0.3\textheight]{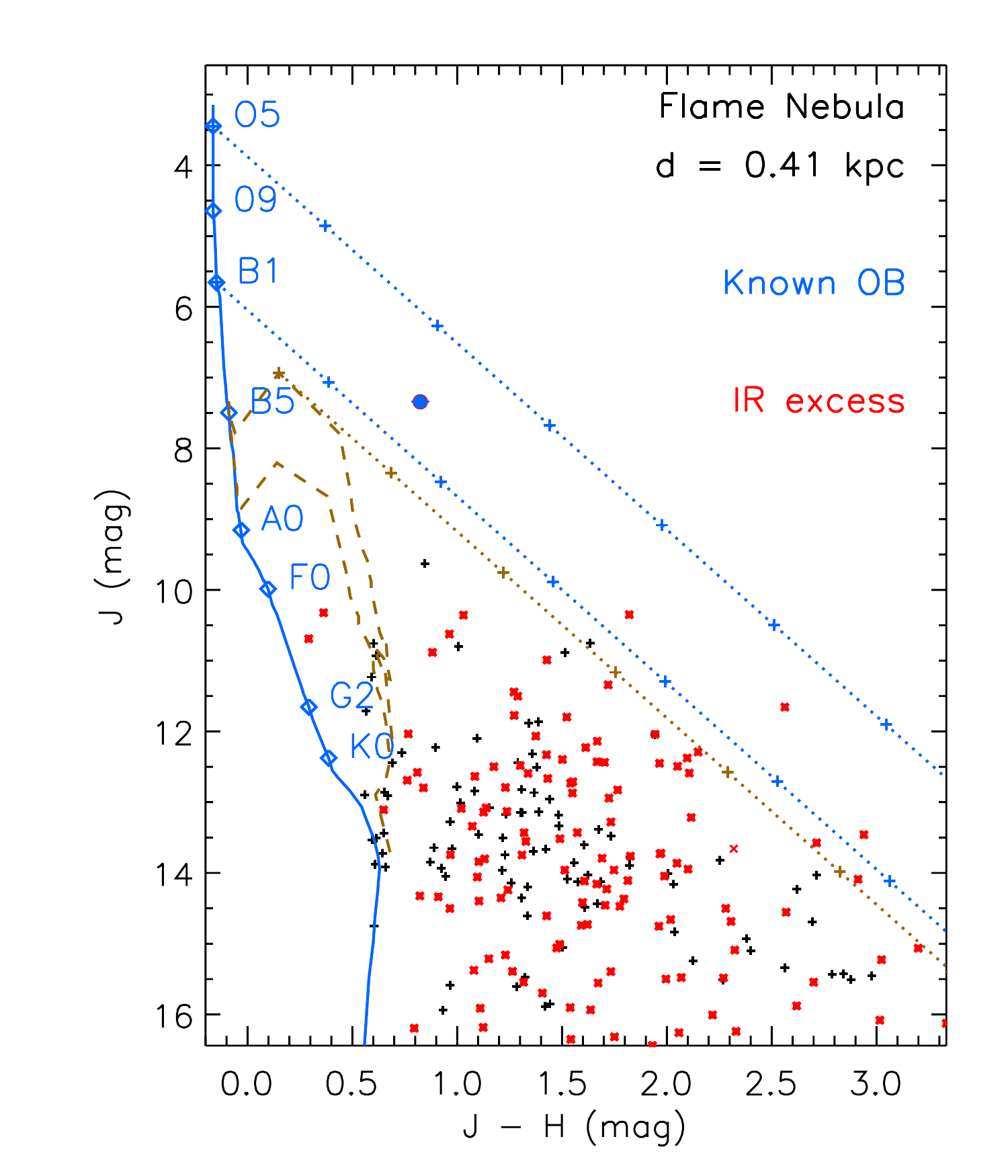}

~ \\
Fig.~\ref{fig:each}({\it a}).--- The Flame Nebula.

\end{figure*}

\begin{figure*}[p]
\centering
\includegraphics[height=0.45\textheight]{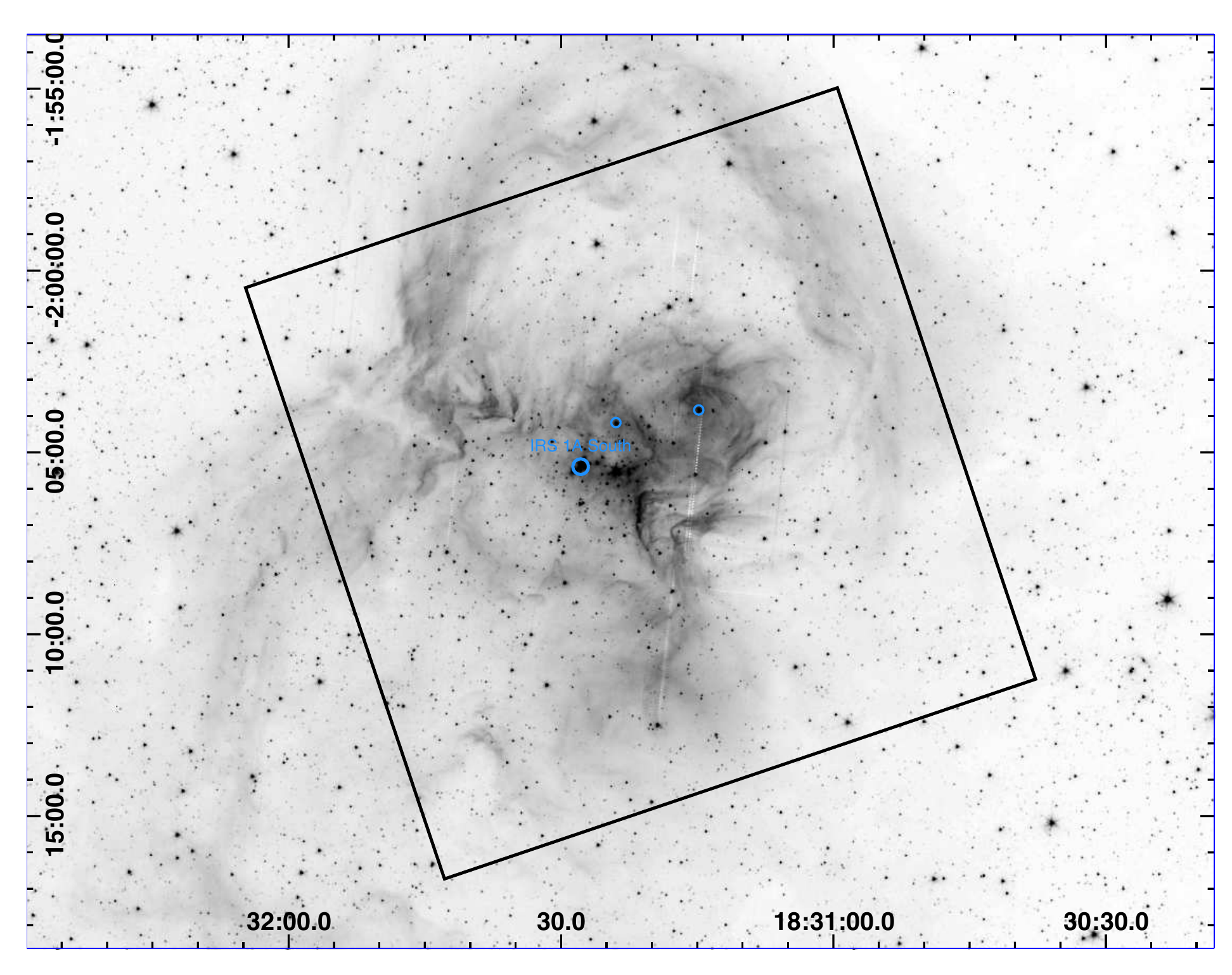}\\
\includegraphics[height=0.4\textheight]{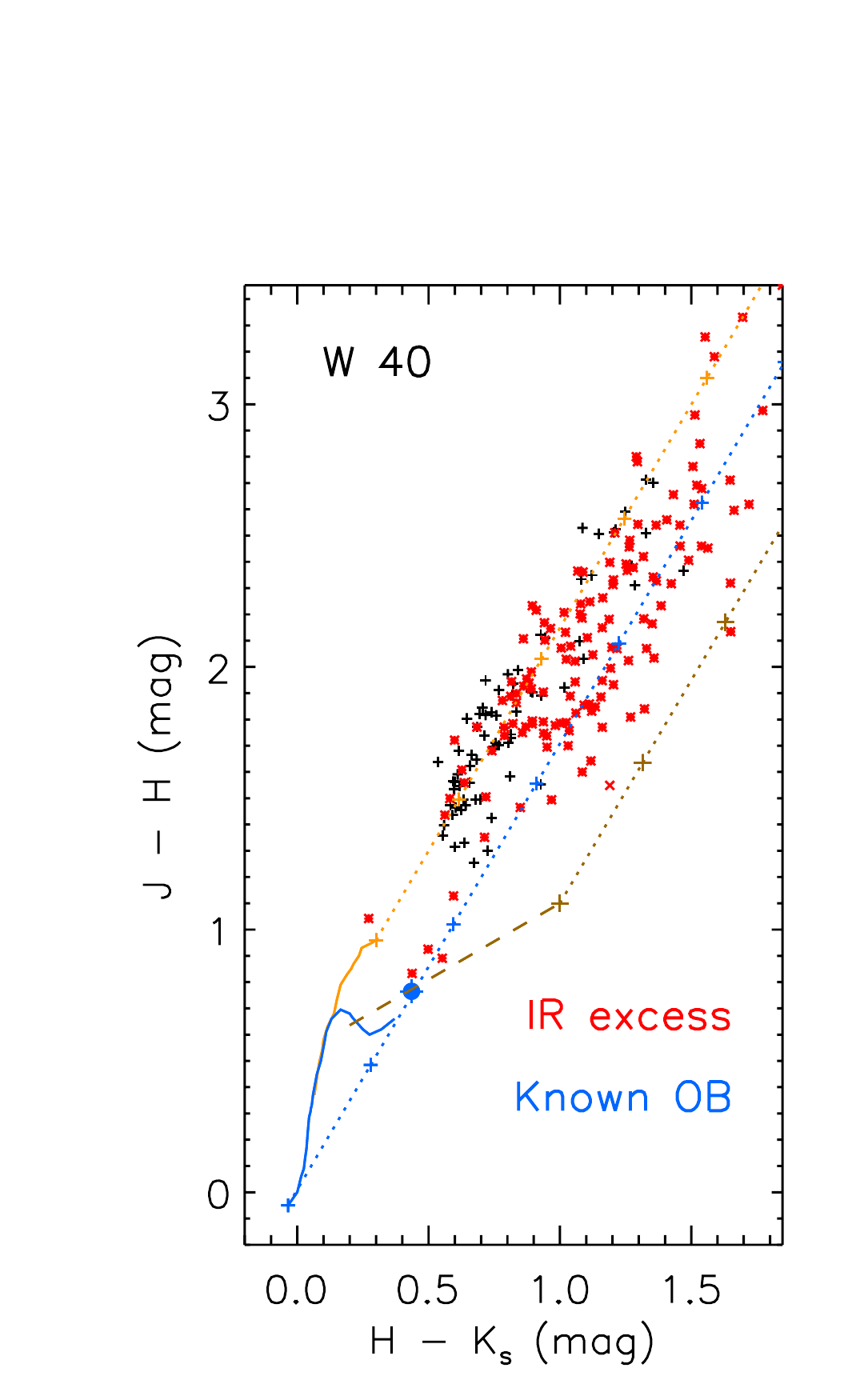}
\includegraphics[height=0.4\textheight]{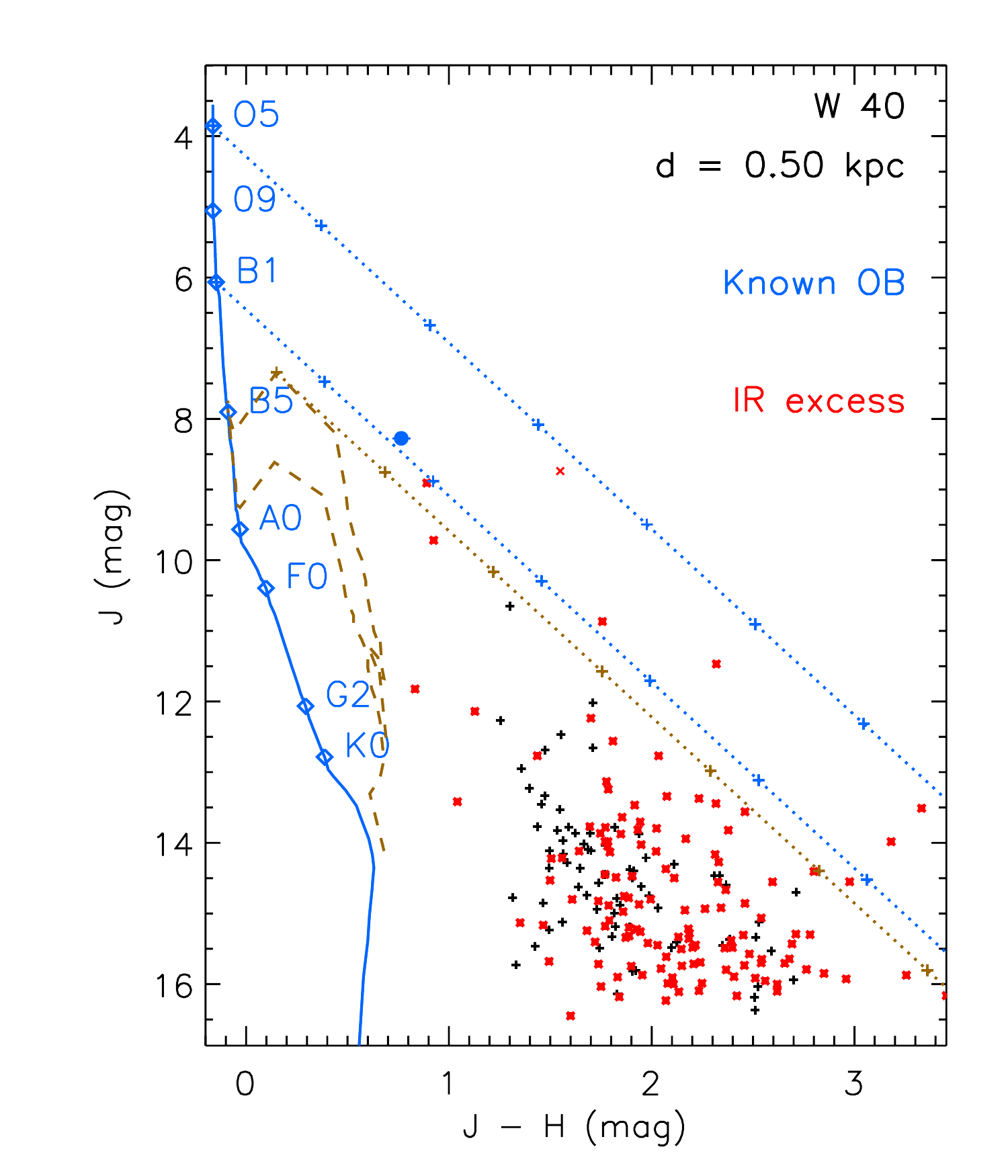}

~ \\
Fig.~\ref{fig:each}({\it b}).--- W40.

\end{figure*}

\begin{figure*}[p]
\centering
\includegraphics[height=0.45\textheight]{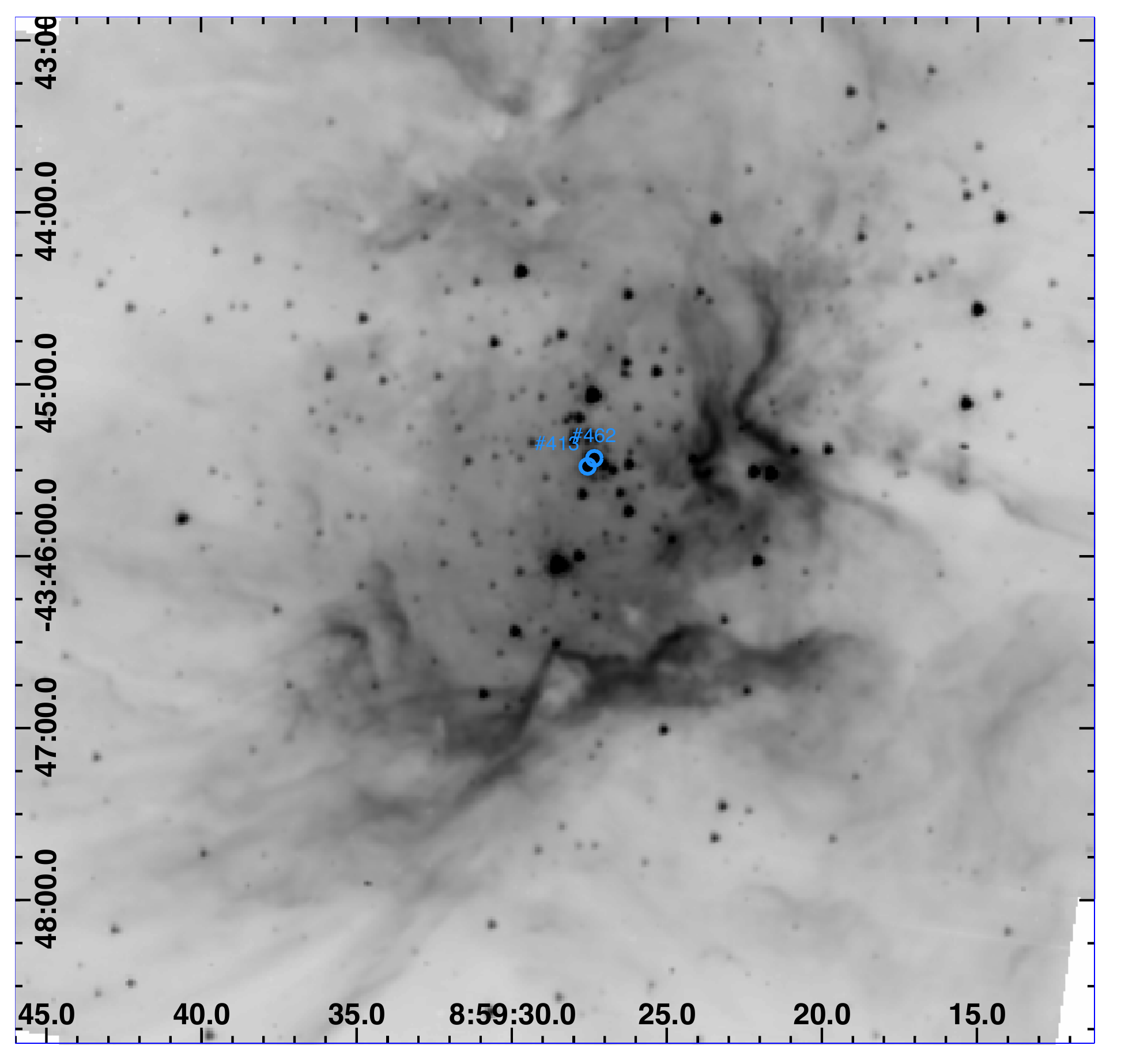}\\
\includegraphics[height=0.4\textheight]{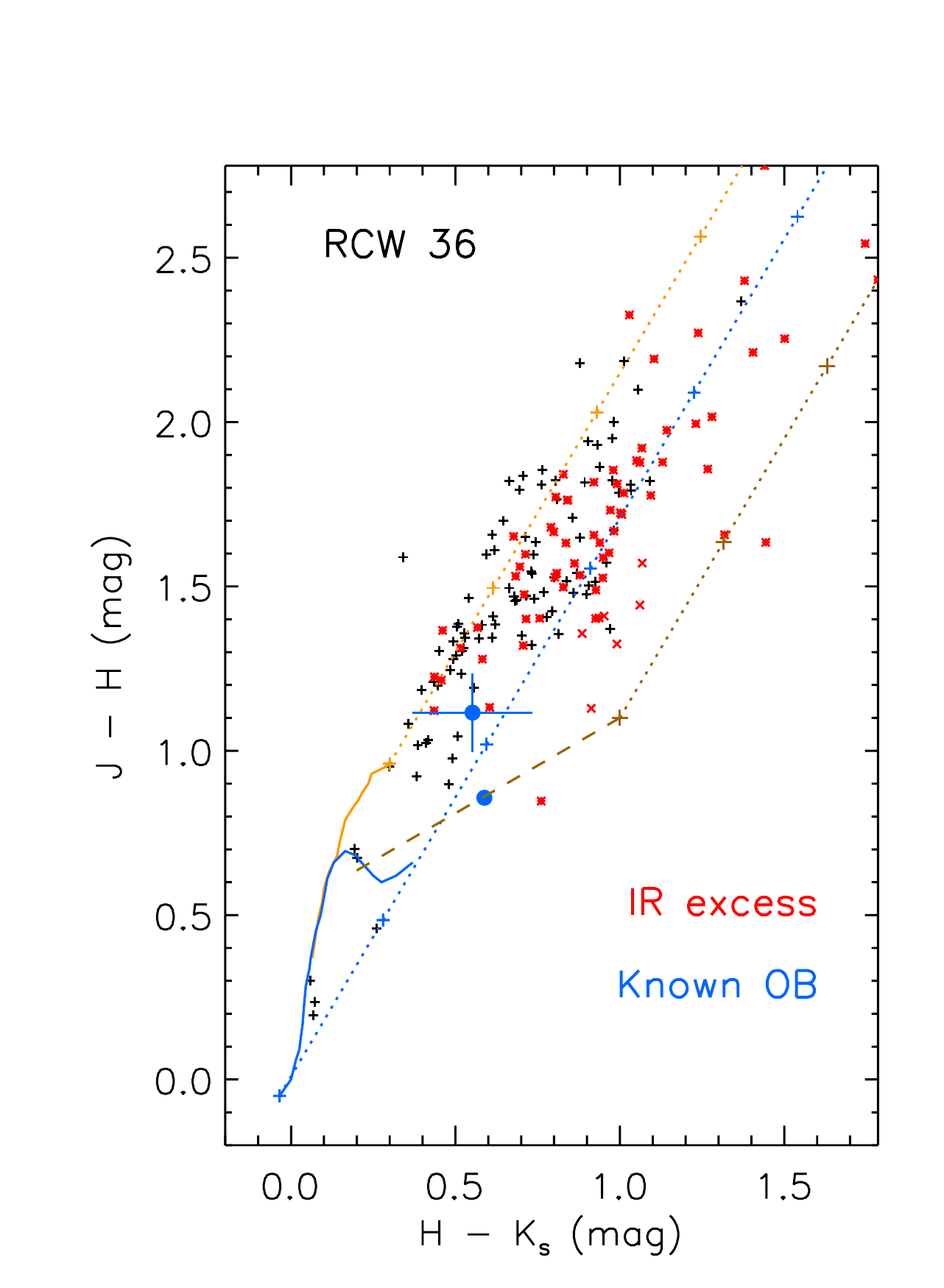}
\includegraphics[height=0.4\textheight]{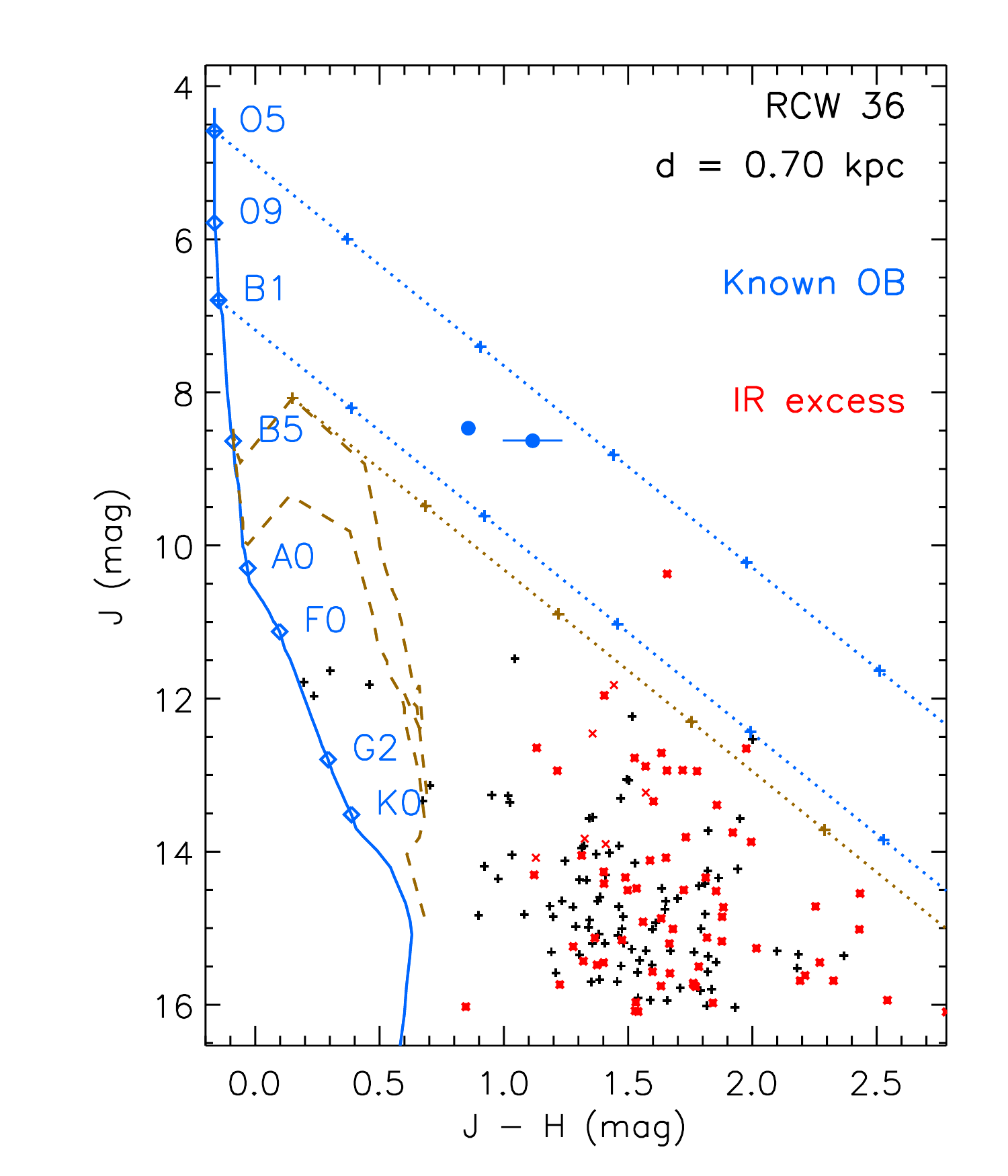}

~ \\
Fig. \ref{fig:each}({\it c}).--- RCW 36.

\end{figure*}

\begin{figure*}[p]
\centering
\includegraphics[height=0.45\textheight]{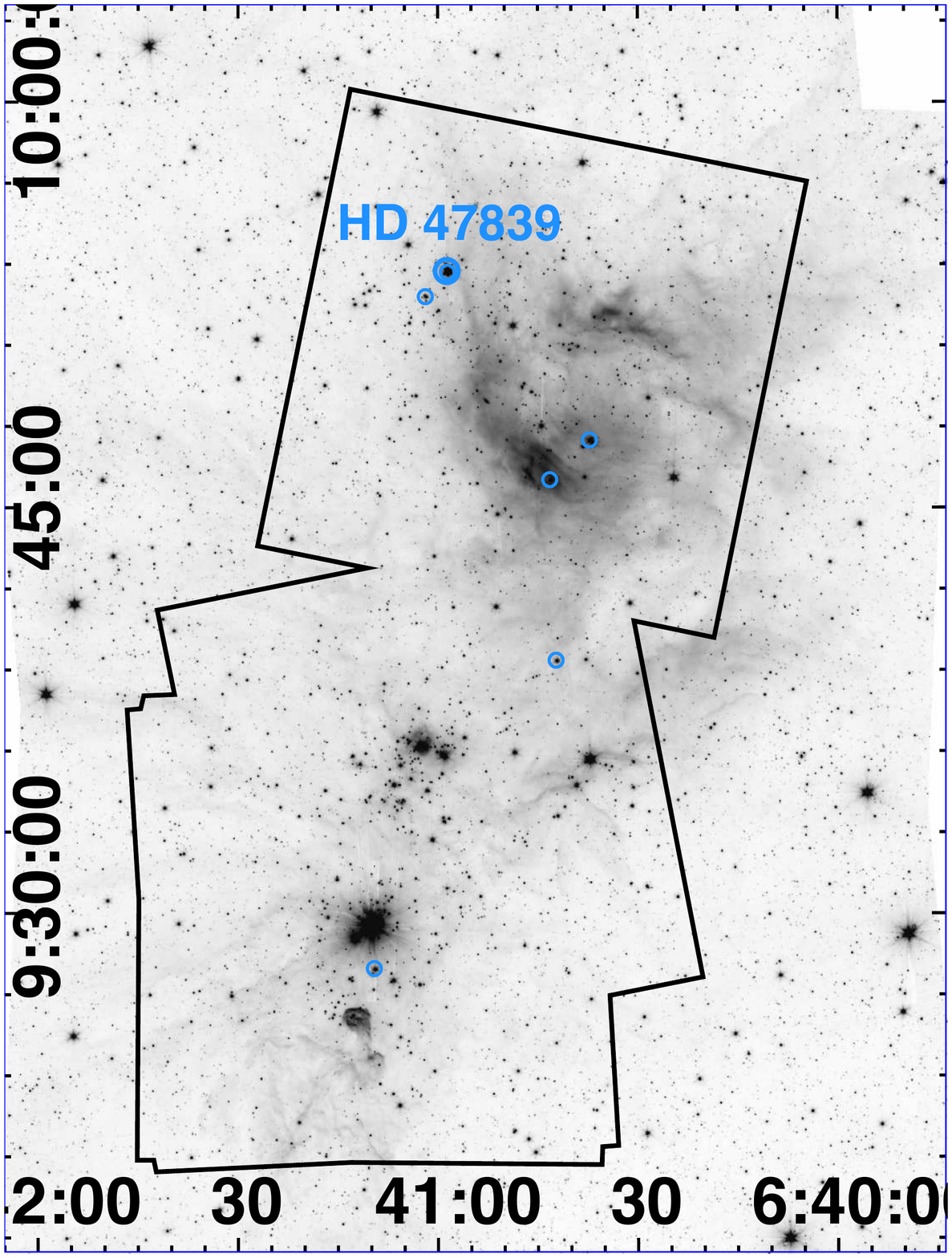}\\
\includegraphics[height=0.4\textheight]{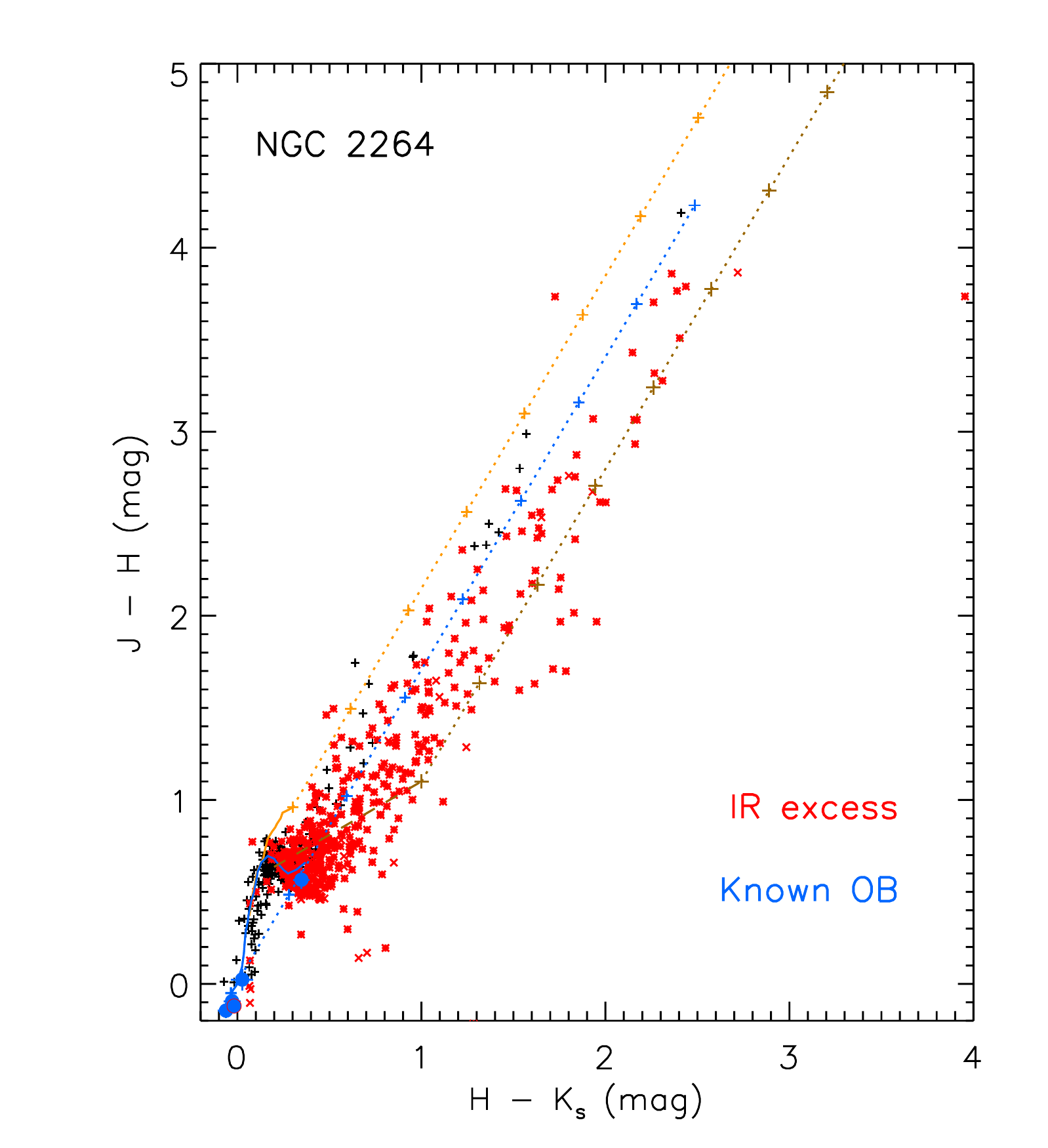}
\includegraphics[height=0.4\textheight]{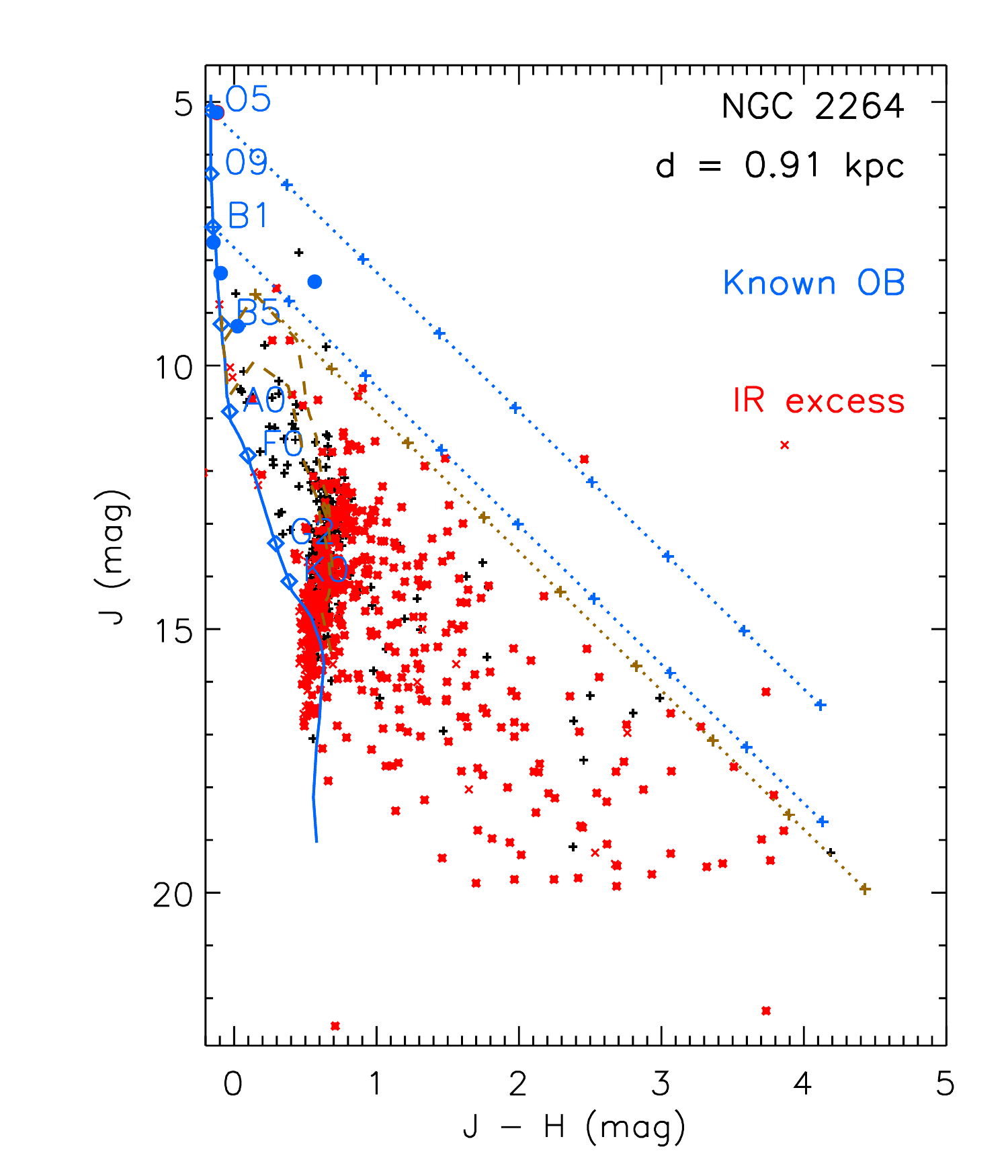}

~ \\
Fig. \ref{fig:each}({\it d}).--- NGC 2264.

\end{figure*}

\begin{figure*}[p]
\centering
\includegraphics[height=0.45\textheight]{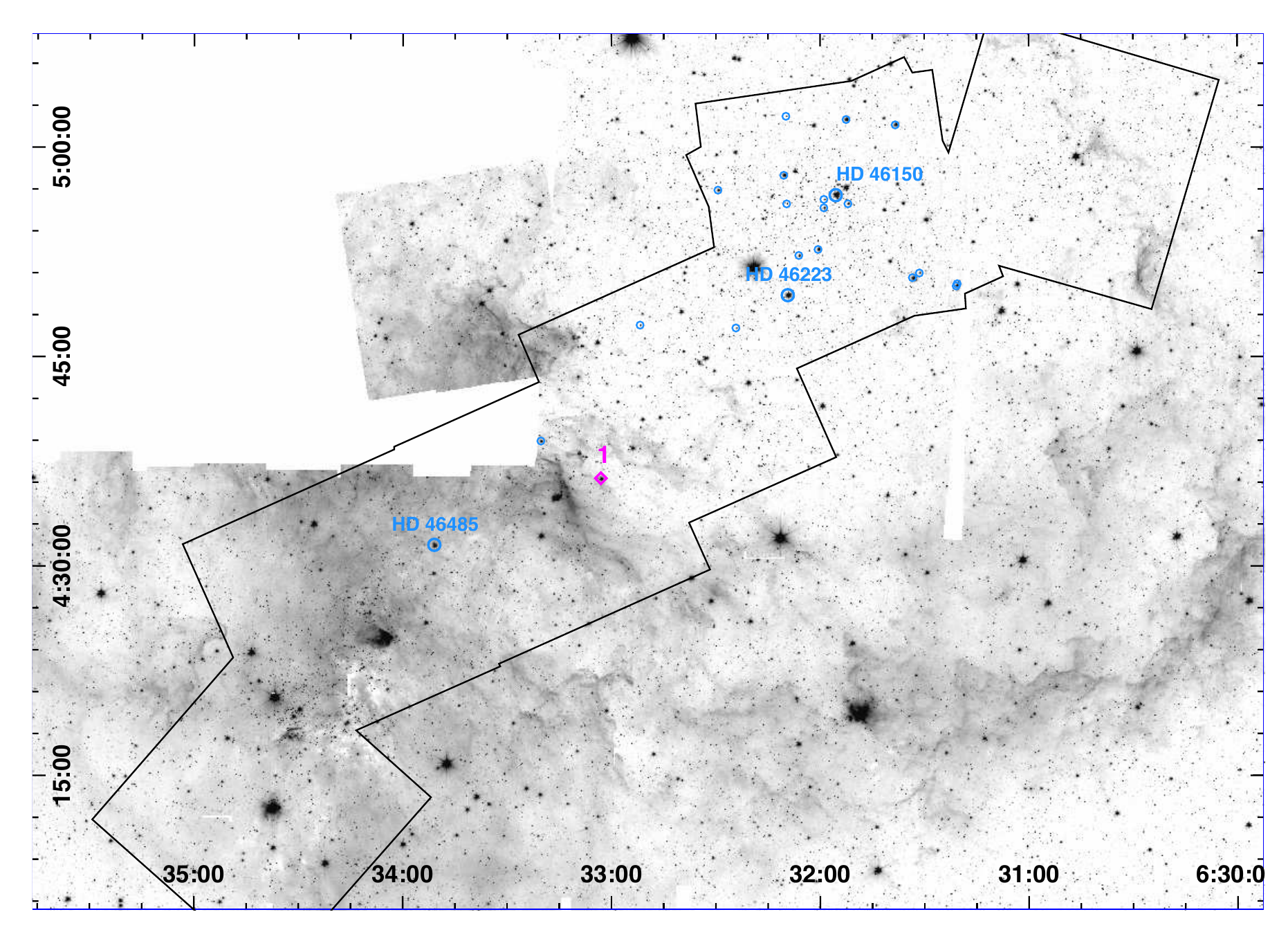}\\
\includegraphics[height=0.4\textheight]{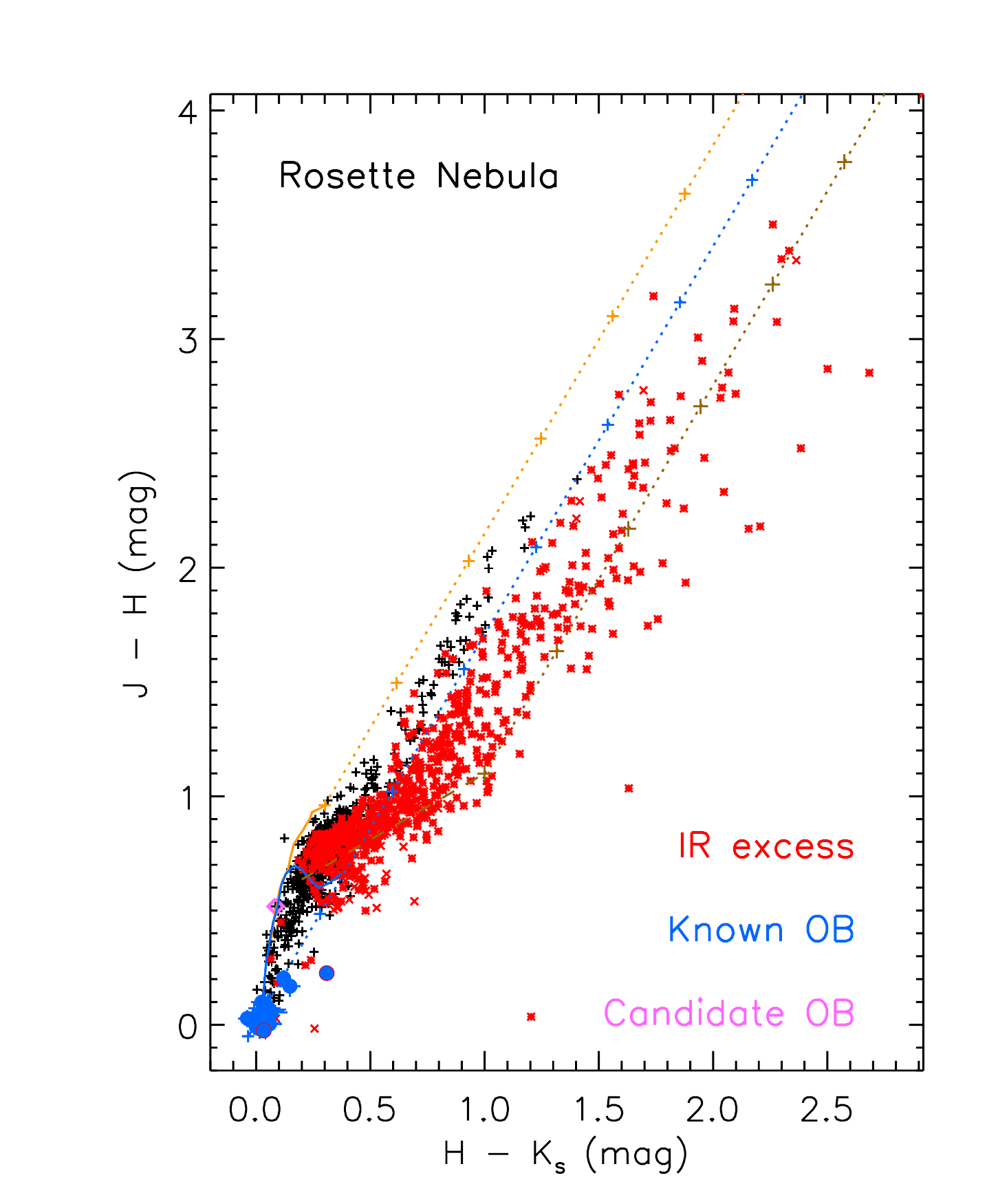}
\includegraphics[height=0.4\textheight]{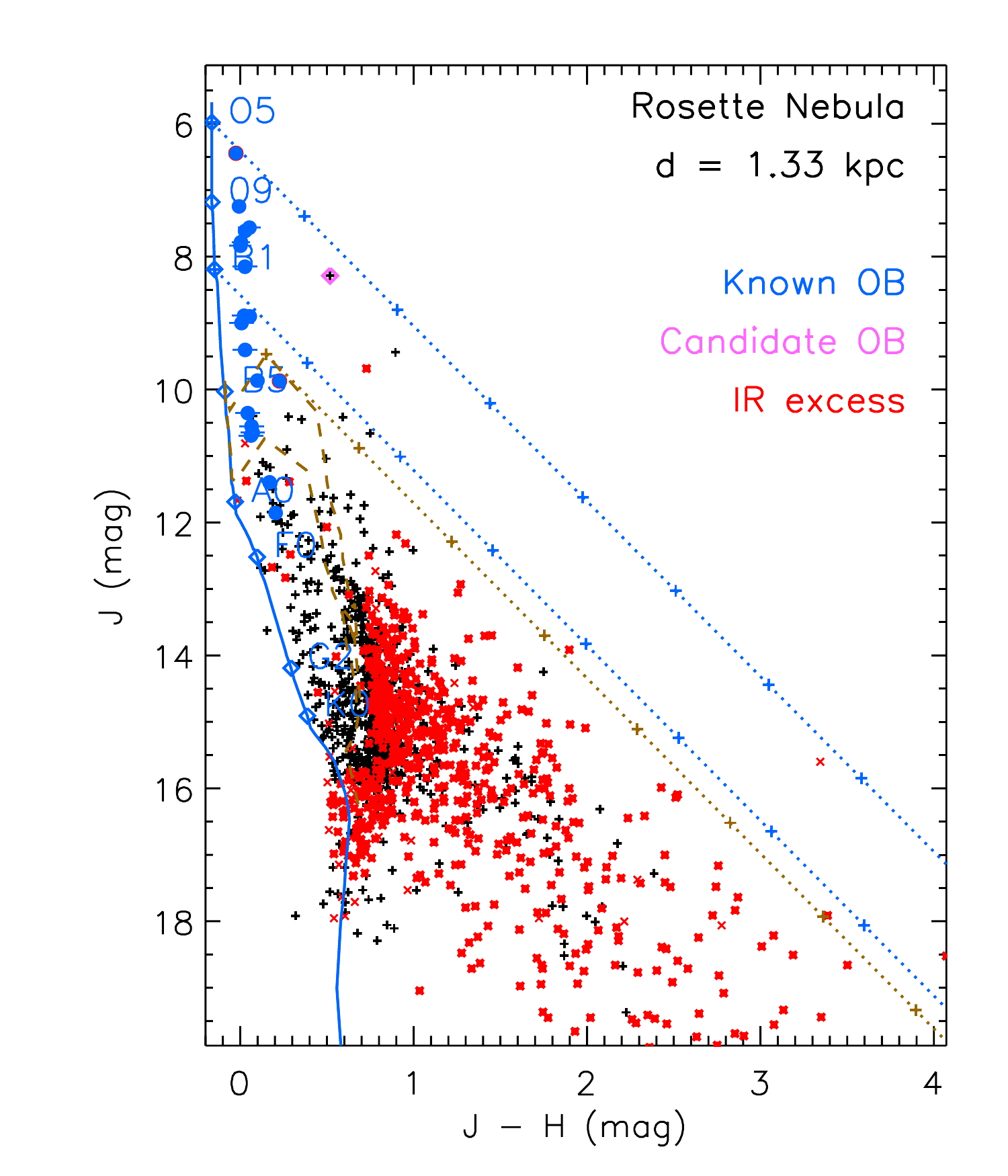}

~ \\
Fig. \ref{fig:each}({\it e}).--- The Rosette Nebula and Molecular Cloud.

\end{figure*}

\begin{figure*}[p]
\centering
\includegraphics[height=0.45\textheight]{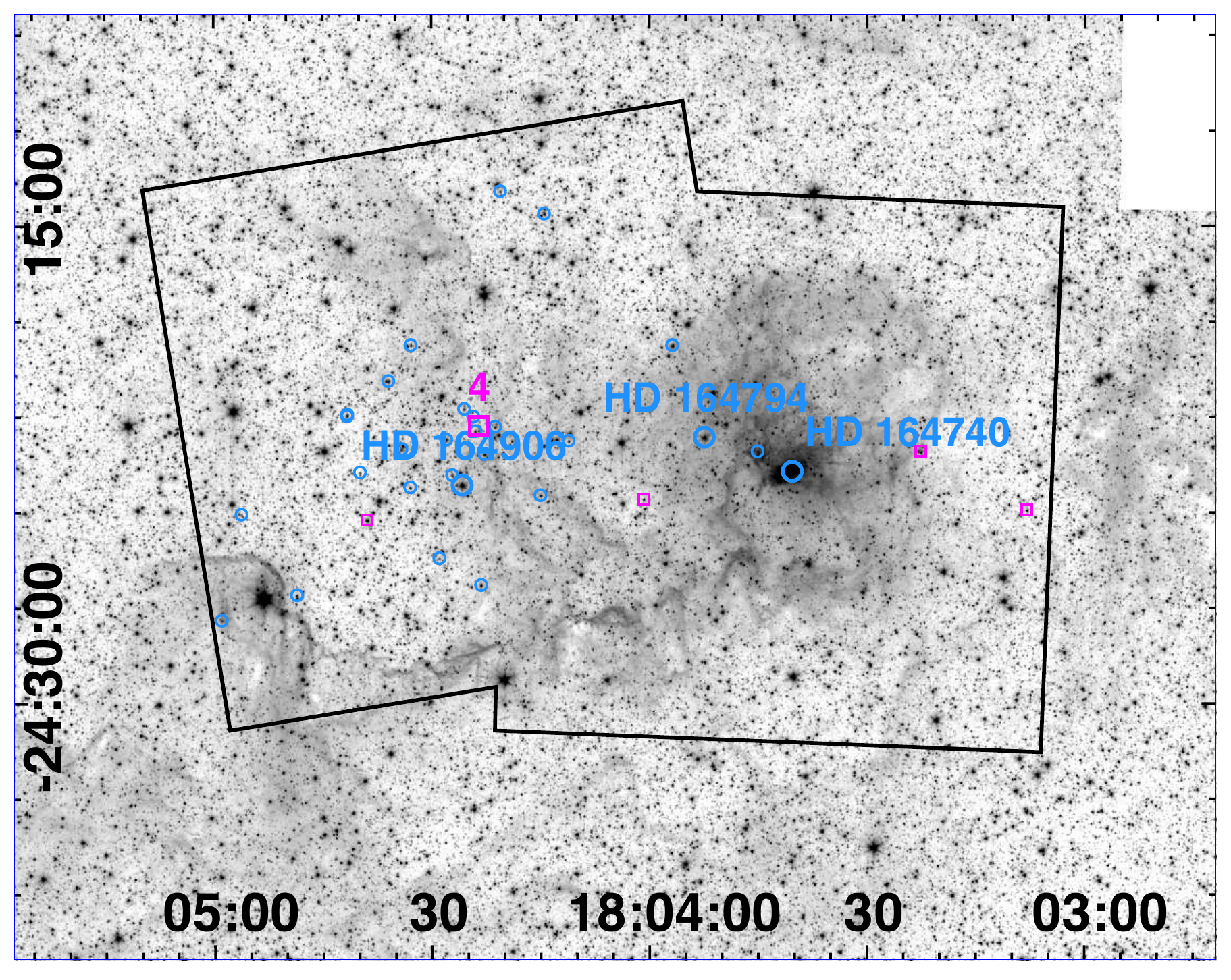}\\
\includegraphics[height=0.4\textheight]{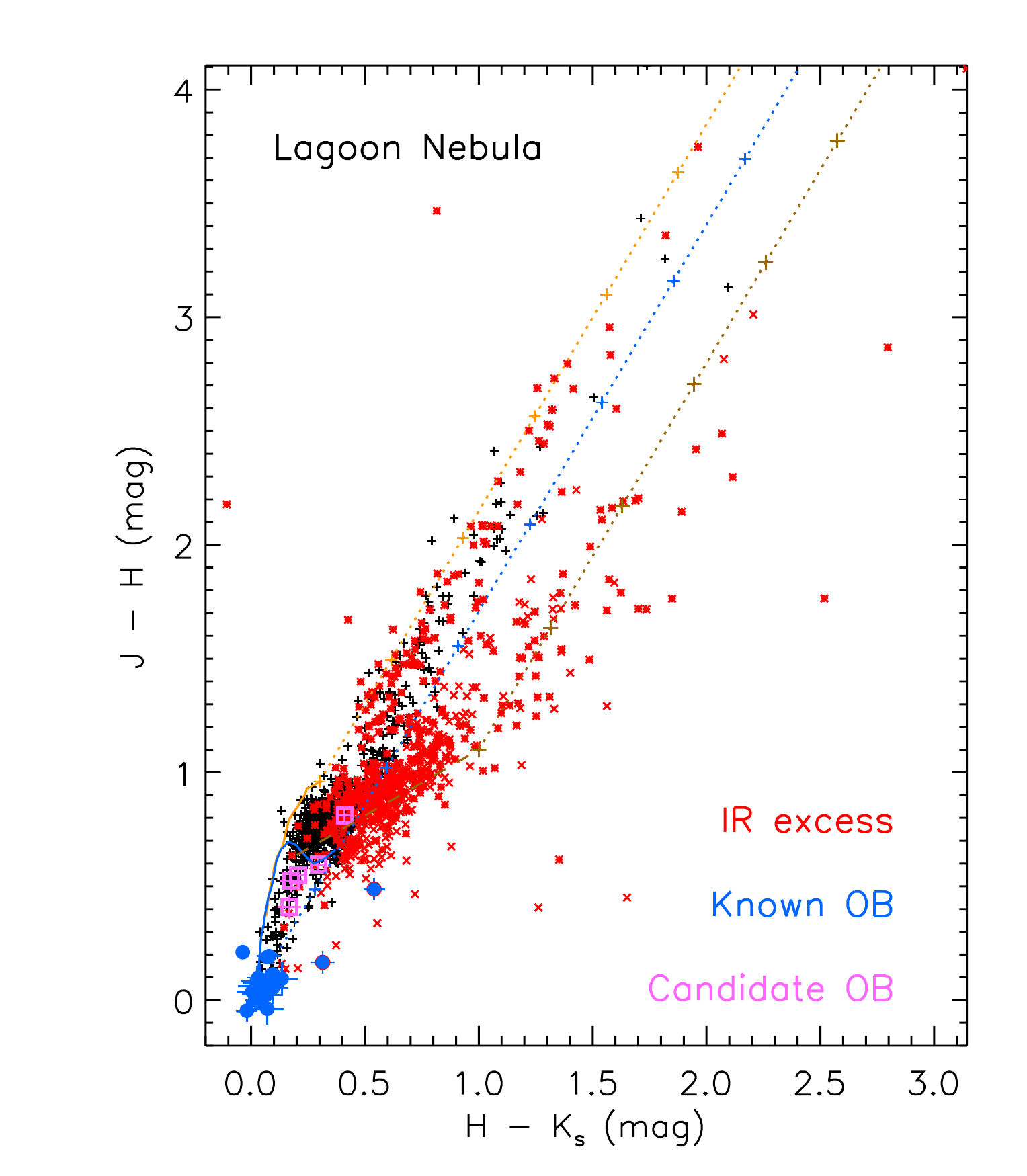}
\includegraphics[height=0.4\textheight]{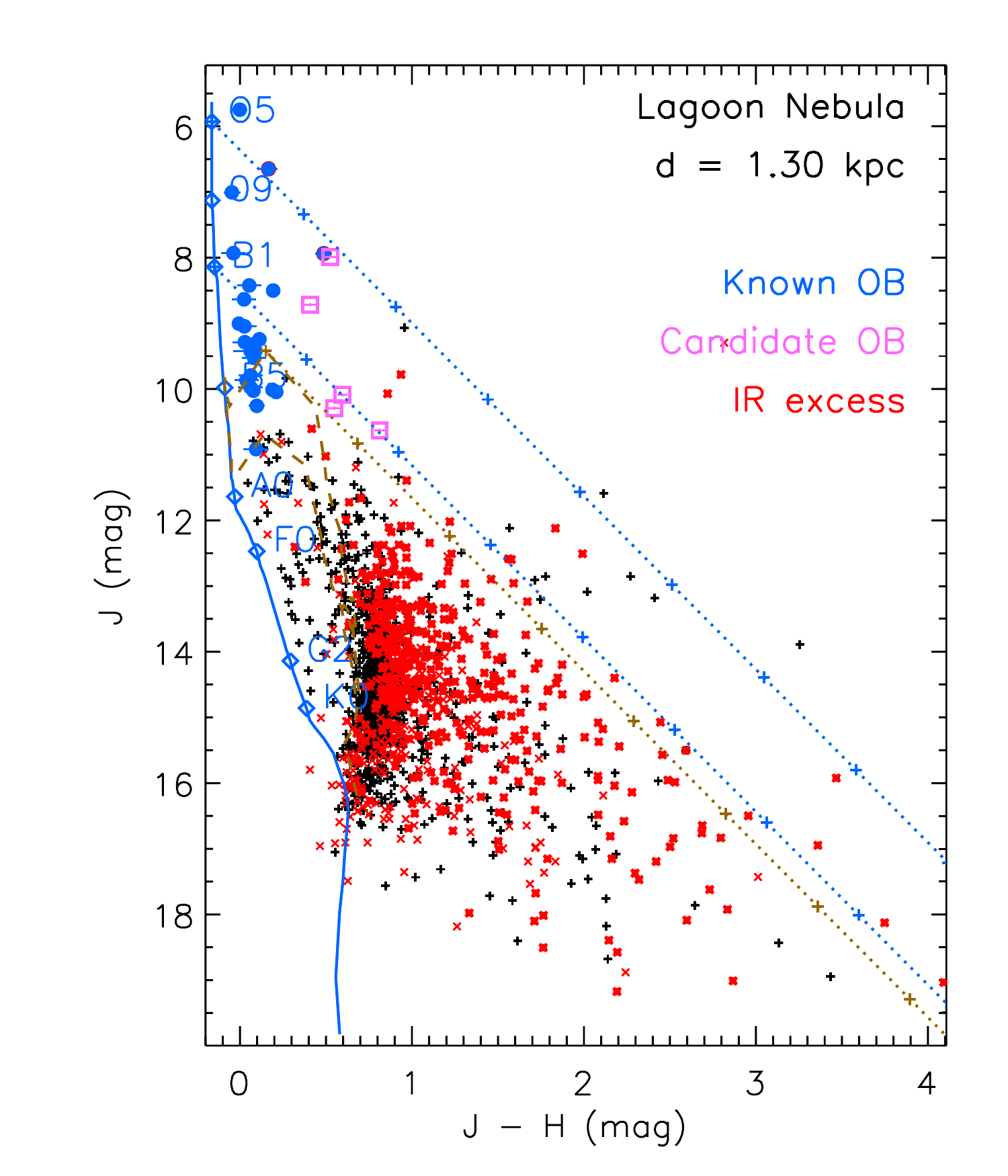}

~ \\
Fig. \ref{fig:each}({\it f}).--- The Lagoon Nebula (M8).

\end{figure*}

\begin{figure*}[p]
\centering
\includegraphics[height=0.45\textheight]{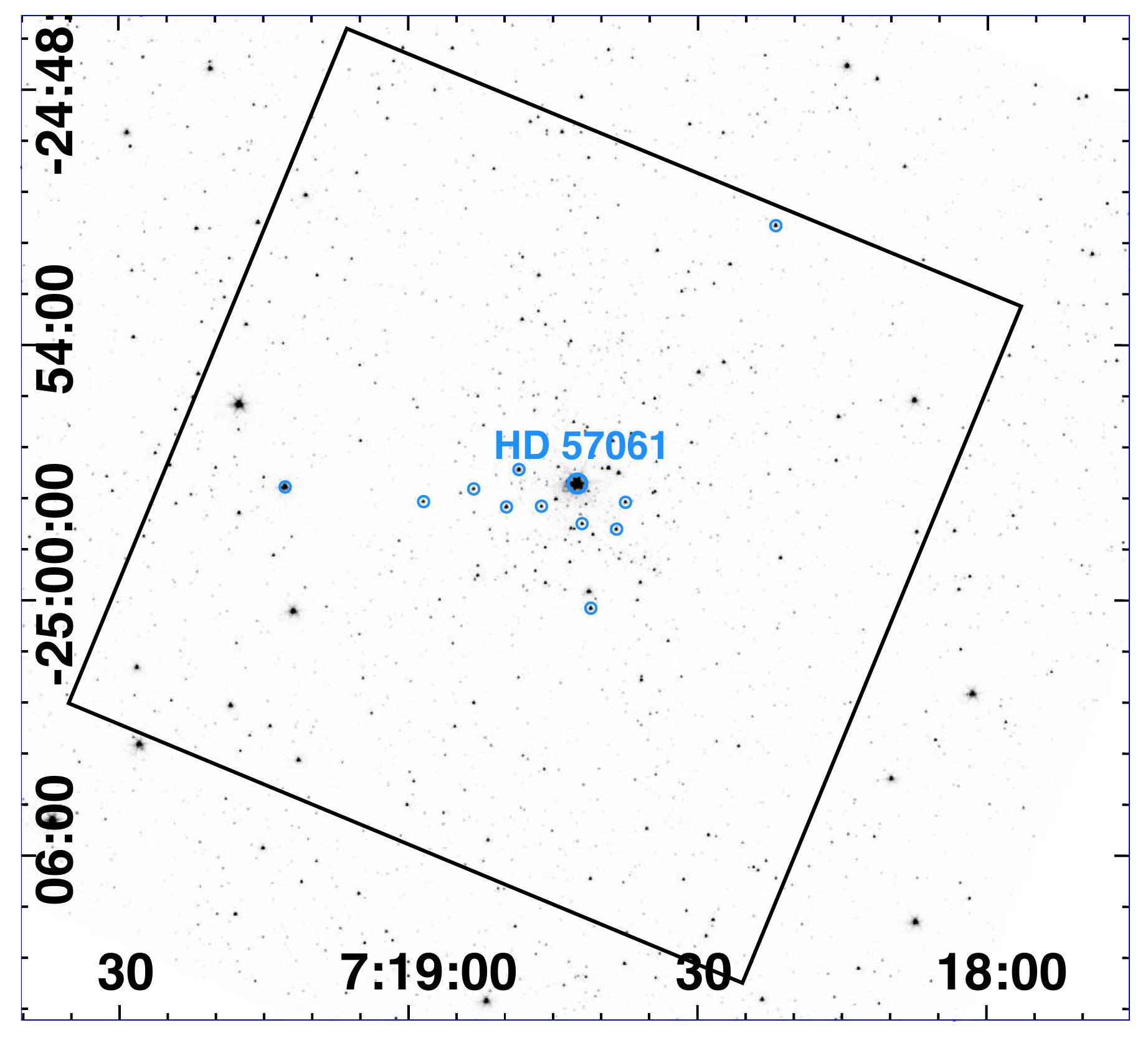}\\
\includegraphics[height=0.4\textheight]{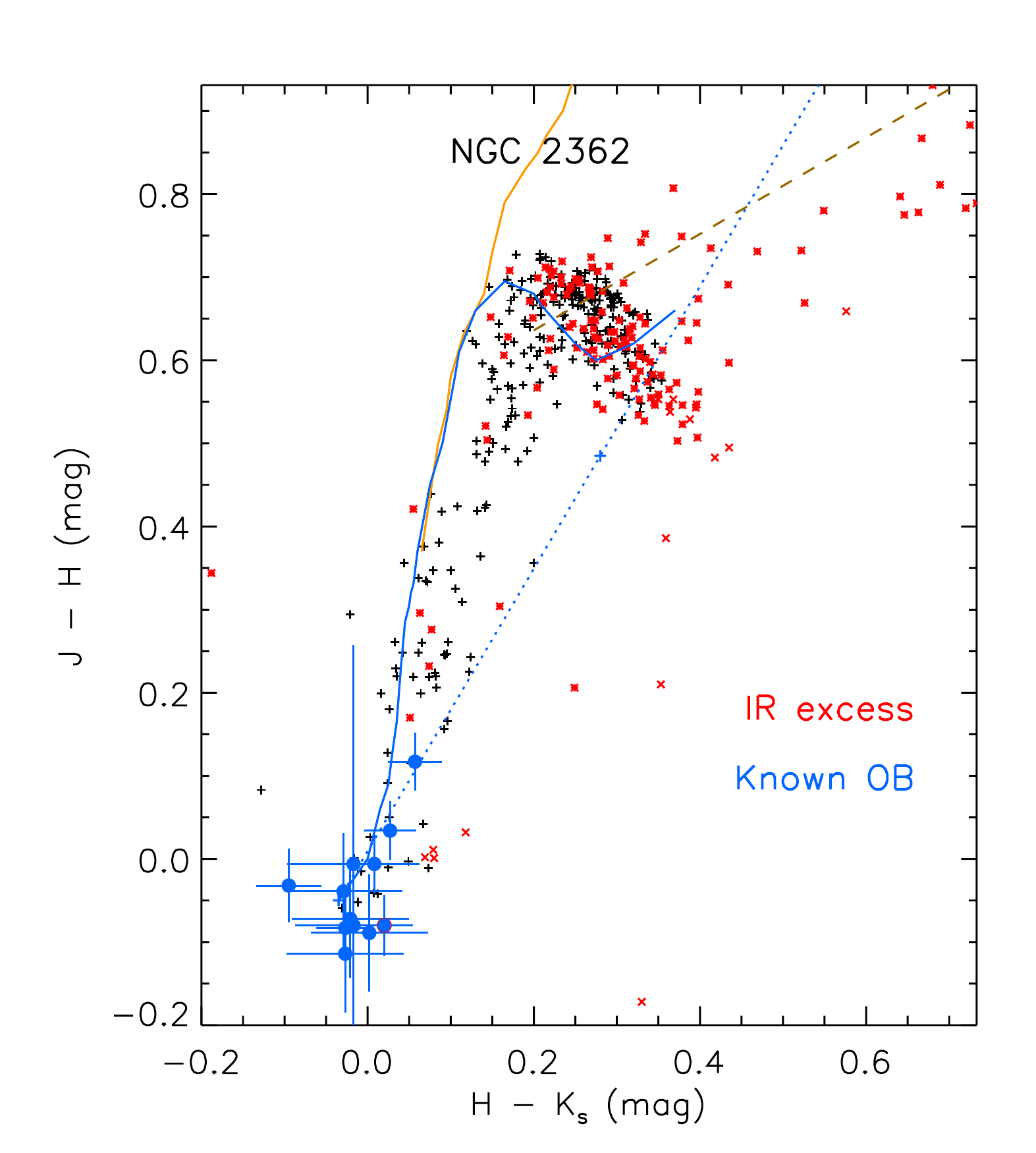}
\includegraphics[height=0.4\textheight]{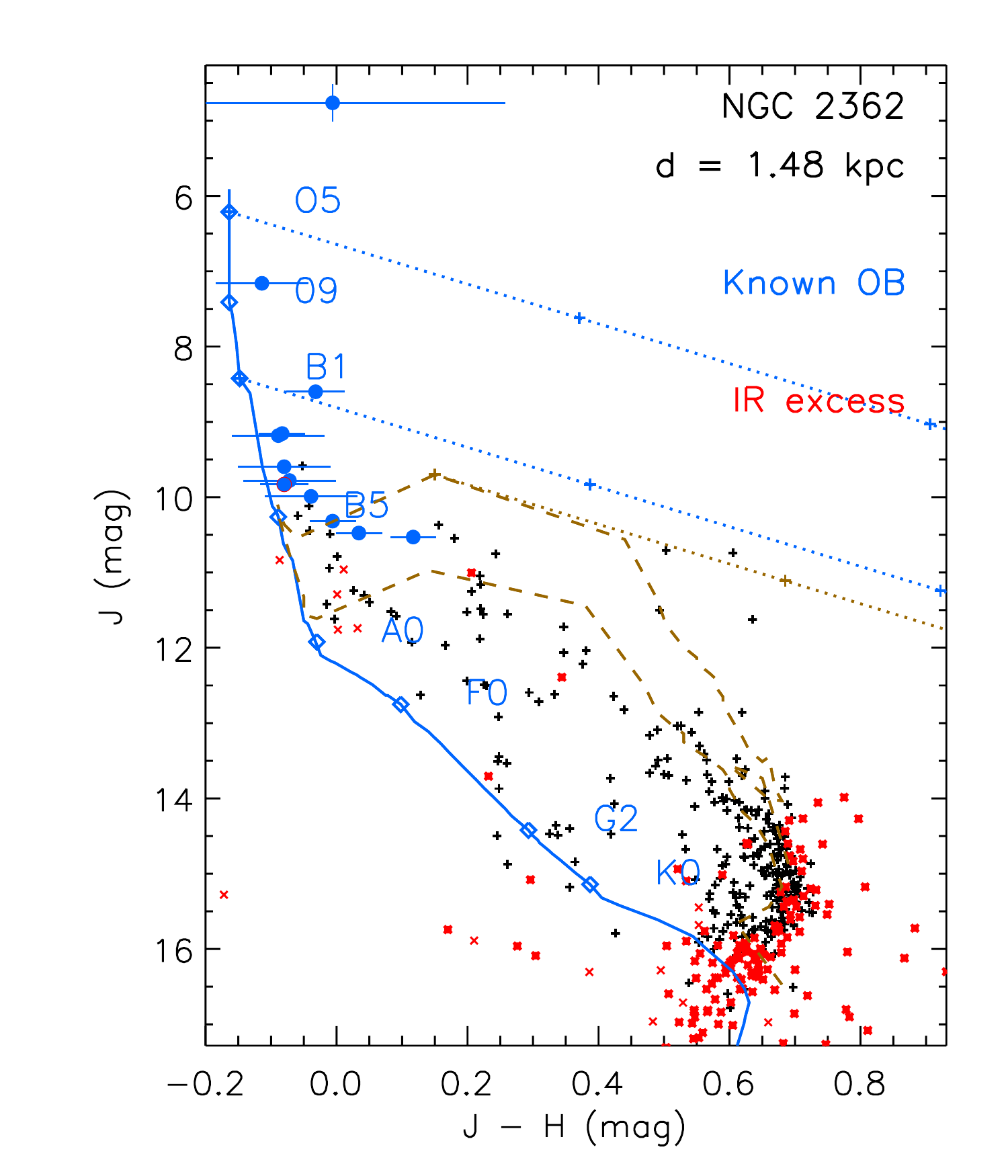}

~ \\
Fig. \ref{fig:each}({\it g}).--- NGC 2362.

\end{figure*}

\begin{figure*}[p]
\centering
\includegraphics[height=0.45\textheight]{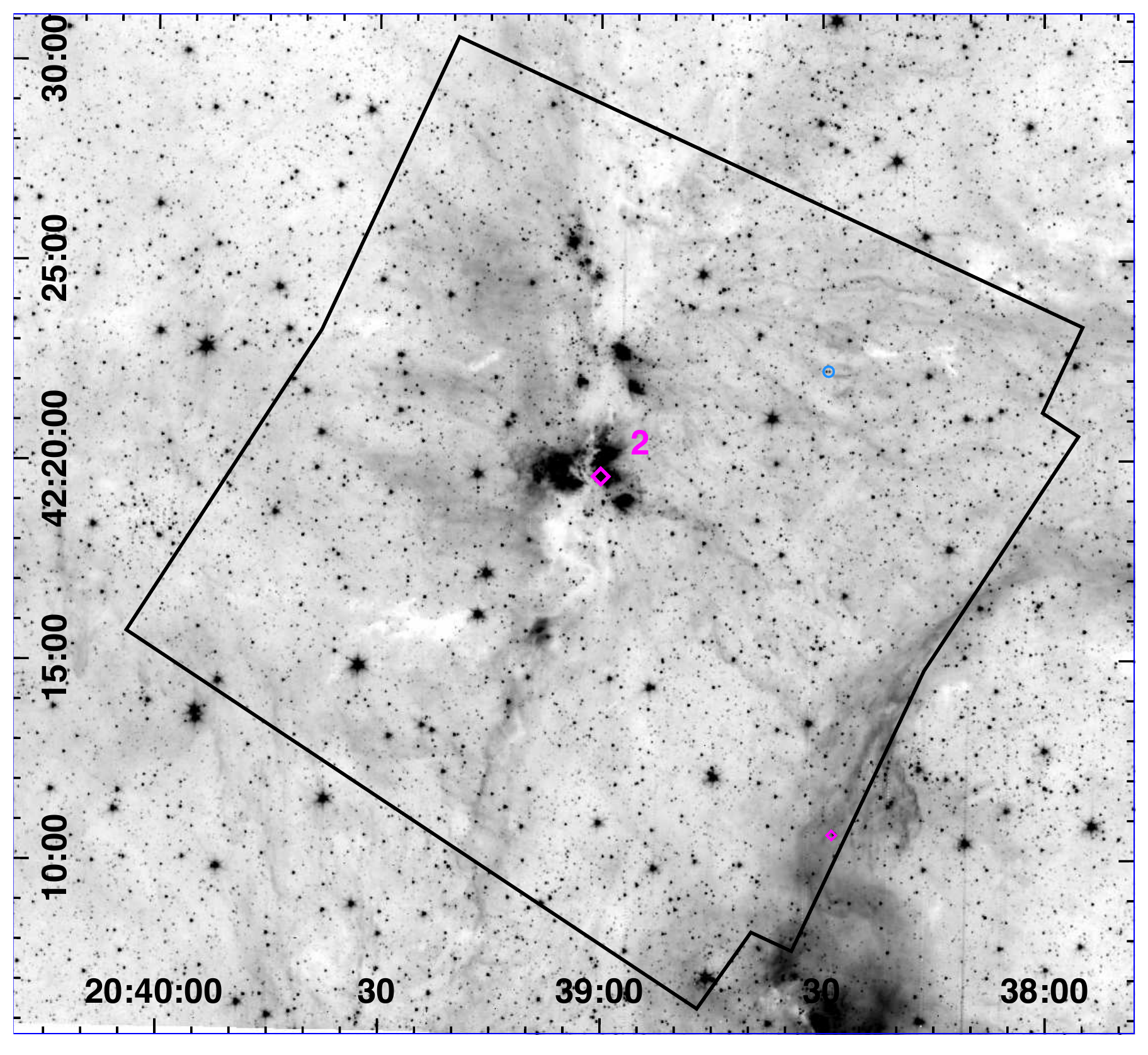}\\
\includegraphics[height=0.4\textheight]{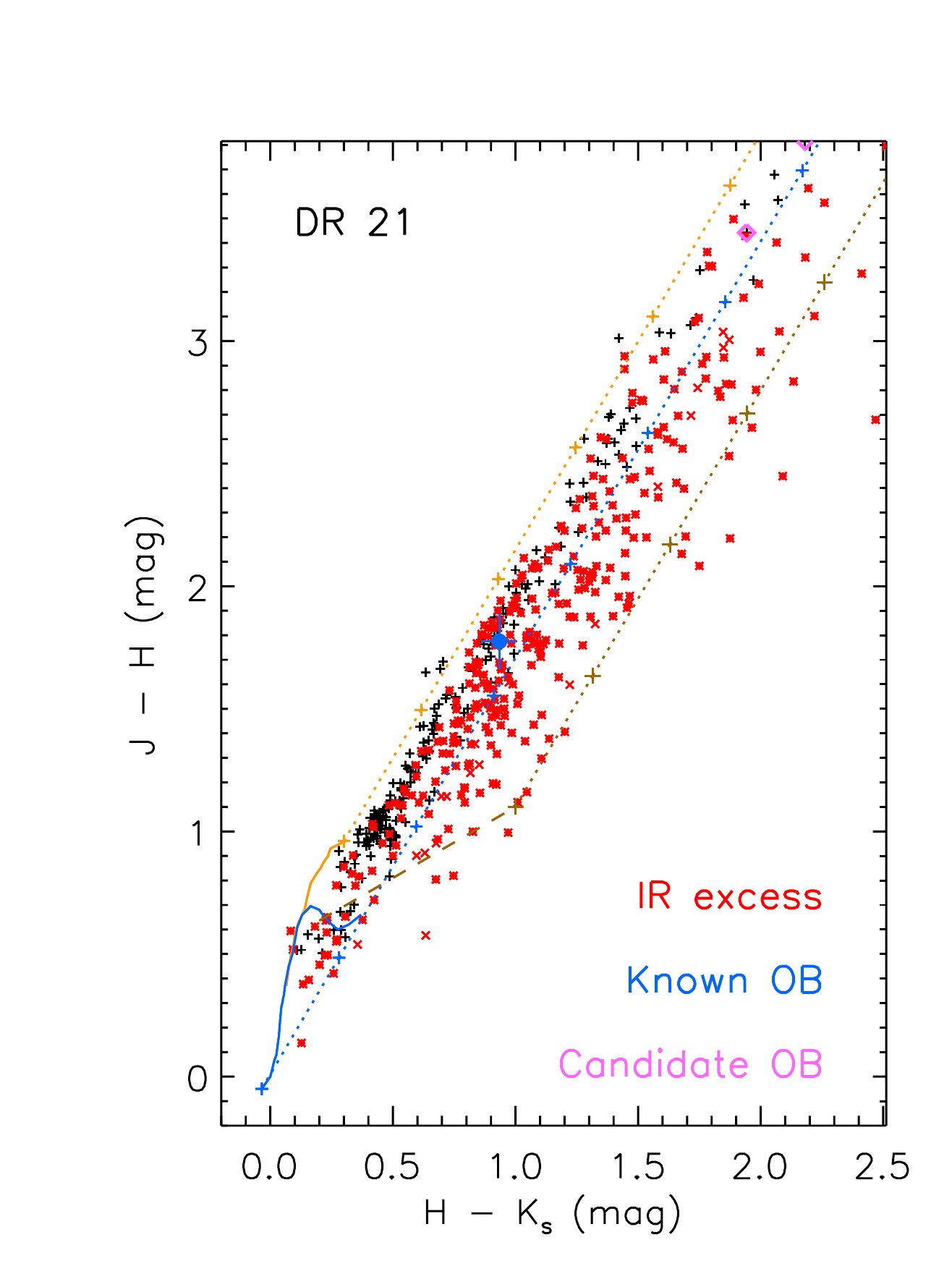}
\includegraphics[height=0.4\textheight]{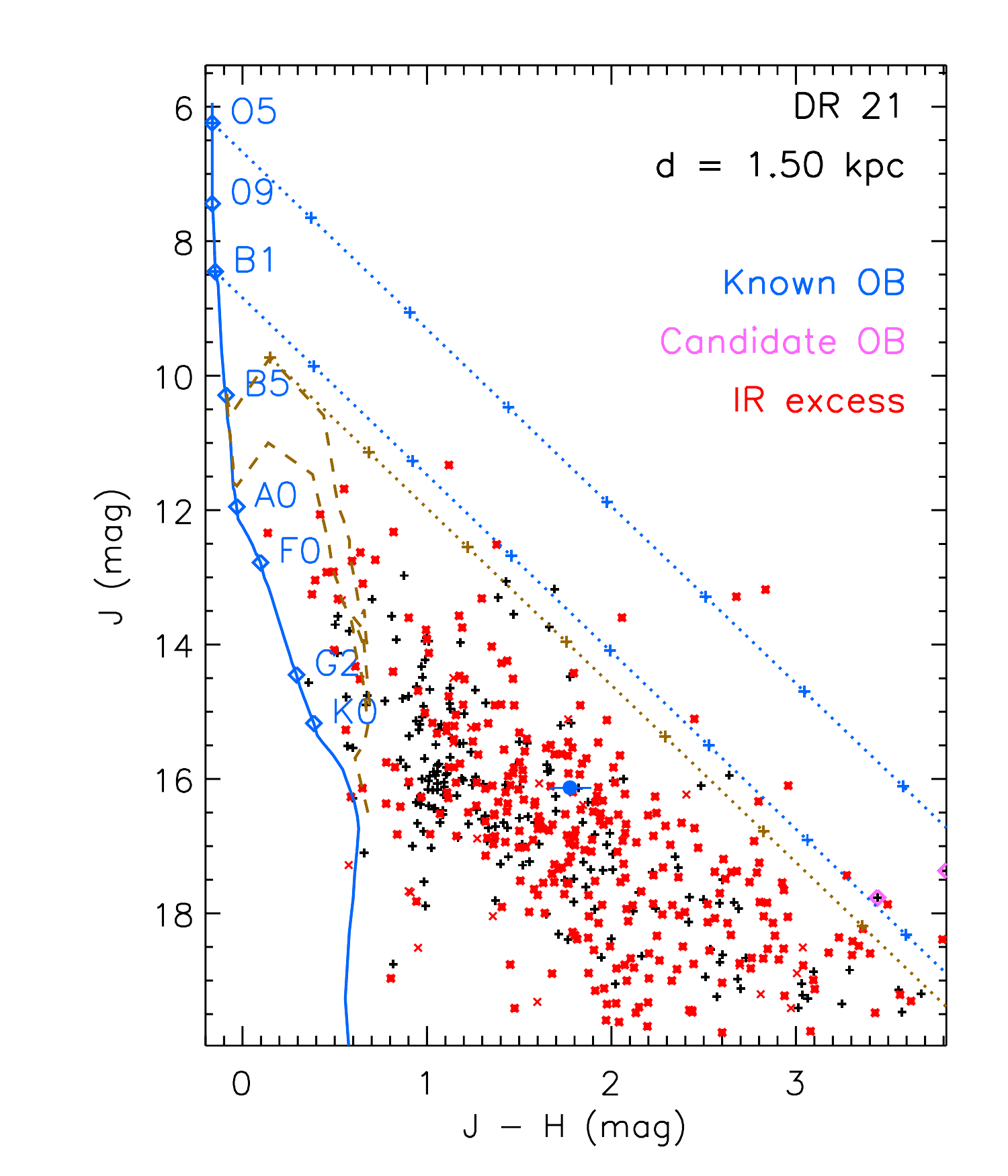}

~ \\
Fig. \ref{fig:each}({\it h}).---DR 21 in Cygnus X.

\end{figure*}

\begin{figure*}[p]
\centering
\includegraphics[height=0.45\textheight]{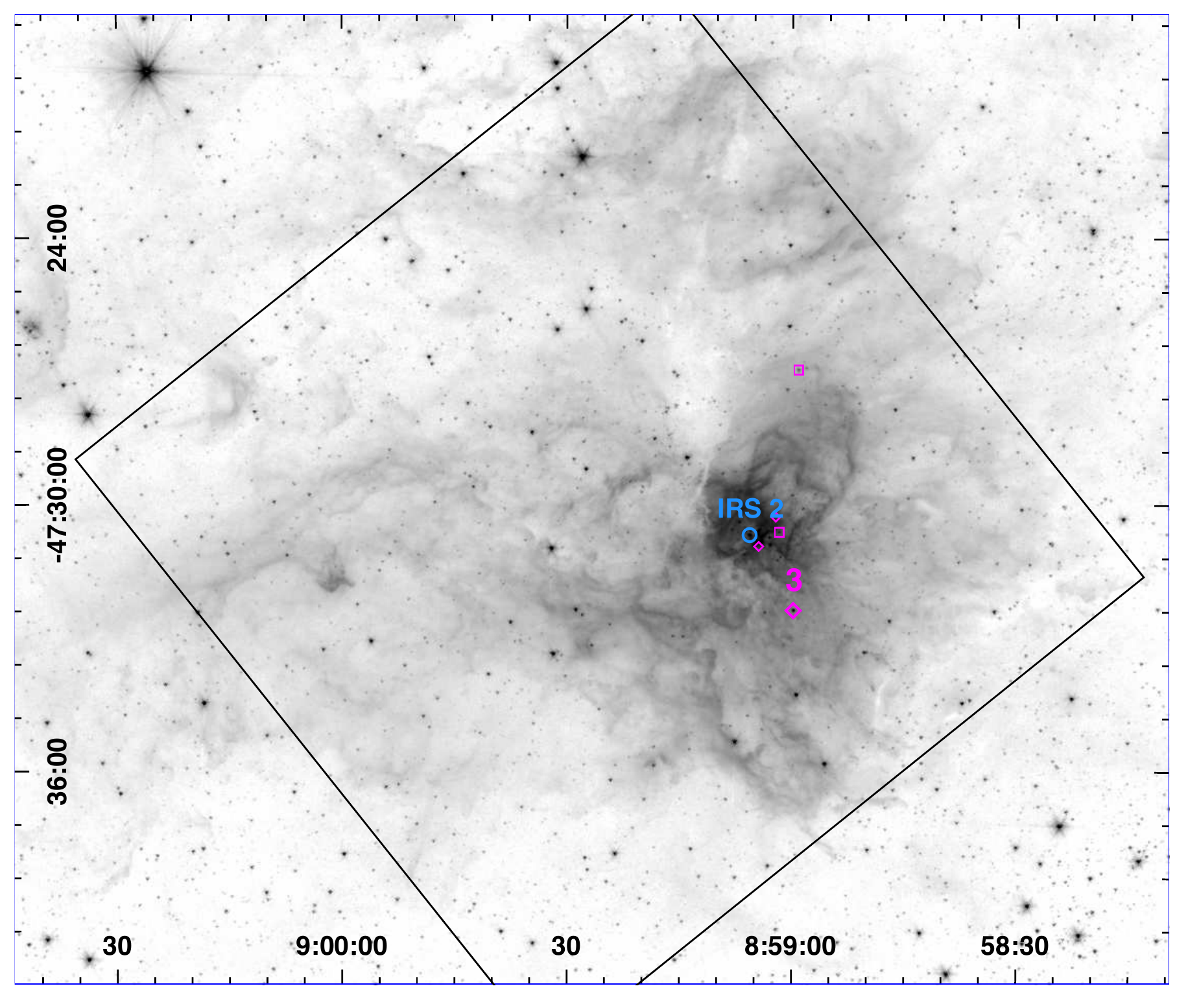}\\
\includegraphics[height=0.4\textheight]{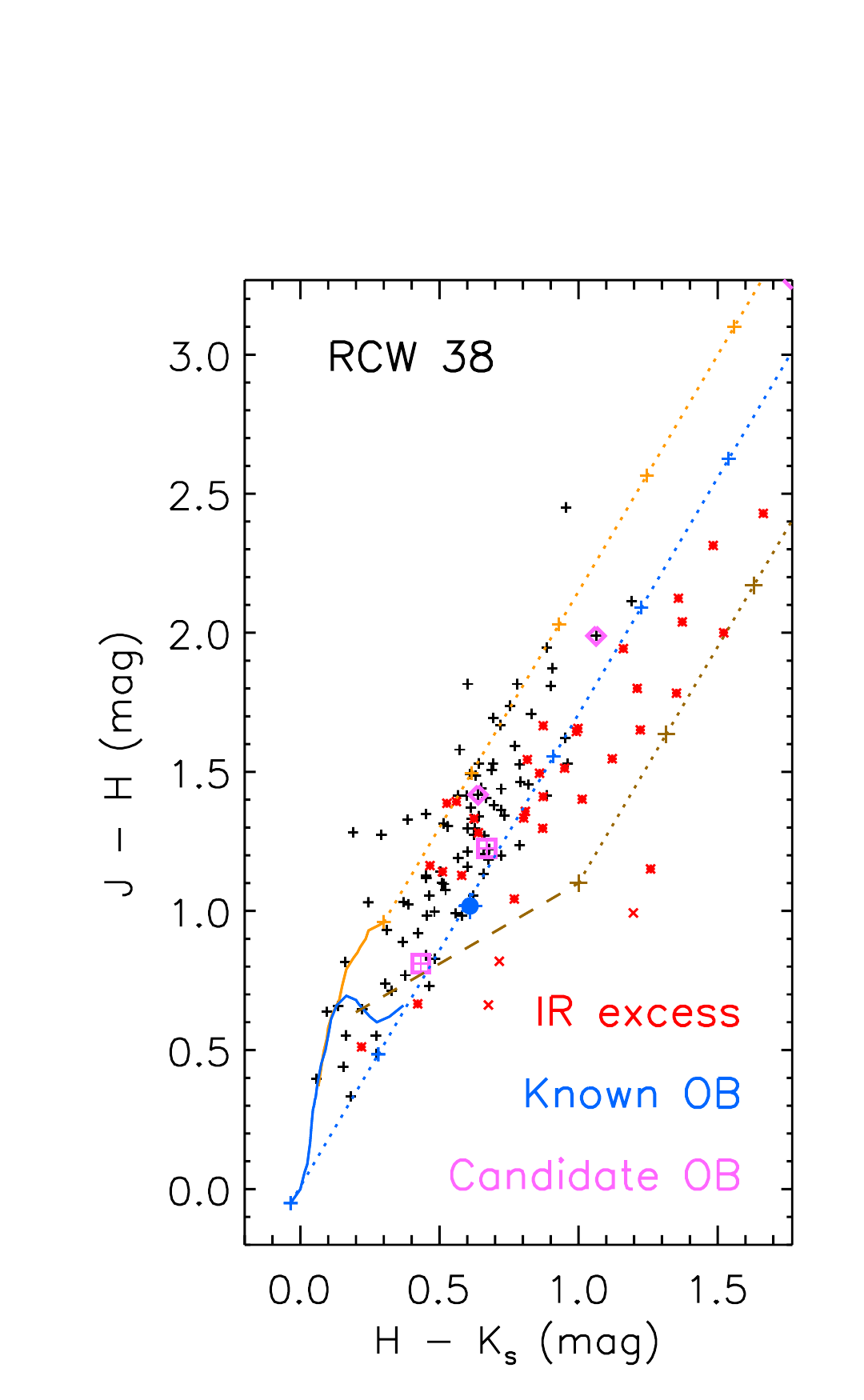}
\includegraphics[height=0.4\textheight]{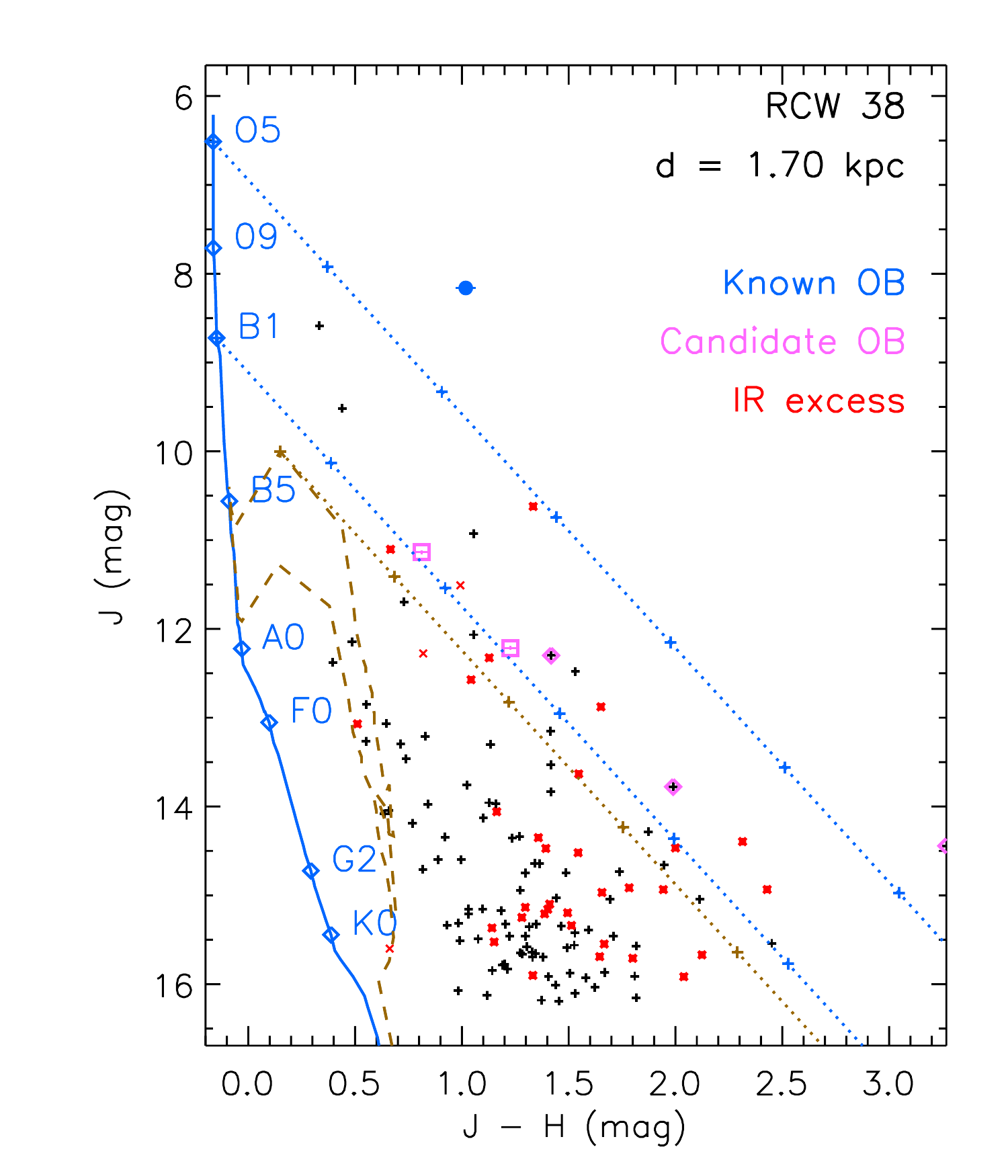}

~ \\
Fig. \ref{fig:each}({\it i}).--- RCW 38.

\end{figure*}

\begin{figure*}[p]
\centering
\includegraphics[height=0.45\textheight]{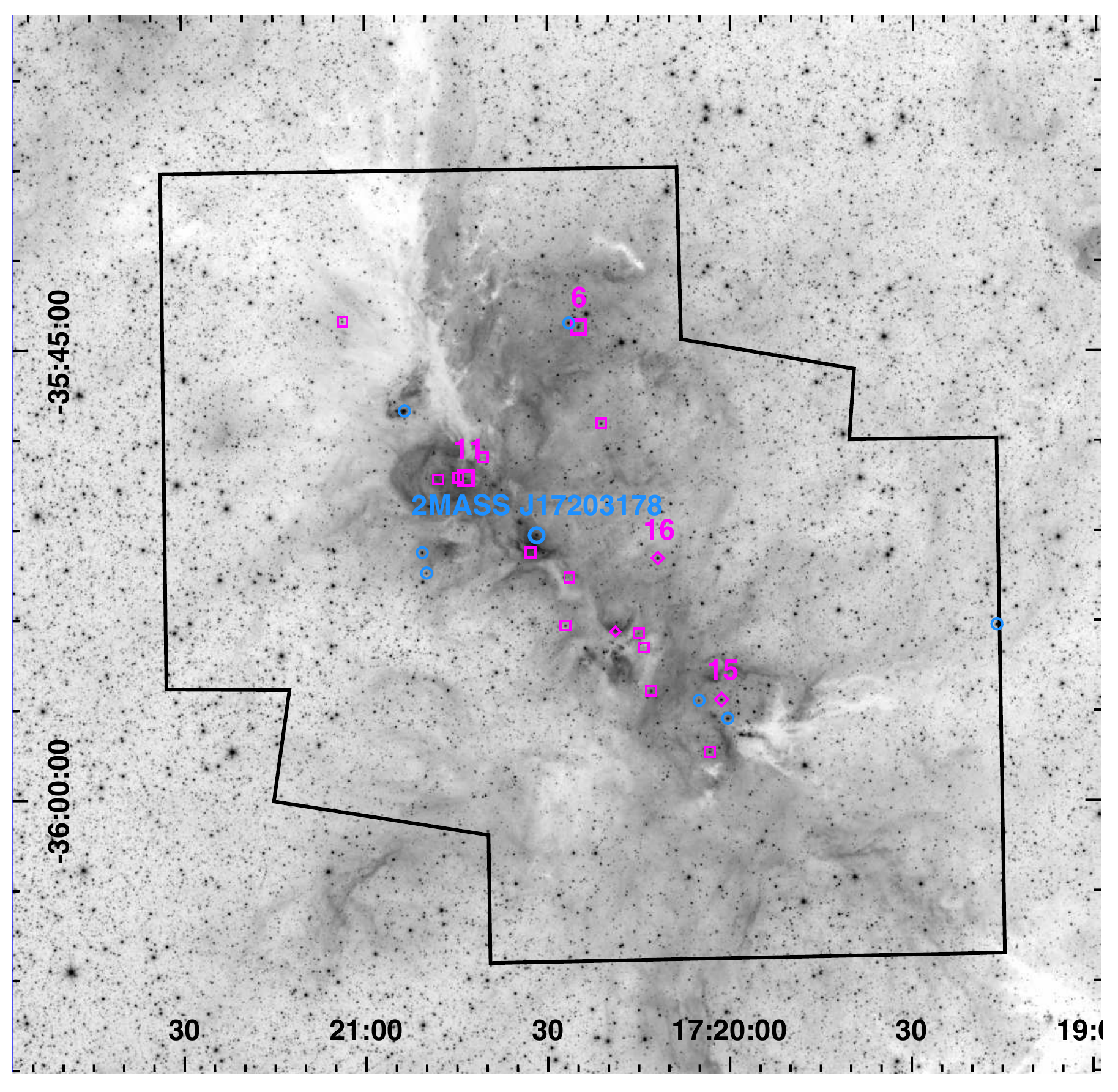}\\
\includegraphics[height=0.4\textheight]{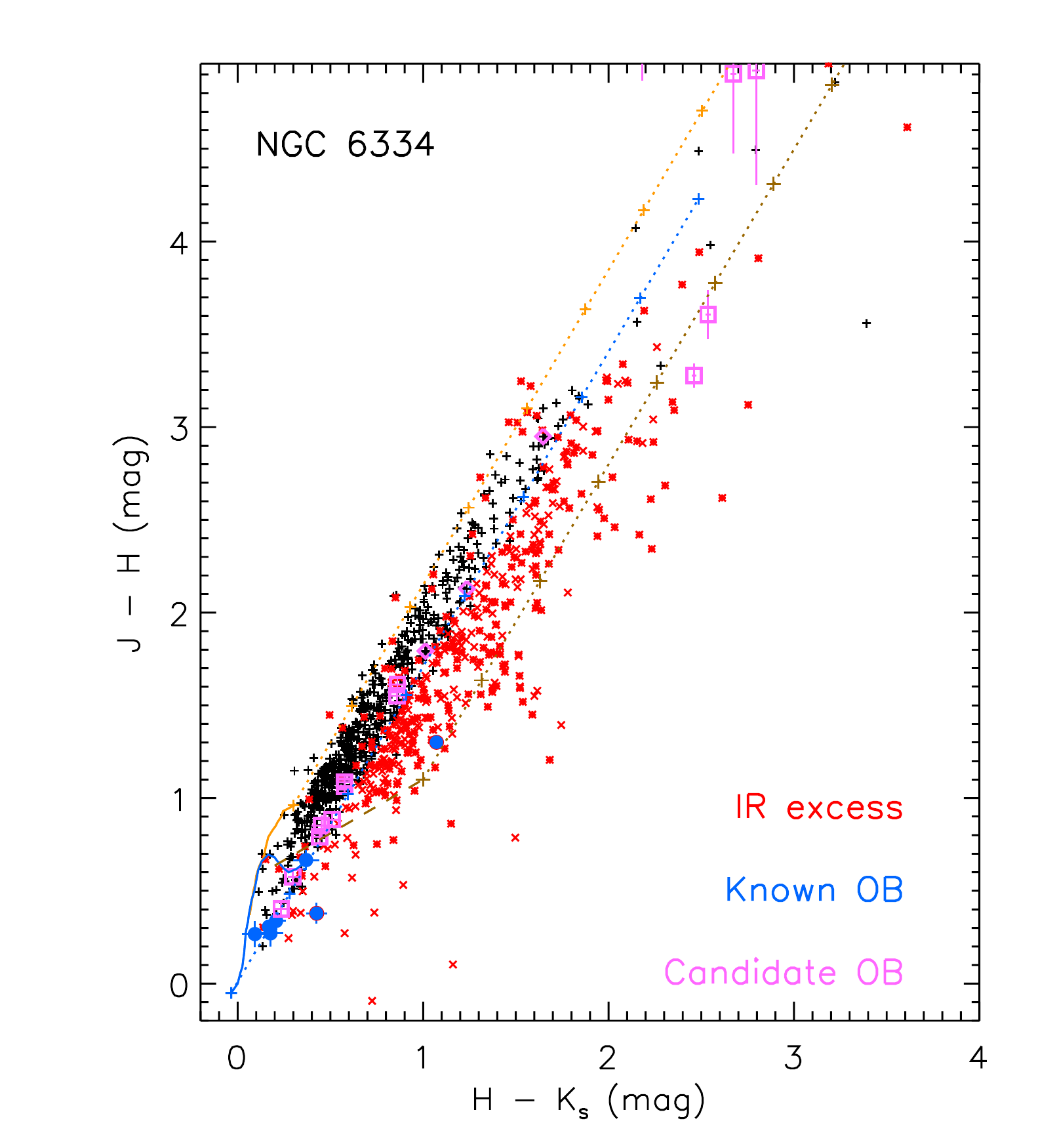}
\includegraphics[height=0.4\textheight]{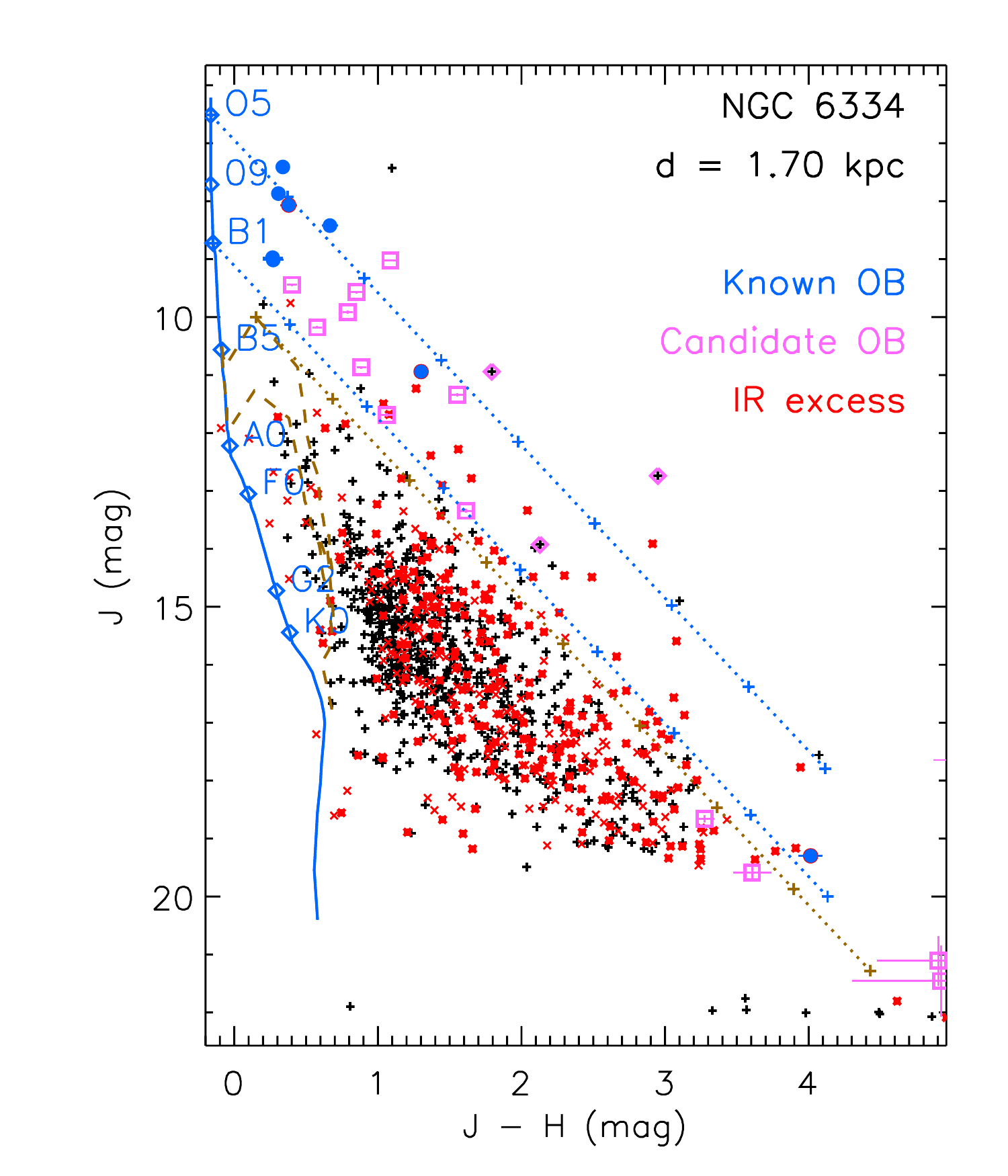}

~ \\
Fig. \ref{fig:each}({\it j}).--- NGC 6334 (the Cat's Paw Nebula).

\end{figure*}





\begin{figure*}[p]
\centering
\includegraphics[height=0.45\textheight]{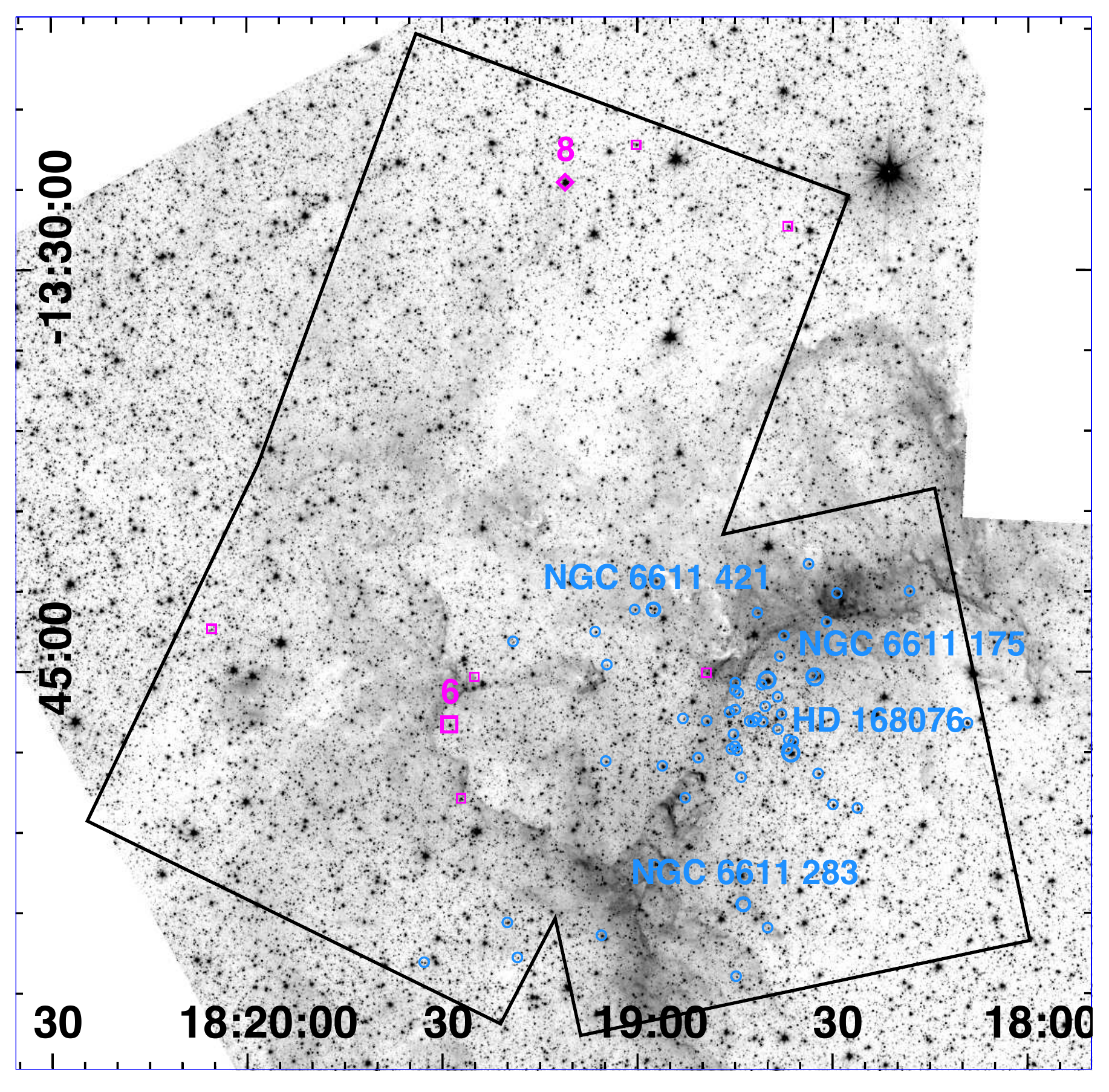}\\
\includegraphics[height=0.4\textheight]{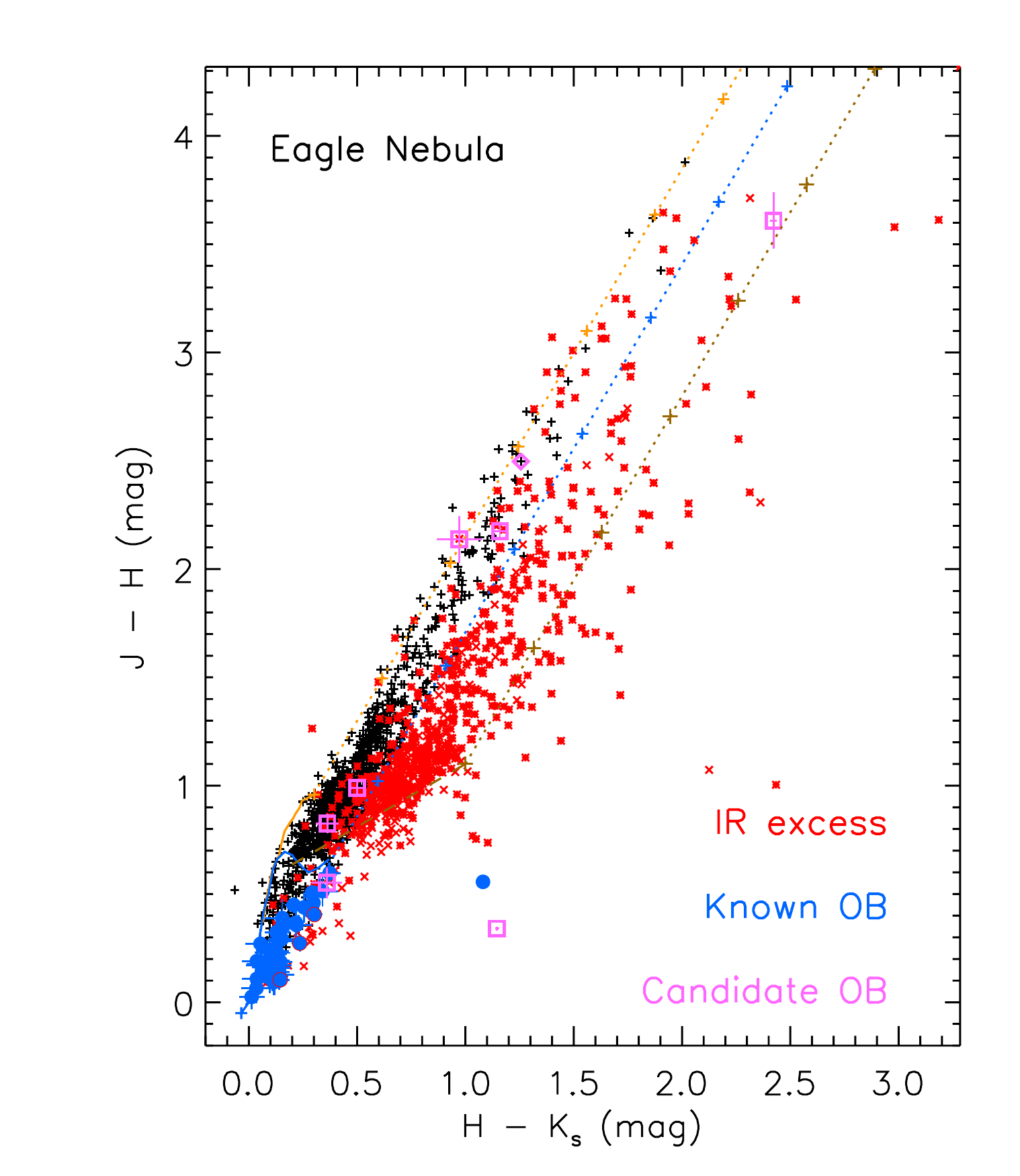}
\includegraphics[height=0.4\textheight]{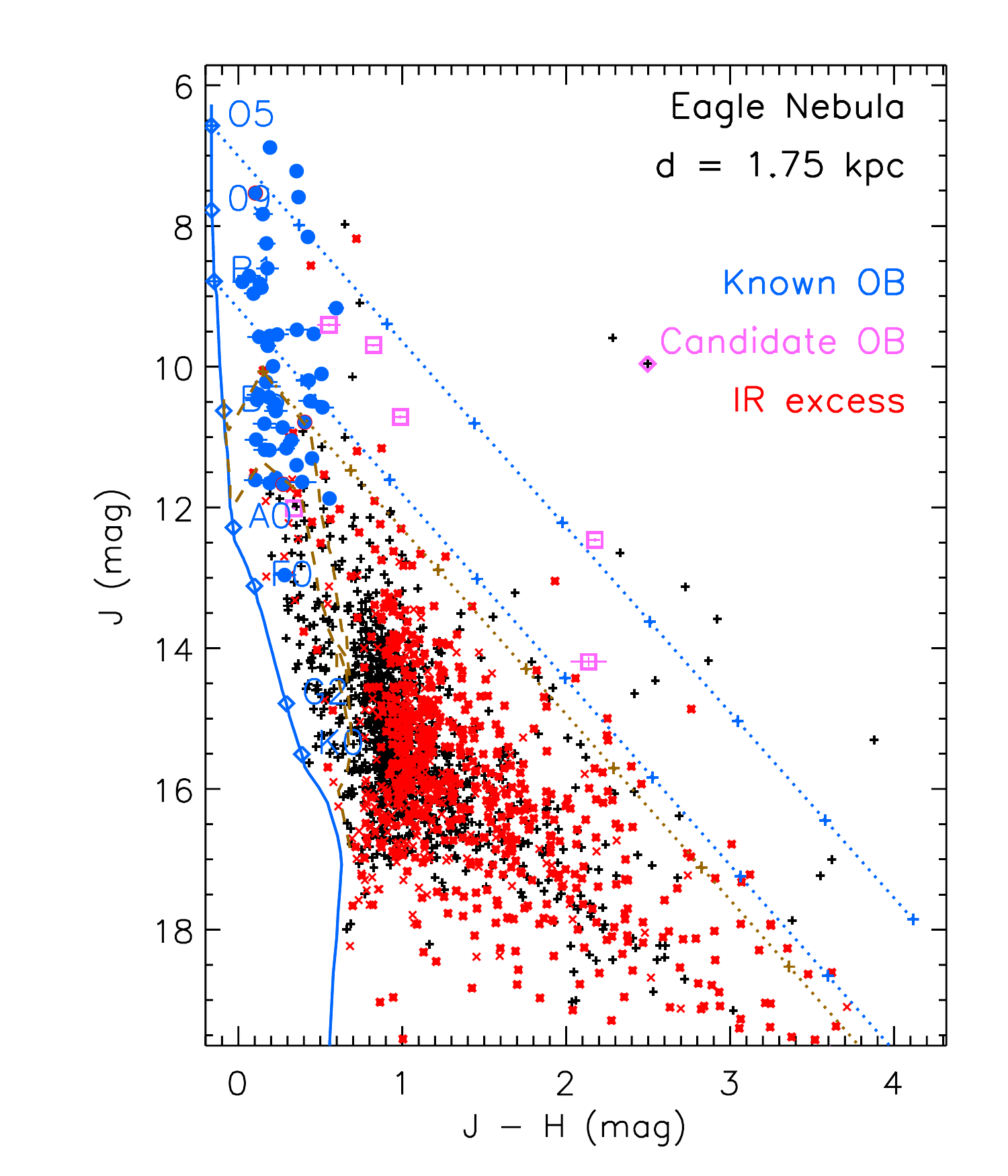}

~ \\
Fig. \ref{fig:each}({\it l}).--- The Eagle Nebula (M16).

\end{figure*}

\begin{figure*}[p]
\centering
\includegraphics[height=0.45\textheight]{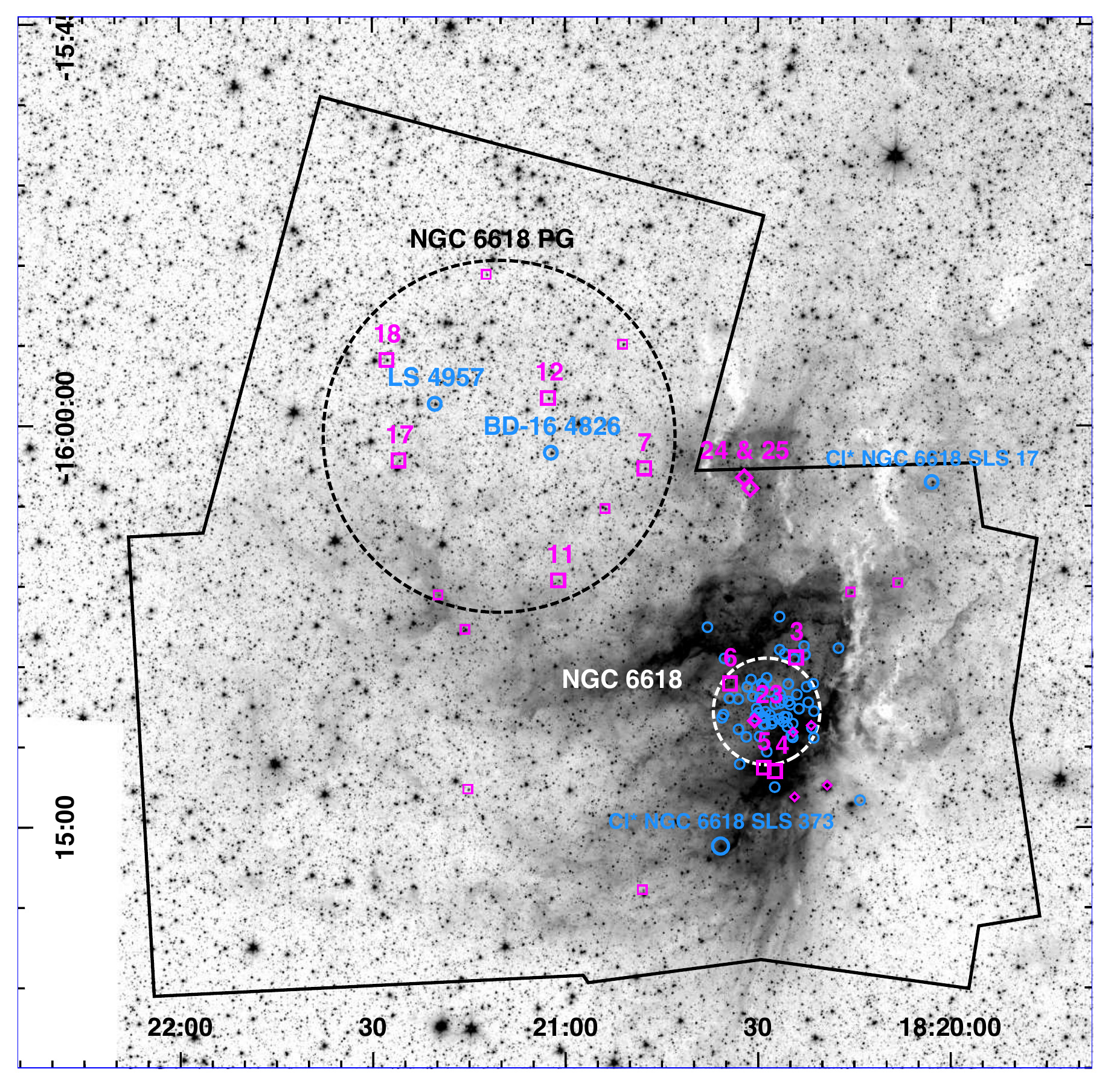}\\
\includegraphics[height=0.4\textheight]{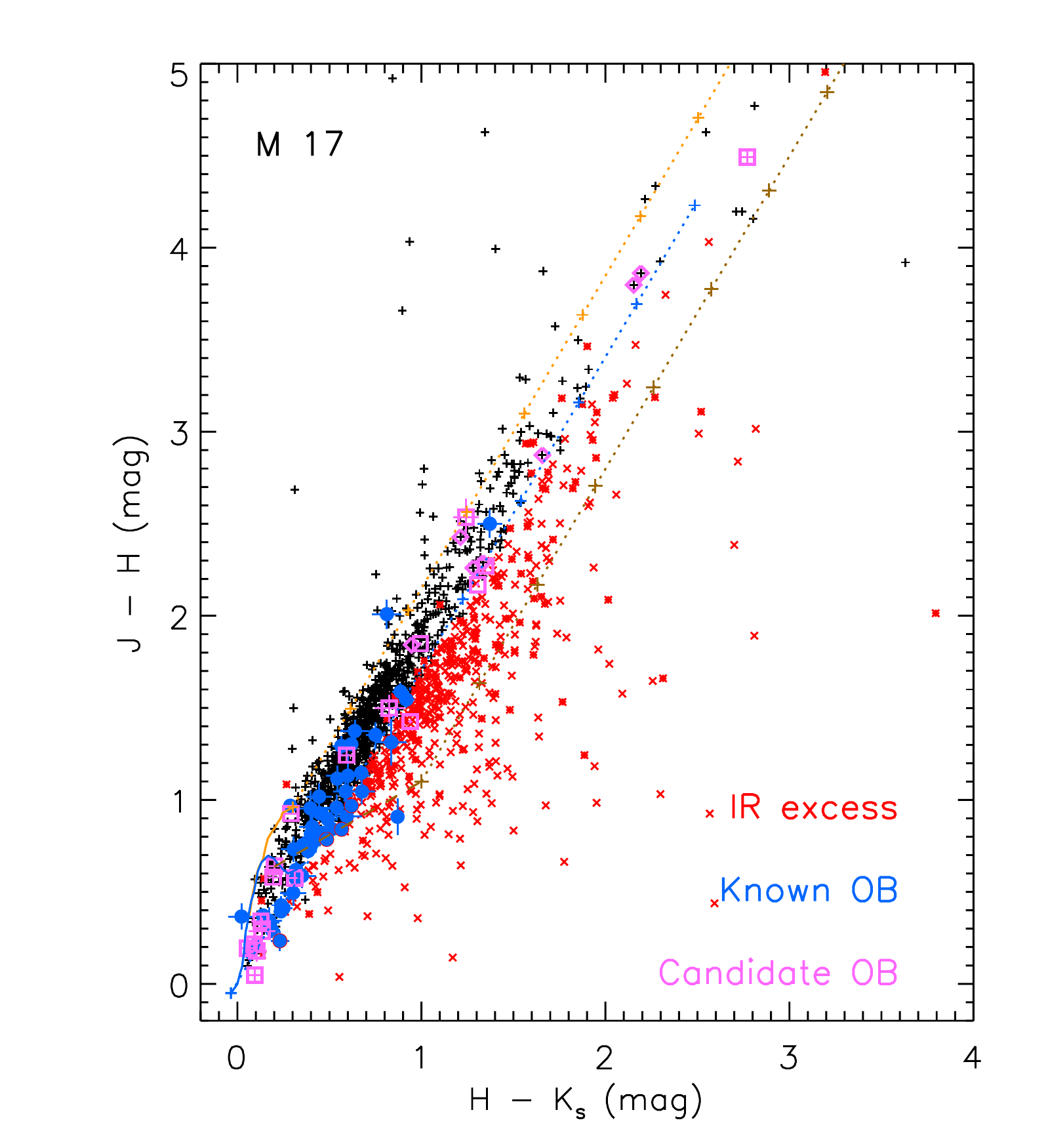}
\includegraphics[height=0.4\textheight]{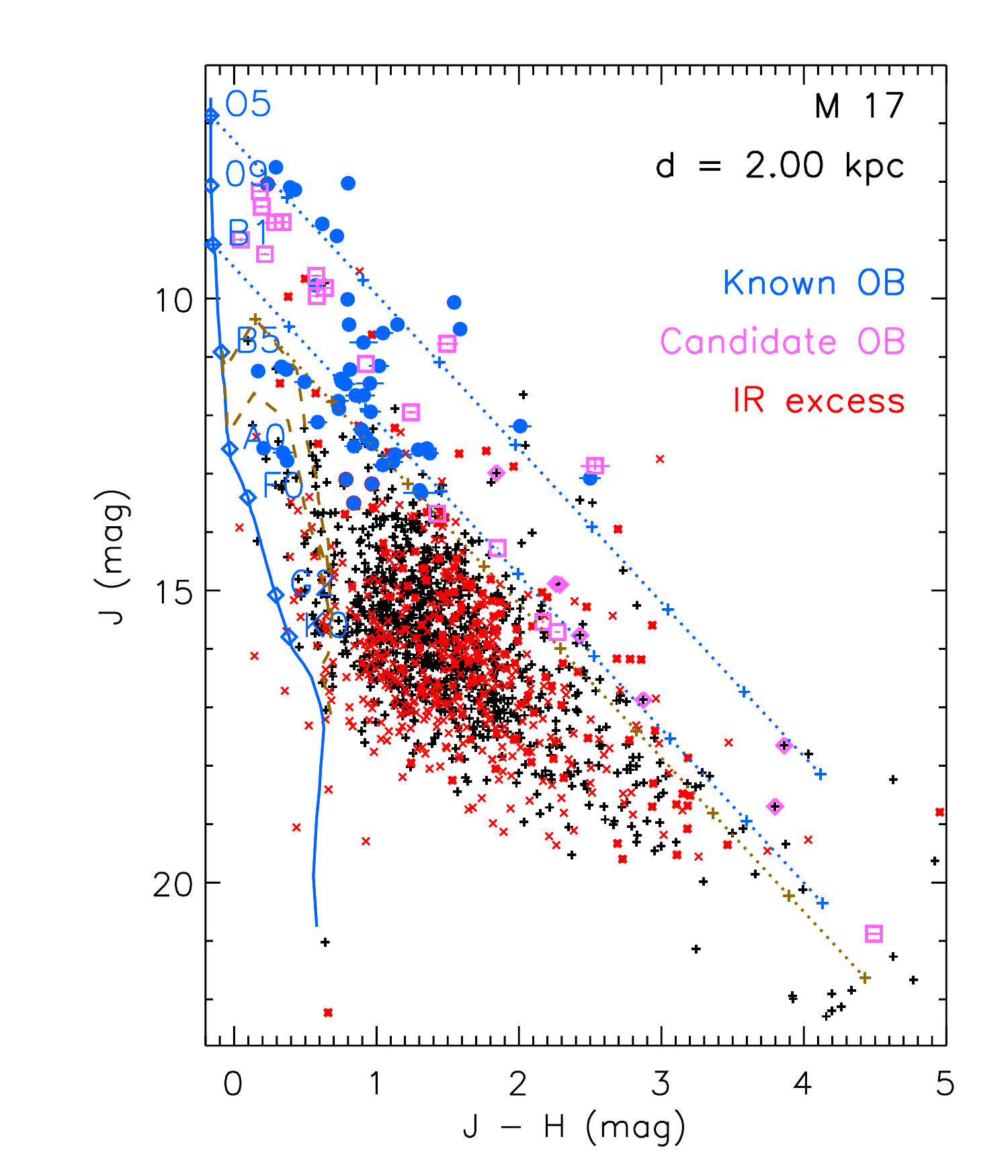}

~ \\
Fig. \ref{fig:each}({\it m}).--- M17 (the Omega Nebula).

\end{figure*}

\begin{figure*}[p]
\centering
\includegraphics[height=0.45\textheight]{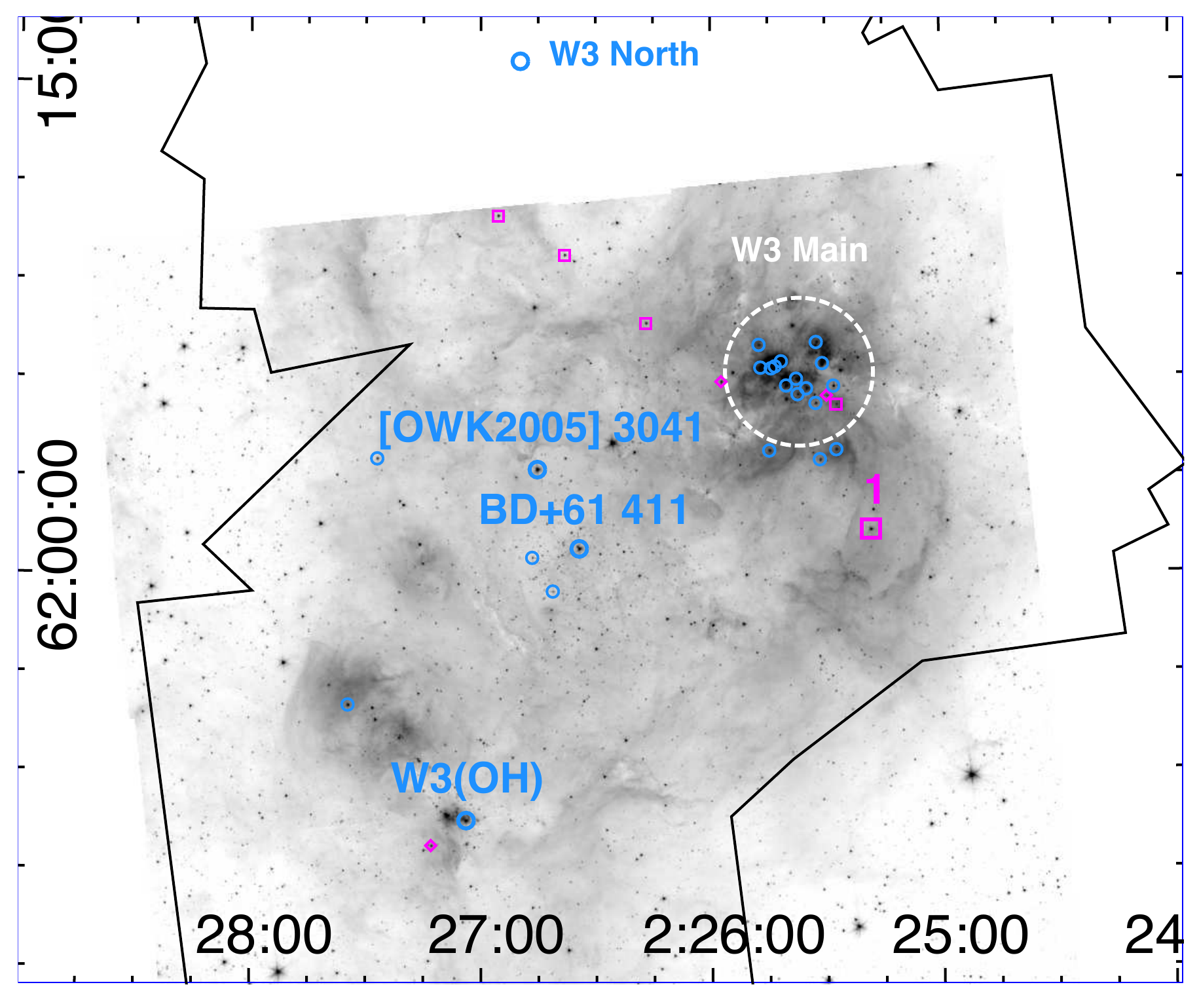}\\
\includegraphics[height=0.4\textheight]{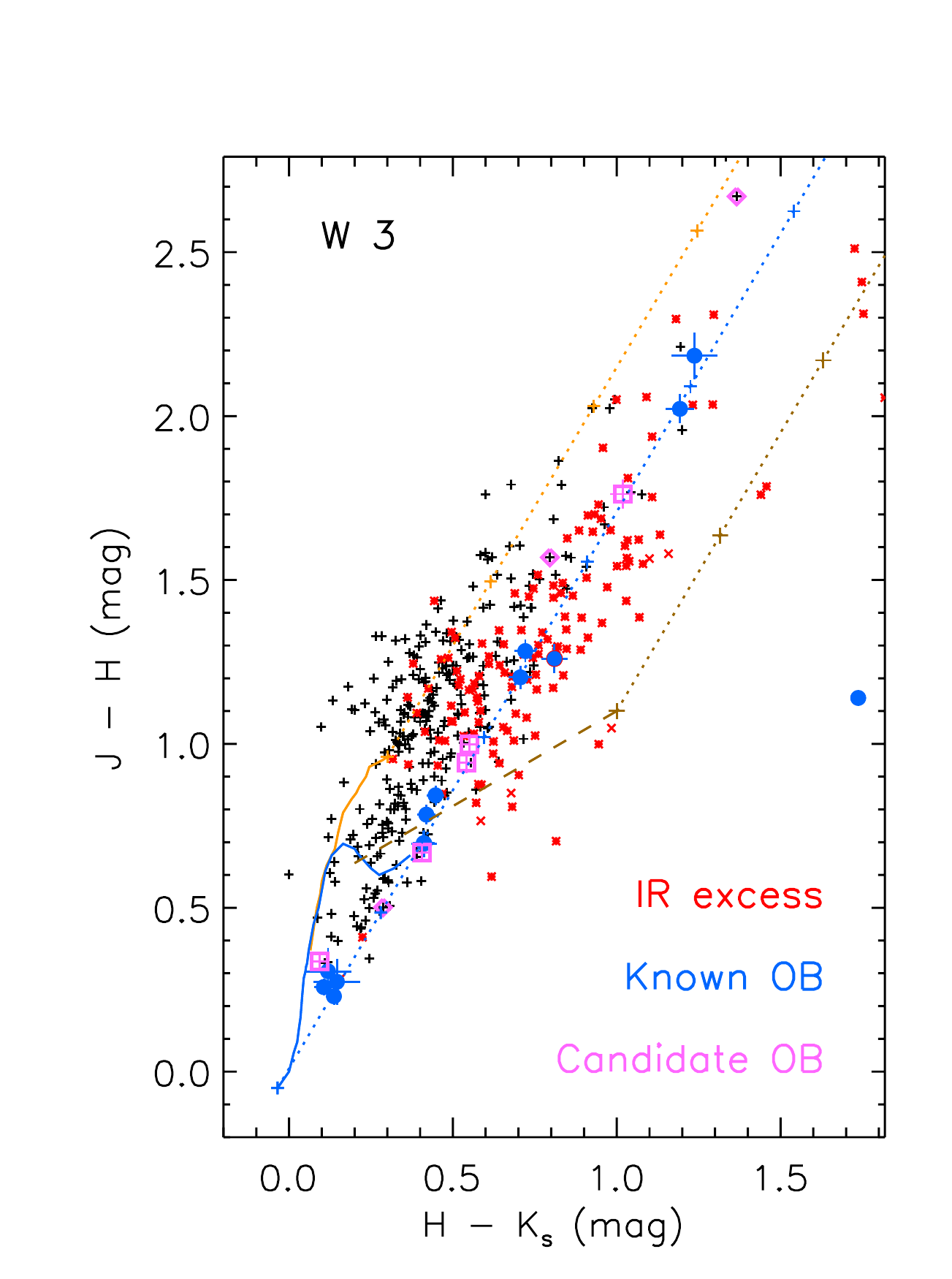}
\includegraphics[height=0.4\textheight]{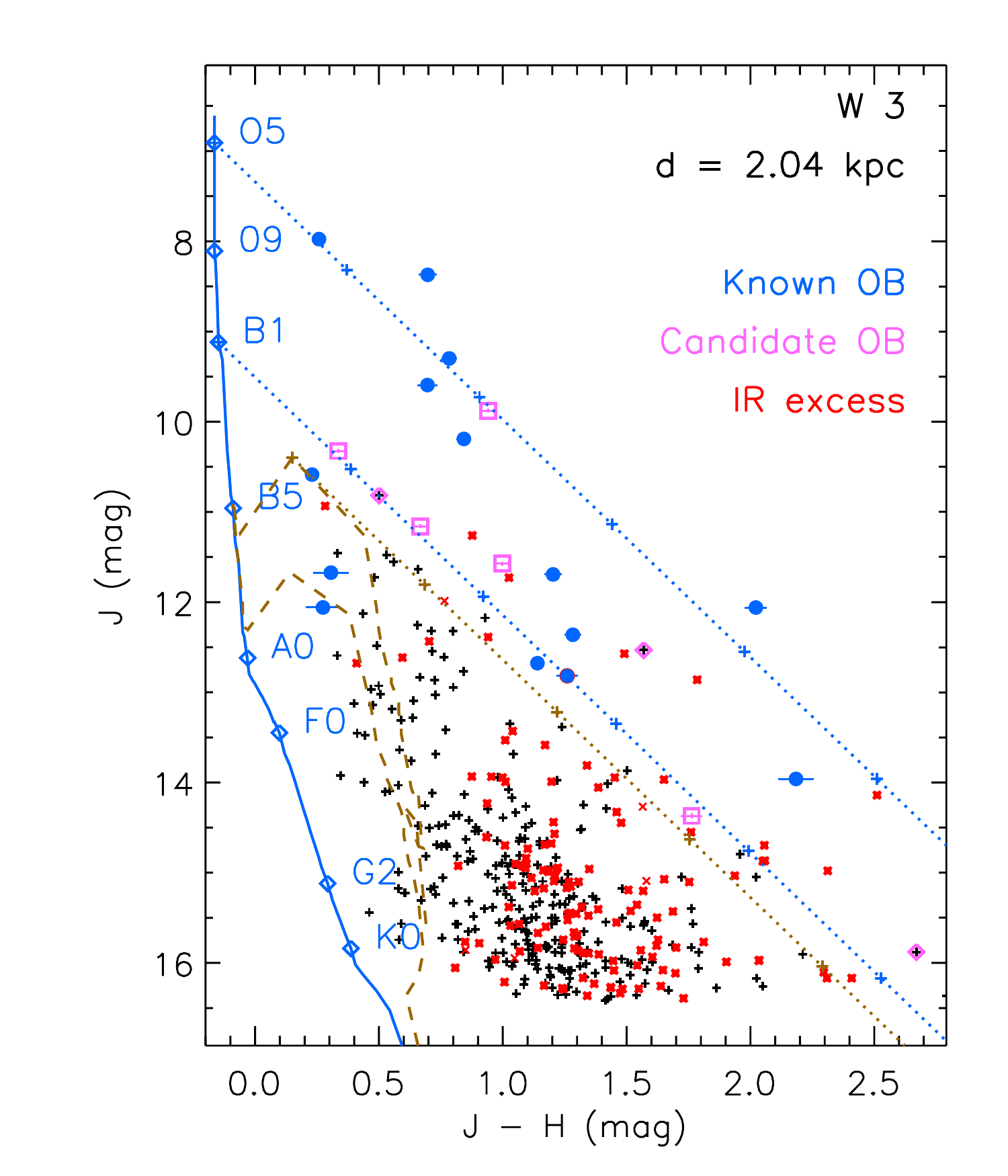}

~ \\
Fig. \ref{fig:each}({\it n}).--- W3.

\end{figure*}

\begin{figure*}[p]
\centering
\includegraphics[height=0.45\textheight]{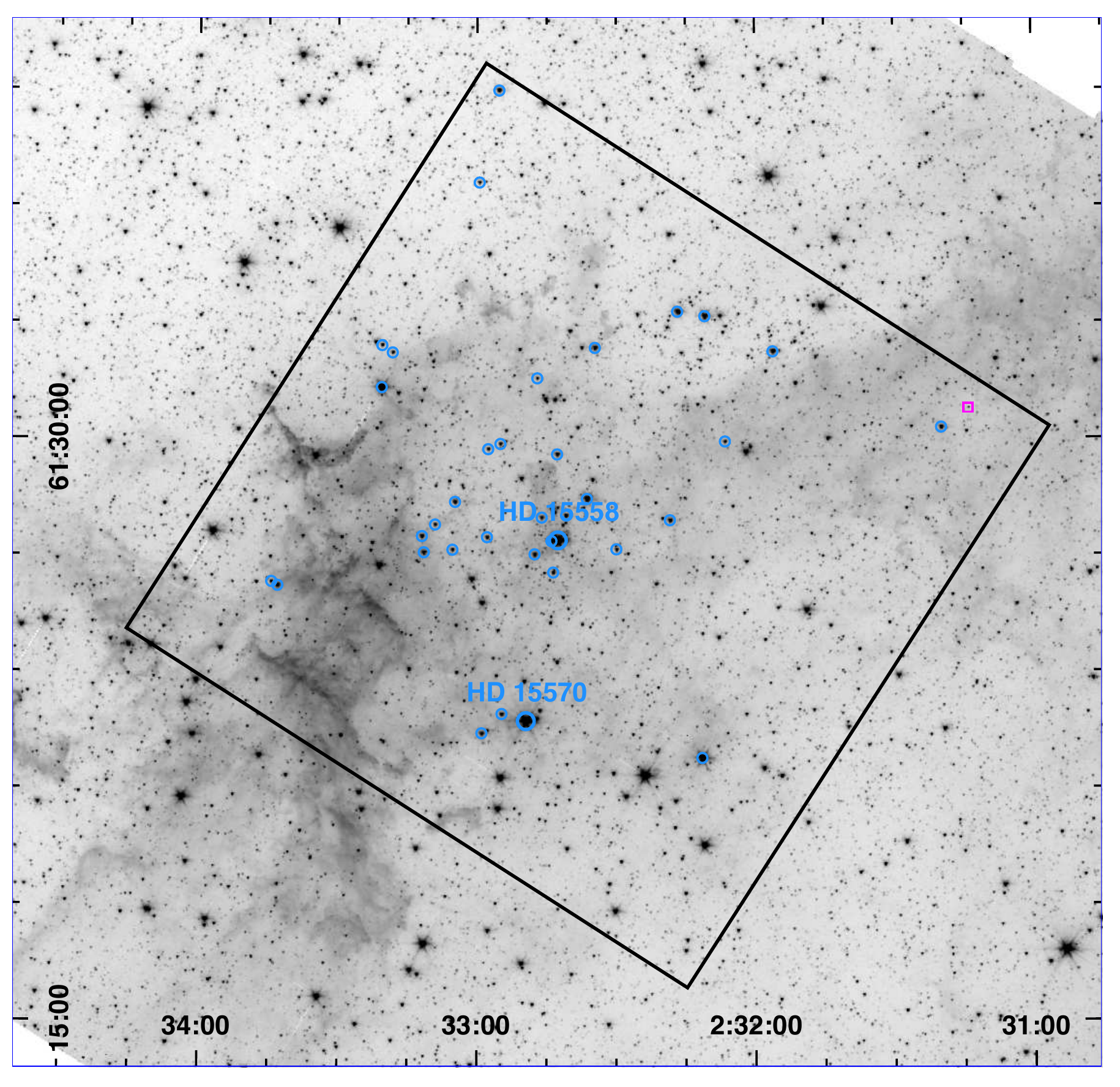}\\
\includegraphics[height=0.4\textheight]{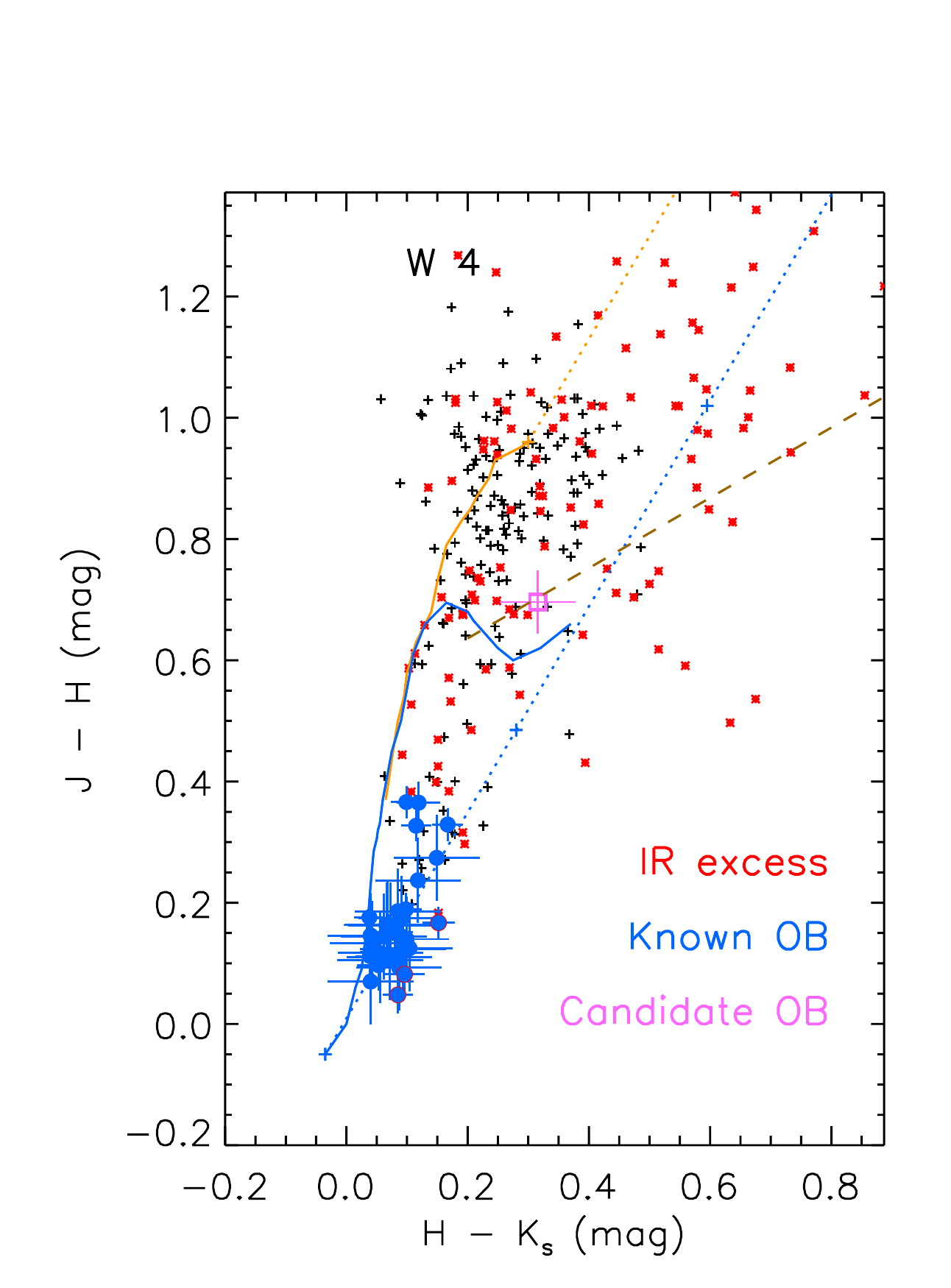}
\includegraphics[height=0.4\textheight]{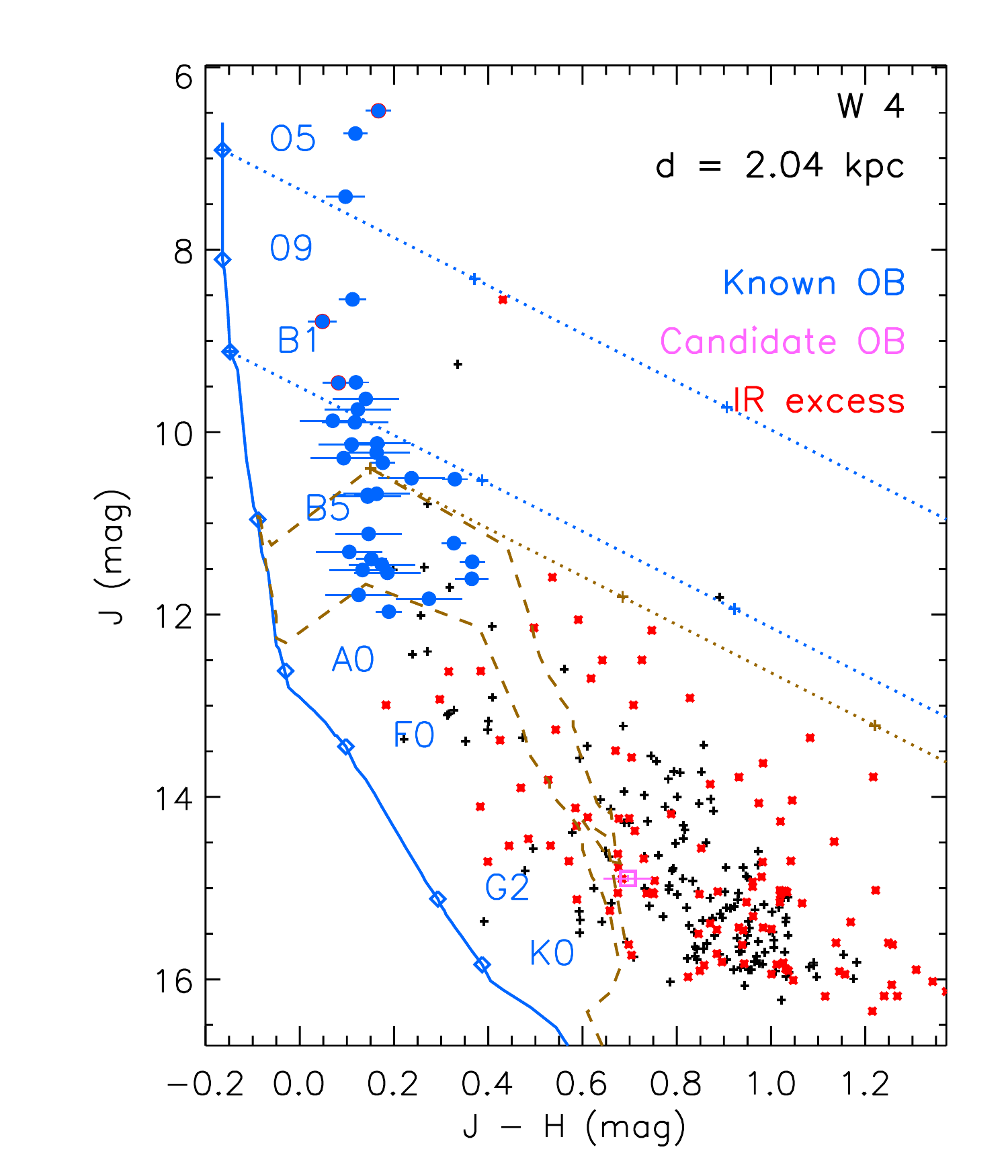}

~ \\
Fig. \ref{fig:each}({\it o}).--- W4.

\end{figure*}

\begin{figure*}[p]
\centering
\includegraphics[height=0.45\textheight]{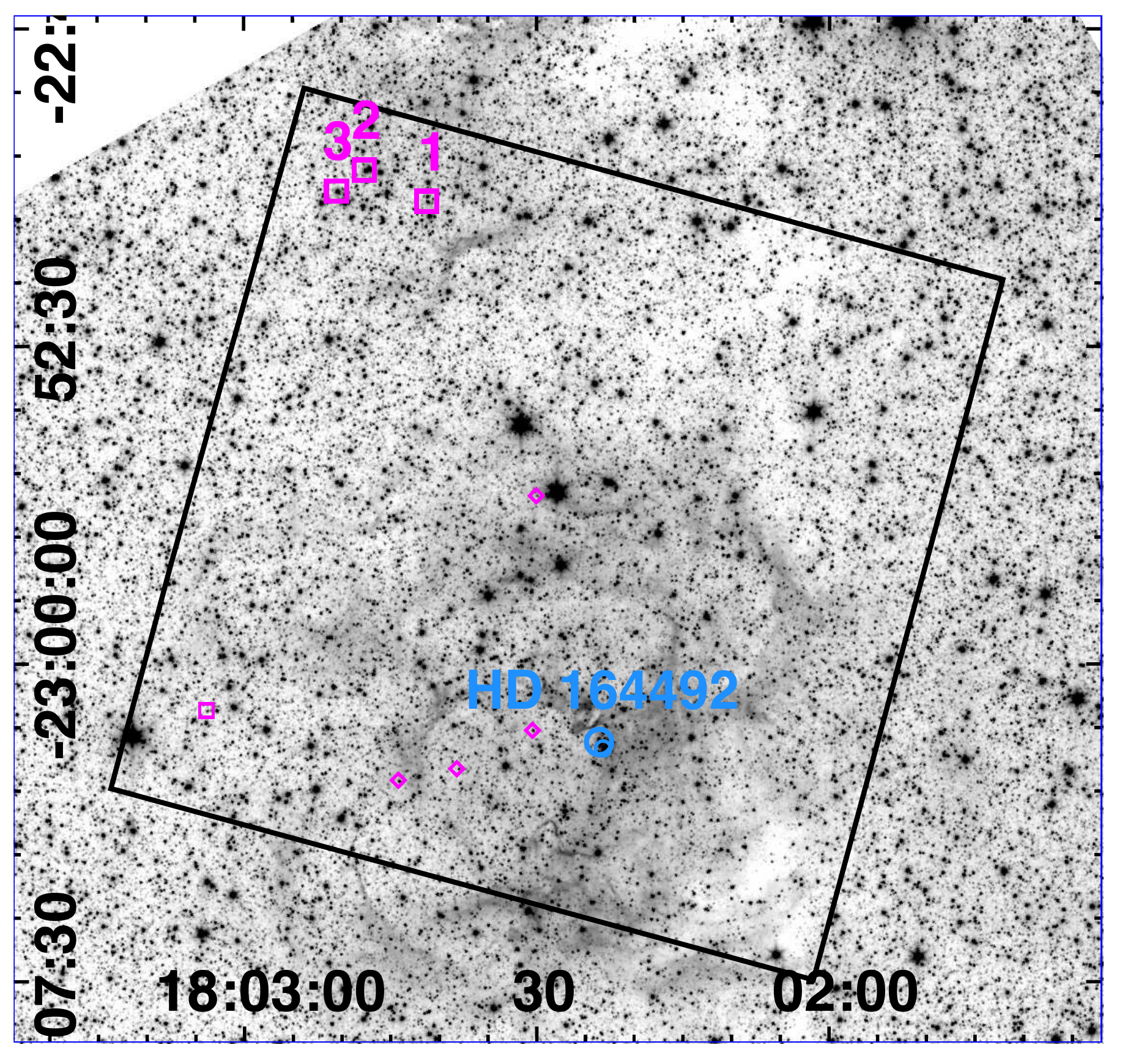}\\
\includegraphics[height=0.4\textheight]{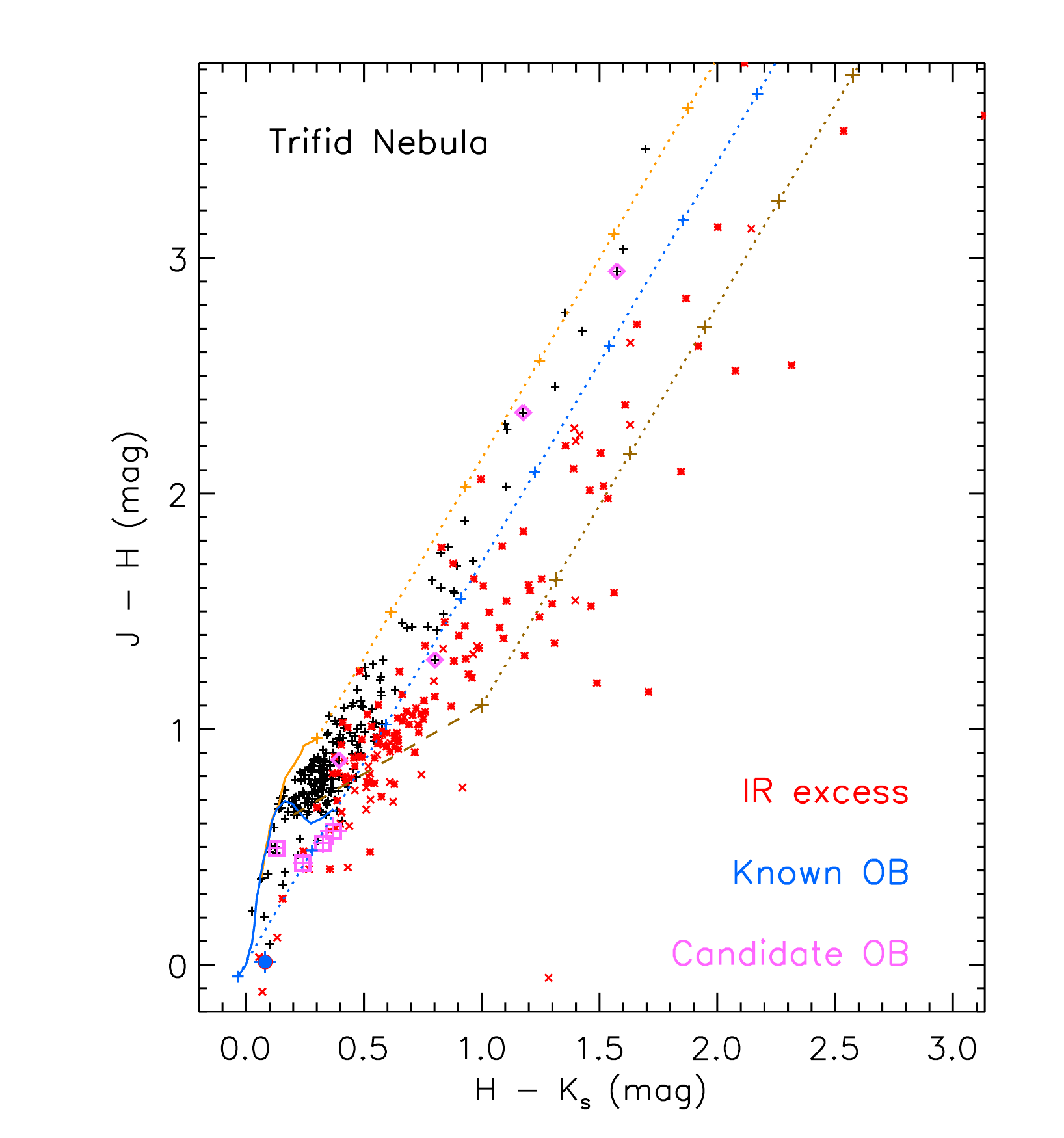}
\includegraphics[height=0.4\textheight]{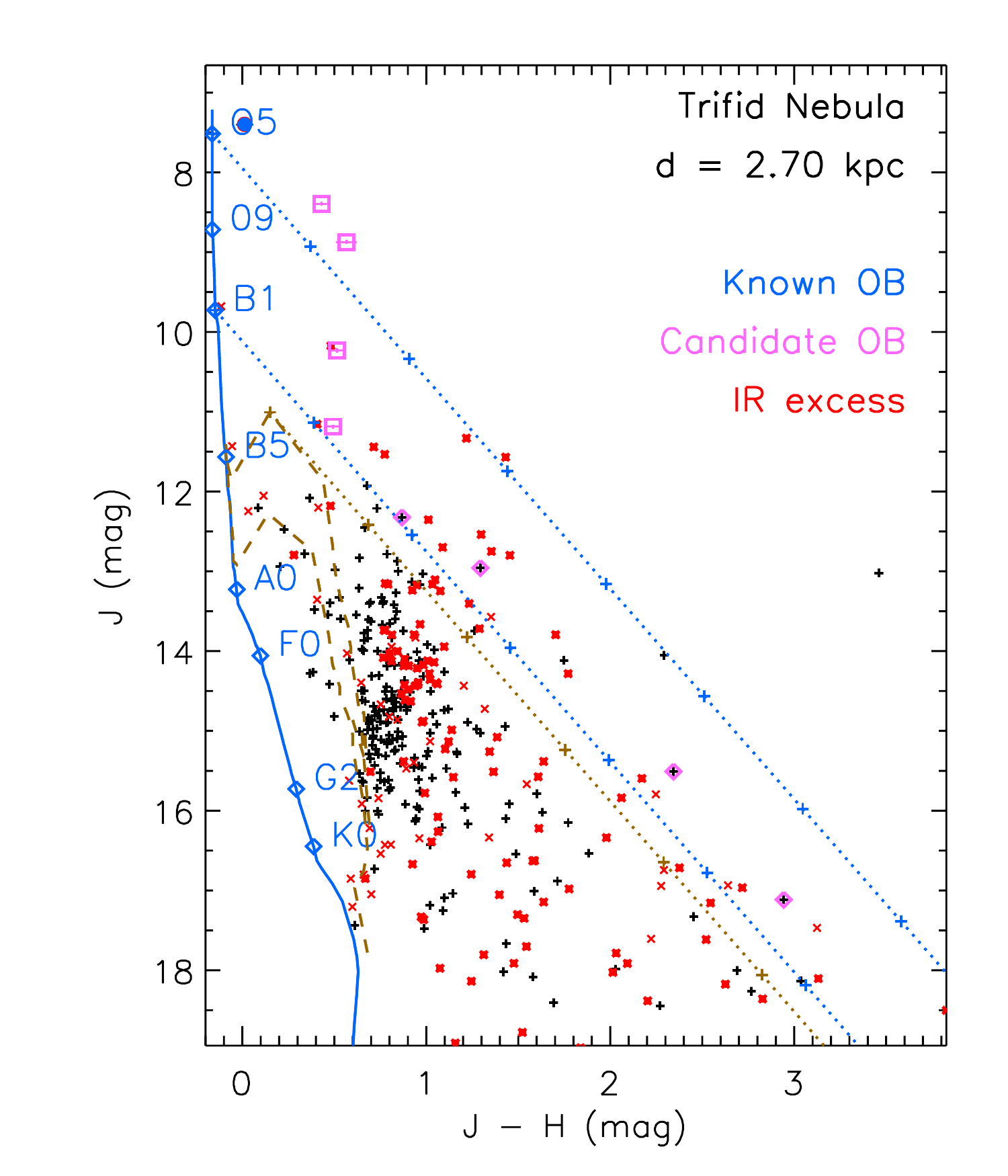}

~ \\
Fig. \ref{fig:each}({\it p}).--- The Trifid Nebula (M20).

\end{figure*}

\begin{figure*}[p]
\centering
\includegraphics[height=0.45\textheight]{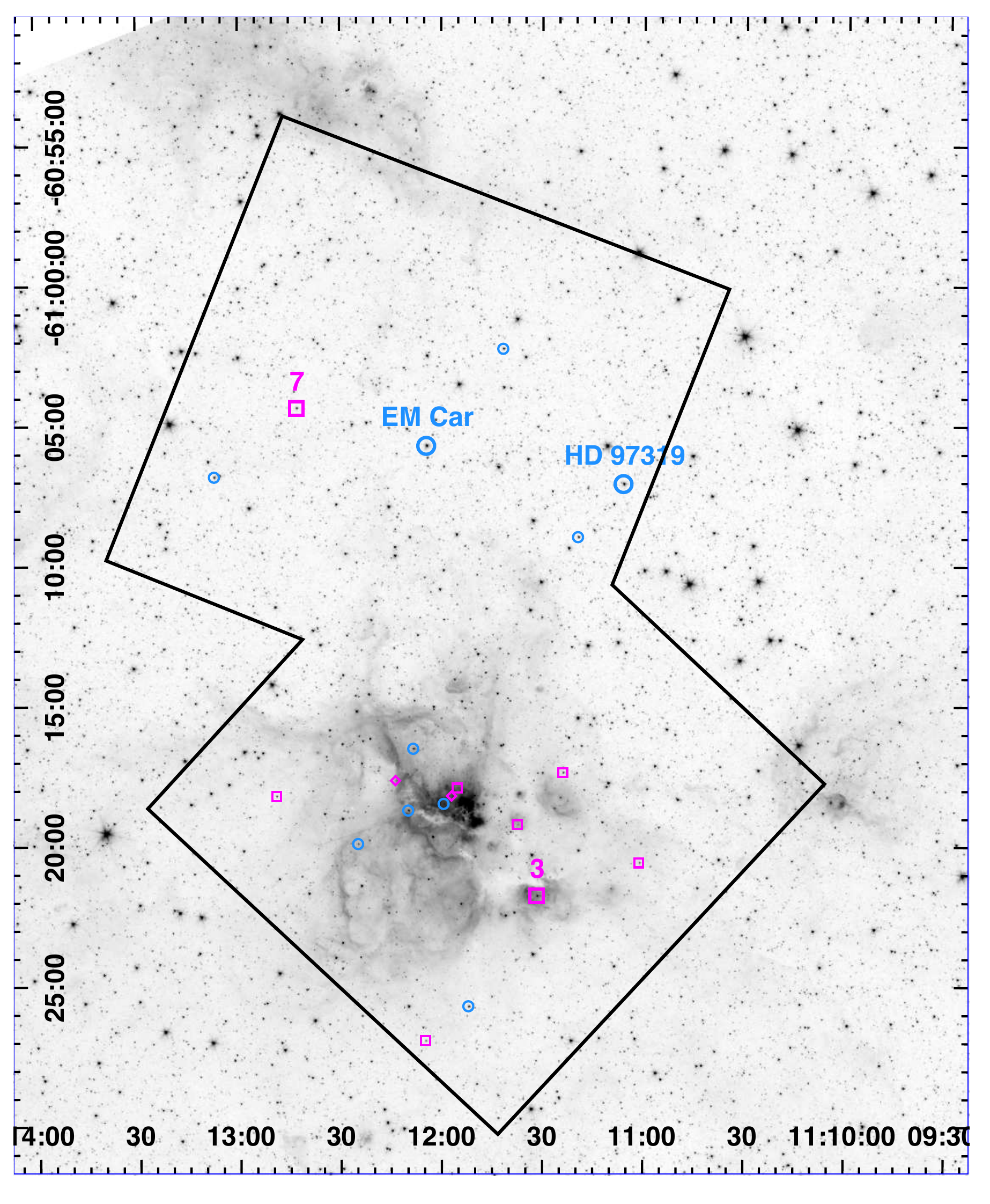}\\
\includegraphics[height=0.4\textheight]{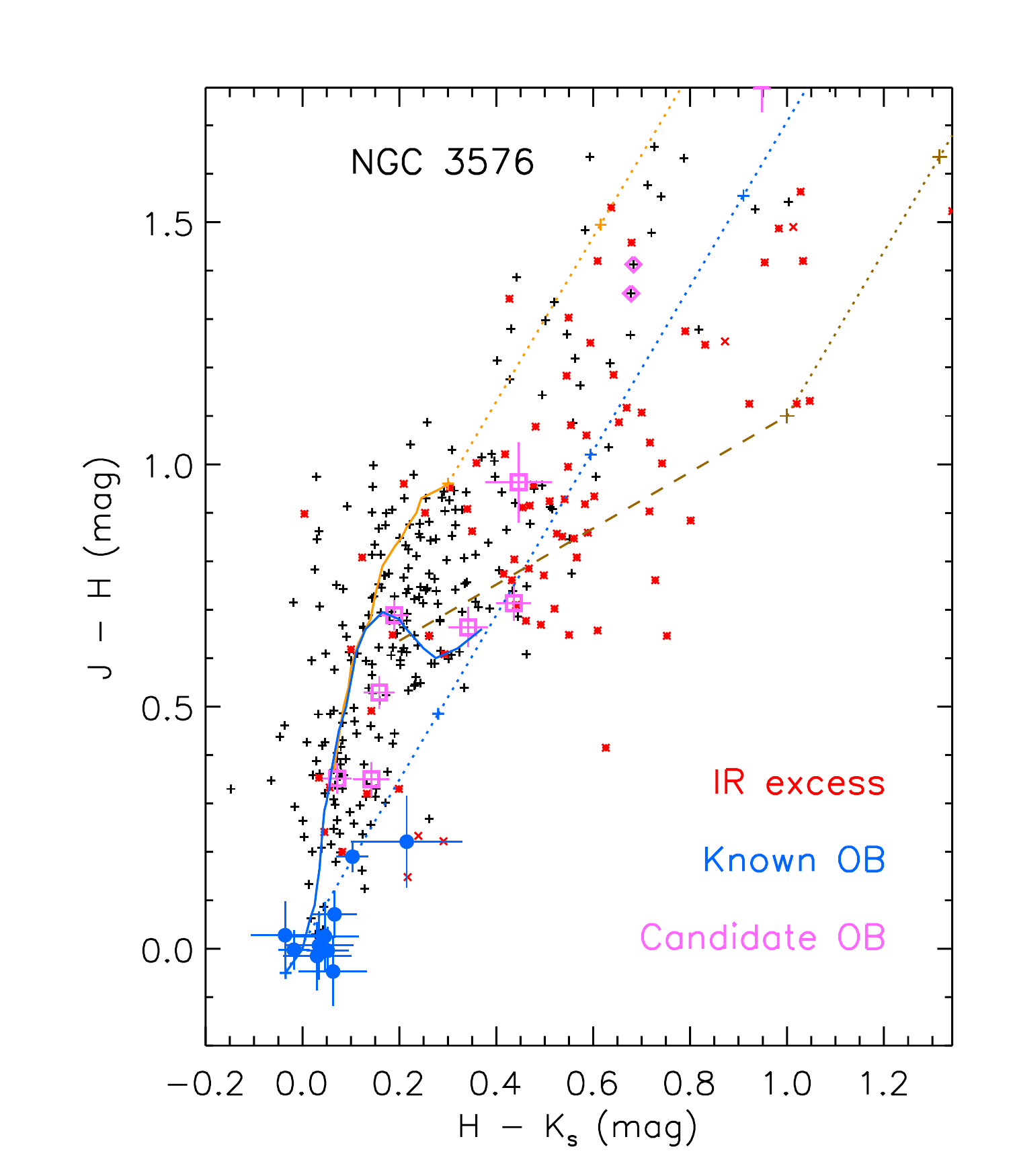}
\includegraphics[height=0.4\textheight]{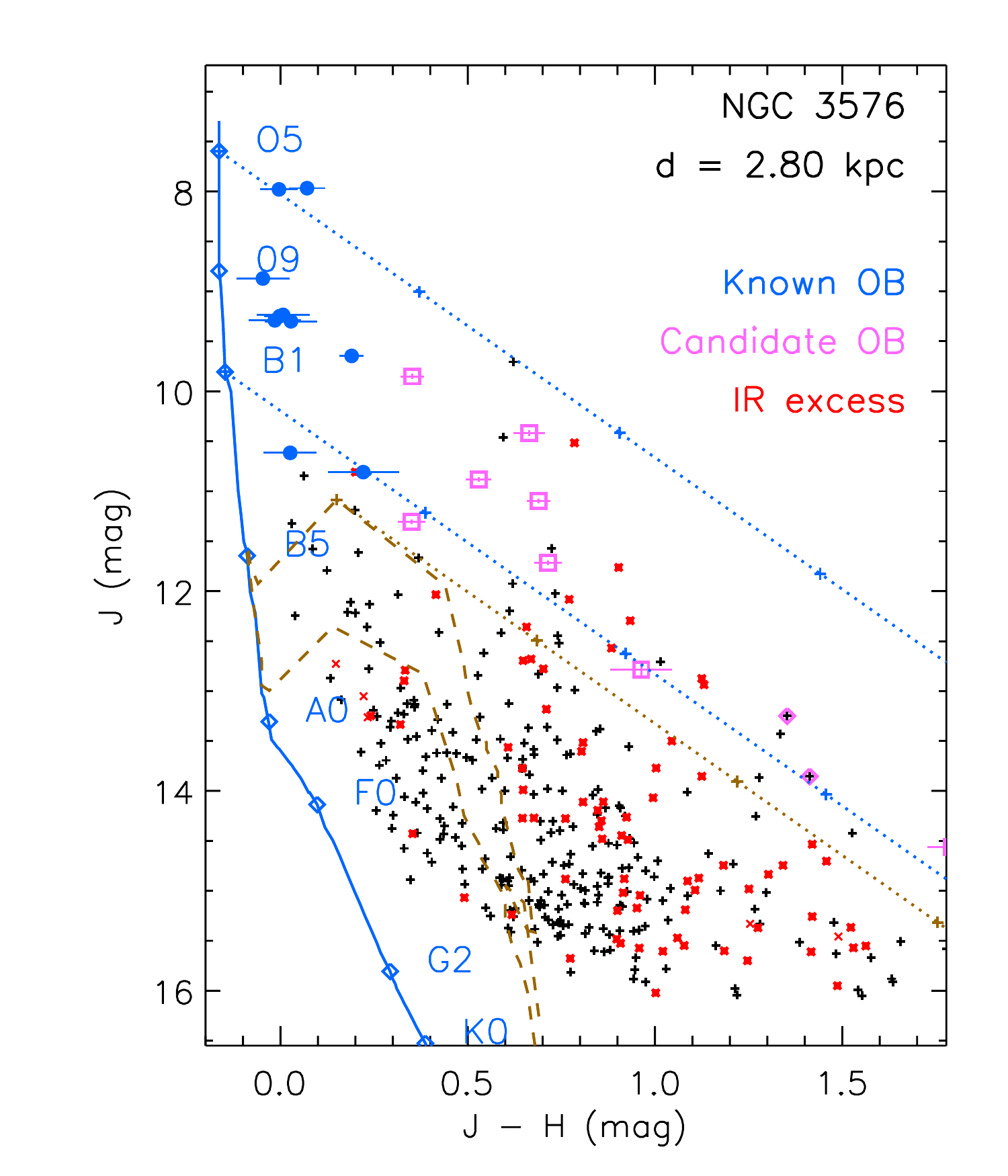}

~ \\
Fig. \ref{fig:each}({\it q}).--- NGC 3576.

\end{figure*}

\begin{figure*}[p]
\centering
\includegraphics[height=0.45\textheight]{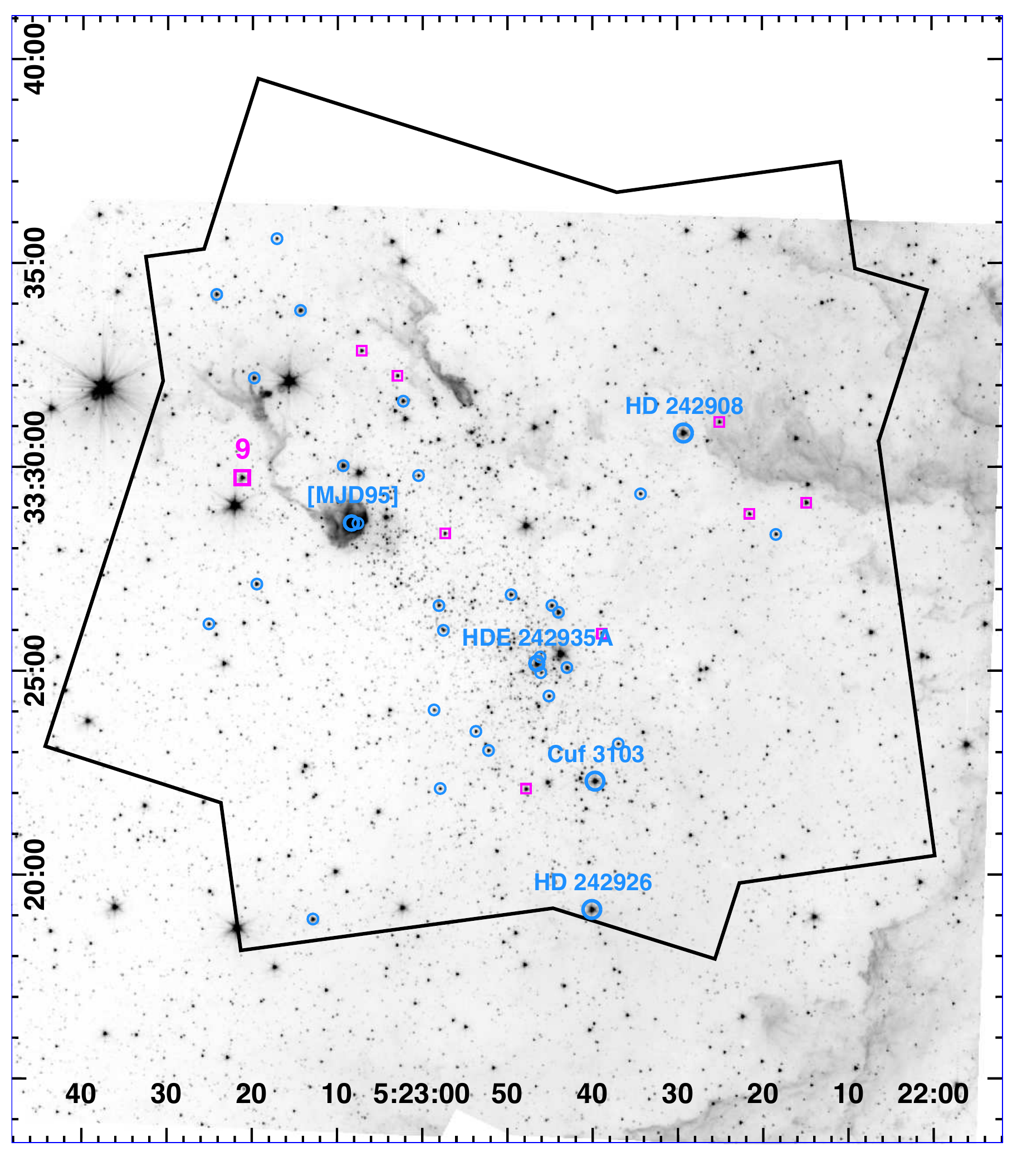}\\
\includegraphics[height=0.4\textheight]{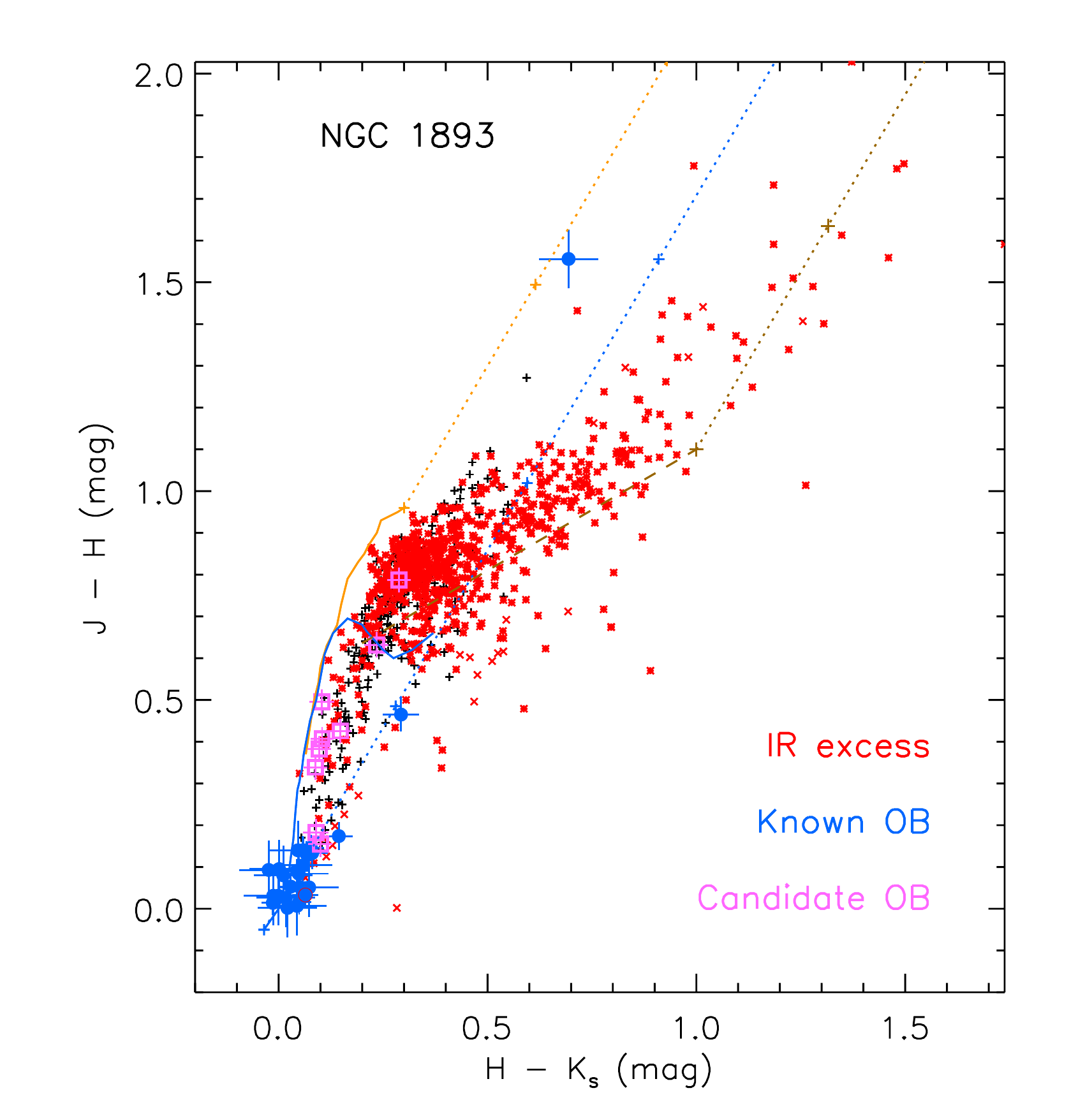}
\includegraphics[height=0.4\textheight]{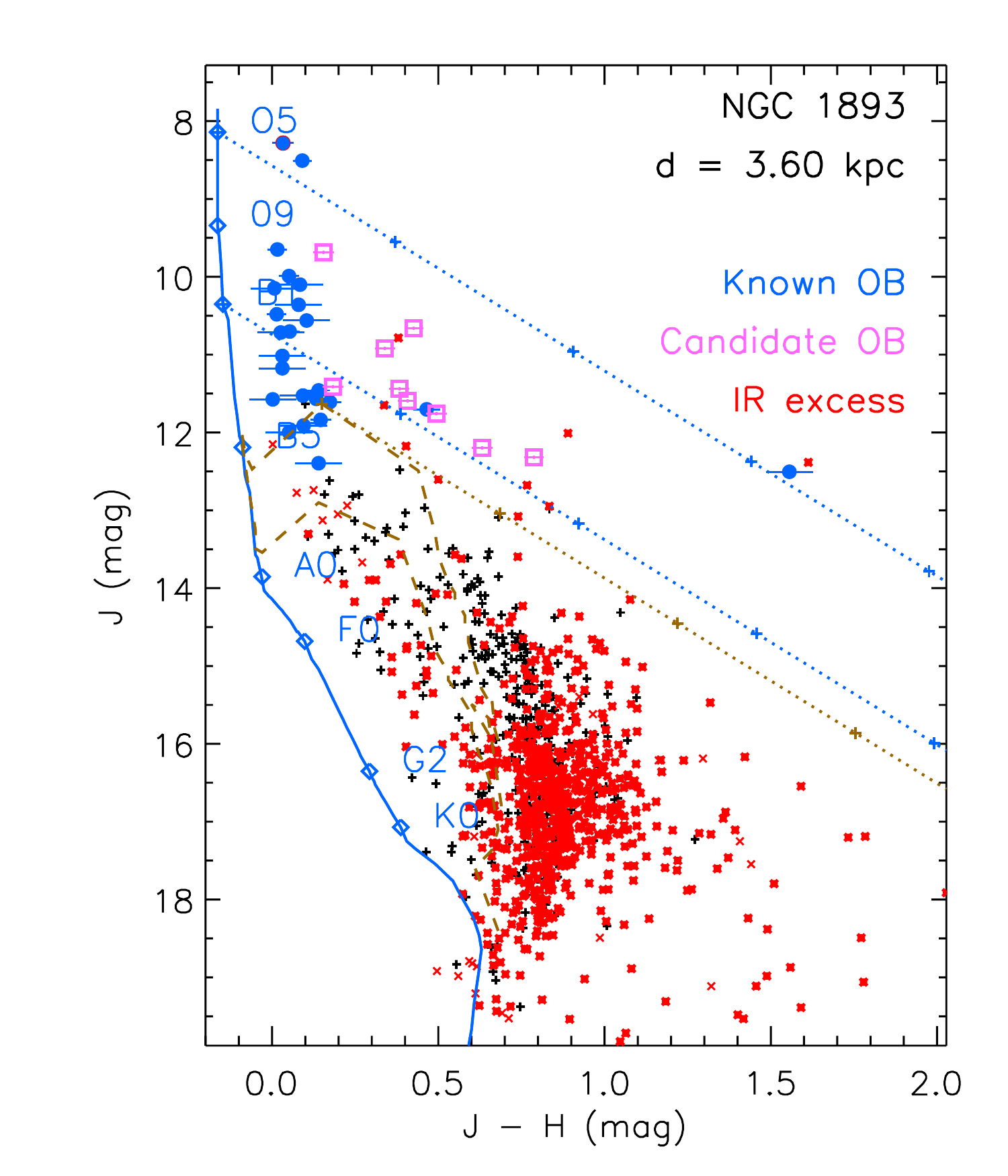}

~ \\
Fig. \ref{fig:each}({\it r}).--- NGC 1893.

\end{figure*}





\end{document}